\documentclass[12pt]{article}
\pdfoutput=1
\makeatletter\def\input@path{{../}}\makeatother
\usepackage{putex}
\usepackage[linktoc=page]{hyperref}
\hypersetup{
    pdfauthor={F. Ambrosino, A. Konechny, I. Runkel, and G. M. T. Watts},
    colorlinks=true,  breaklinks=true,citecolor=blue, urlcolor=blue,
    pdfstartview={FitH}, 
pdftitle={Defect Flows from functional relations in the asymptotic semiring}, % title
linkcolor=blue, % color of internal links (change box color with linkbordercolor)
citecolor=blue, % color of links to bibliography
filecolor=magenta, % color of file links}
}

\usepackage{putex}

\usepackage{geometry}

\usepackage{graphicx} 
\usepackage{amsfonts,amsmath,amssymb,amsthm}
\usepackage{graphics} 
\usepackage{cancel}
\usepackage{tikz}
    \usetikzlibrary{cd}
\usepackage[T1]{fontenc}
\usepackage[utf8]{inputenc}
\usepackage{enumerate}
\usepackage{array}

\usepackage[svgnames]{xcolor}
\usepackage[compress]{cite} % [1-3] instead of [1,2,3].

\usepackage{graphicx}

\usepackage[export]{adjustbox}

\usepackage{subcaption}

\usepackage[english]{babel}

\usepackage{titlesec}
\titleformat*{\subsection}{\bfseries\boldmath}
\titleformat*{\section}{\bfseries\boldmath}

\usetikzlibrary{arrows.meta,decorations.markings}

\newcommand\be{\begin{equation}}
\newcommand\ee{\end{equation}}

\newcommand{\myarrow}[1][.6]{postaction={decorate,decoration={
                markings,
                mark=at position .6 with {\arrow{Stealth}}}}}

\theoremstyle{definition}
\newtheorem{theorem}{Theorem}[section]
\newtheorem{lemma}[theorem]{Lemma}
\newtheorem{proposition}[theorem]{Proposition}
\newtheorem{corollary}[theorem]{Corollary}
\newtheorem{definition}[theorem]{Definition}
\newtheorem{remark}[theorem]{Remark}
\newtheorem{example}[theorem]{Example}

\numberwithin{equation}{section}

\usepackage{orcidlink}

\definecolor{hlorange}{RGB}{247,183,107}
\definecolor{hlgreen}{RGB}{130,230,130}
\newcommand{\Gbox}[1]{\vcenter{\hbox{\setlength{\fboxrule}{1.2pt}\setlength{\fboxsep}{5pt}%
  \fcolorbox{hlorange}{white}{$\displaystyle #1$}}}}
\newcommand{\Rbox}[1]{\vcenter{\hbox{\setlength{\fboxrule}{1.2pt}\setlength{\fboxsep}{5pt}%
  \fcolorbox{hlgreen}{white}{$\displaystyle #1$}}}}

\graphicspath{{Fig/}}

\usepackage{comment}
\usetikzlibrary{decorations.markings}
\usetikzlibrary{arrows.meta,decorations.markings,calc}

\newcommand\doi[2]        {\href{https://dx.doi.org/#1}{#2}}

\numberwithin{equation}{section}

\usepackage{twemojis}

\newbox\IceBox
\newbox\CastleBox
\newbox\AnchorBox
\newbox\LondonBox

\AtBeginDocument{%
  \setbox\IceBox=\hbox{\twemoji{snowflake}}%
  \setbox\CastleBox=\hbox{\twemoji{castle}}%
  \setbox\AnchorBox=\hbox{\twemoji{anchor}}%
  \setbox\LondonBox=\hbox{\twemoji{ferris wheel}}%
}

\def\ice{%
  \leavevmode\copy\IceBox
}

\def\castle{%
  \leavevmode\copy\CastleBox
}

\def\anchoricon{%
  \leavevmode\copy\AnchorBox
}

\def\london{%
  \leavevmode\copy\LondonBox
}

\newcommand{\Id}{\mathbf{1}}

\newcommand{\cX}{\mathcal{X}}

\newcommand{\princ}[1]{\left[#1\right]}

\date{}

\begin{document}
	
	\institution{Perimeter}{$^{\ice}$ \; 
Perimeter Institute for Theoretical Physics, \cr
     $\quad \;$  Waterloo, Ontario N2L 2Y5, Canada 
    }

 \institution{HeriotWatt}{$^{\castle}$\;\,\, Department of Mathematics, Heriot-Watt University,\cr
 $\quad \;\,$ Edinburgh EH14 4AS, United Kingdom\cr
 $\quad \;\,$ {\em and}\cr
$\quad \;\,$ Maxwell Institute for Mathematical Sciences,\cr
$\quad \;\,$ Edinburgh, United Kingdom}

\institution{HamMath}{$^{\anchoricon}$\;  \,Fachbereich Mathematik,
 Universit\"at Hamburg, \cr
 $\quad \;$ Bundesstra\ss e\ 55,
 20146 Hamburg, Germany }

 \institution{Kings}{$^{\london}$\;\; Department of Mathematics,
 King’s College London,\cr
 $\quad \;\,$ Strand, London WC2R 2LS, United Kingdom}

\title{\textbf{Defect flows from functional relations\\ 
in the asymptotic semiring}
\vspace{1cm}}

\authors{Federico Ambrosino$^{\ice}$\orcidlink{0000-0002-5617-5446}, 
Anatoly Konechny$^{\castle}$\orcidlink{0000-0003-3261-8262}, 
Ingo Runkel$^{\anchoricon}$\orcidlink{0000-0002-0496-3096} and\\  
G\'erard M.\ T.\ Watts$^{\london}$\orcidlink{0000-0002-9066-2838}}

\abstract{
We study perturbations of topological defects by a relevant holomorphic defect field in two-dimensional conformal field theory. 
This produces translation invariant defects and these necessarily flow to topological defects in the infrared. Translation invariant defects form a monoidal category which extends the fusion category of topological defects. We argue that the renormalisation group flow defines a semiring homomorphism from the semiring of translation invariant defects to an ``asymptotic semiring'' which we construct from topological defects. The semiring homomorphism preserves functional relations, and we illustrate in the example of Virasoro minimal models that the resulting relations in the asymptotic semiring constrain the defect flow endpoints. 
The companion paper \cite{phys-paper} contains a detailed study of such defect flows in several minimal models, both numerically and analytically.

}

    \maketitle

\newpage
    
    \setcounter{tocdepth}{2}	

    \tableofcontents

\section{Introduction}
Topological defects in quantum field theory are a far-reaching generalisation of symmetries described by groups, allowing for non-invertible defects and including higher categorical data from defects of higher codimension (see e.g.\ \cite{Carqueville:2023jhb} for a review with references to the original literature). 
Translation invariant defects in turn are a generalisation of topological defects in that they include defects which cannot be arbitrarily deformed, but only rigidly translated.
They still allow for the definition of a non-singular defect fusion, and so translation invariant defects also carry higher categorical structure \cite{Manolopoulos:2009np,Buecher:2012ma,Ambrosino:2025pjj}.
From this perspective, it seems natural to employ translation invariant defects as a further tool to understand properties of quantum field theories.

\medskip

In this paper, we continue to develop the theory and applications of translation invariant defects (TIDs) in two-dimensional quantum field theories. Such defects have been employed before in the study of RG flows 
\cite{Bazhanov:1994ft,Fioravanti:1995cq,Bazhanov:1996dr,Bazhanov:1996aq,Konik:1997gx,  Bazhanov:2001xm,Bachas:2004sy,  Runkel:2007wd,Manolopoulos:2009np,Runkel:2010ym,Buecher:2012ma,Gaiotto:2020fdr,  Ambrosino:2025myh,Ambrosino:2025pjj}.
Specifically, we focus on renormalisation group flows of TIDs within a given two-dimensional conformal field theory (CFT). These necessarily start and end at a topological line defect \cite{Ambrosino:2025myh} (provided the start- and endpoint exist at all). The question we are asking in this paper is: 
\begin{quote}
Given a topological defect and a relevant perturbation to a TID, can we use the additional categorical structure on TIDs to constrain or even predict the topological defect at the end of the flow?
\end{quote}
In brief, our approach is to compute the fusion rules of TIDs from the tensor product in the corresponding category, and to argue that the RG flow defines a (semi)ring homomorphism from the fusion rule semiring of TIDs to what we call the ``asymptotic semiring'', and which is built from topological defects. Functional relations in the fusion rules of TIDs then have to hold exactly in the asymptotic semiring as well, which strongly constrains the flow endpoints.

Functional relations have been used extensively in the context of integrable models and the thermodynamic Bethe ansatz, both on the lattice and in the continuum, see e.g.\ the reviews \cite{Dorey:2007zx,Kuniba:2010ir}. Our approach is slightly different as we do not first derive integral equations, solve these, and then study the limiting behaviour of the solutions. Instead, we directly derive algebraic relations which have to hold in the infrared and which we then try to solve.

\medskip

Let us review the results of this paper in more detail.
The situation we consider specifically is a two-dimensional rational conformal field theory with diagonal modular invariant. Neither rationality nor choosing the diagonal modular invariant is essential, but it simplifies the treatment. In this case, the category $\mathcal{C}$ of representations of the chiral algebra is a braided fusion category (in fact a modular fusion category), and the same category $\mathcal{C}$ describes the topological line defects which commute with the chiral symmetry of the CFT.

\smallskip

\noindent
\textbf{Holomorphically perturbed defects.}
The TIDs we investigate are obtained by perturbing a topological defect $T$ with a relevant holomorphic defect field $\psi$ with conformal weights $(h,0)$, $h<1$. 
We note that not all TIDs are obtained in this way. For example one can consider perturbations by marginal holomorphic currents (see \cite{Bachas:2004sy}) or by irrelevant currents (see \cite[Sec.\,7.5]{Ashok:2026cyi}; they also appear implicitly e.g.\ in \cite[Eqn.\,(3.20)]{Dijkgraaf:1996iy}). Of course, one can also take anti-holomorphic defect fields, or both (cf.\ the discussion below (13) in \cite{Ambrosino:2025pjj}).
Coming back to the relevant holomorphic defect field $\psi$,
let $X \in \mathcal{C}$ be the representation of the chiral algebra in which $\psi$ transforms. Throughout the construction, we fix $\psi$ and $X$. The perturbation is then determined by an element $\eta$ in the multiplicity space of the field $\psi$ on the defect $T$. Depending on $T$, this multiplicity space can be zero-, one-, or higher-dimensional. For example, if $T = \mathbf{1}$ is the trivial defect, the multiplicity space is zero (unless $\psi$ is the identity field), while for $T = \mathbf{1} \oplus X$, it is at least two-dimensional (assuming $X$ is self-dual). This is most transparent in the 3d\,TFT description of 2d\,CFT correlators \cite{Fuchs:2002cm, Frohlich:2006ch, Runkel:2007wd, Ambrosino:2025pjj}, and we briefly review the relevant parts in Section~\ref{sec:chiral-TFT-defects}. That such a perturbation results in a TID has been known for some time \cite{Konik:1997gx,Bachas:2004sy,Runkel:2007wd} and is explained in the present setting e.g.\ in \cite[App.\,A.2]{Buecher:2012ma} (set $\Phi,\psi_-=0$ there). 

The categorical description of holomorphically perturbed defects used below relies on the assumption that the iterated integrals of $\psi$ in the perturbative expansion converge without the need for regularisation. 
Indeed, if this is not the case qualitatively different behaviour can arise, see e.g.\ \cite{Gaiotto:2020fdr} (and the end of Section 4 of \cite{Ambrosino:2025pjj} for a comparison) and \cite{Ambrosino:2026kfu}.
If the OPE of $\psi$ with itself contains the identity field, convergence requires the weight $h$ of $\psi$ to satisfy
\begin{equation}\label{eq:intro-h-constraint}
    h < \frac12 \ ,
\end{equation}
rather than just $h<1$ for a relevant perturbation. However, this condition is in general not sufficient, e.g.\ for non-unitary models.

The category $\mathcal{C}_X$ of such TIDs obtained by holomorphically perturbing topological
defects has a very clean algebraic description. The objects of $\mathcal{C}_X$ are pairs $(T,\eta)$ where $\eta : X \otimes T \to T$ describes the element in the multiplicity space. Morphisms in $\mathcal{C}_X$ are just morphisms in $\mathcal{C}$ compatible with $\eta$. This category was introduced in \cite{Manolopoulos:2009np} and we describe it in detail in Section~\ref{sec:cat-of-pert-def}. 
For practical purposes and computer implementation, in Section~\ref{sec:quiver} we give an explicit description of $\mathcal{C}_X$ in terms of representations of a certain quiver, similar to \cite{schaumann2023fusionquivers}. This in particular provides an explicit formula for the tensor product in $\mathcal{C}_X$, that is, for the fusion of holomorphically perturbed defects. 

Technically, since $\mathcal{C}_X$ is not semisimple, the appropriate notion of fusion rules
is the Grothendieck ring $\mathrm{Gr}(\mathcal{C}_X)$. It has a $\mathbb{Z}$-basis given by (isomorphism classes of) simple objects in $\mathcal{C}_X$. Below we instead need the fusion 
rule semiring 
\begin{equation}
    \mathrm{Gr}(\mathcal{C}_X)_{\ge 0}\ ,
\end{equation}
where one restricts to $\mathbb{Z}_{\ge 0}$-linear combinations of these basis elements, see Example~\ref{ex:semirings}.

\smallskip

\noindent
\textbf{Lee-Yang example, part 1.}
An example which we carry along in the development is the Lee-Yang CFT, that is, the minimal model $M(2,5)$ of central charge $c=-\frac{22}5$. This example is also studied in detail in the companion paper \cite{phys-paper}.
Let $\tau$ be the irreducible Virasoro representation of lowest weight $h=-\frac15$, and write $\mathcal{D}_\tau$ for the operator on the bulk state space corresponding to the topological line defect labelled $\tau$. The defect $\tau$ carries a unique-up-to-scalar non-trivial relevant  quasi-primary holomorphic field $\psi$ (of weight $h=-\frac15$), and we denote by $\mathcal{D}_\tau(\lambda)$ the TID obtained by perturbing $\tau$ with $\lambda\psi$,
\begin{equation}
    \mathcal{D}_\tau(\lambda) = \mathcal{D}_\tau \exp\Bigl( \lambda \, L^{h-1} \! \int_0^L \!\! \psi(x)\,dx \Bigr) \ ,
\end{equation}
where $L$ denotes the circumference of the cylinder on which we place the perturbed defect, and the factor $L^{h-1}$ is included to make $\lambda$ dimensionless.
The fusion of TIDs gives the functional relation (see \cite{Runkel:2007wd} and Example~\ref{ex:LY-fun-rel})
\begin{equation}\label{eq:LY-fun-rel_intro}
    \mathcal{D}_\tau(e^{2 \pi i / 5} \lambda) \, \mathcal{D}_\tau(e^{-2 \pi i / 5} \lambda) \,=\,
    \mathrm{id} + \mathcal{D}_\tau(\lambda)   \ ,
    \qquad \lambda \in \mathbb{C} \ ,
\end{equation}
among the corresponding defect operators. Note that for $\lambda=0$, we get the unperturbed, topological defect $\mathcal{D}_\tau = \mathcal{D}_\tau(0)$, and the above formula reduces to the fusion rule $\mathcal{D}_\tau \mathcal{D}_\tau = \mathrm{id} + \mathcal{D}_\tau$ for the operator of the topological defect $\tau$.

\smallskip

\noindent
\textbf{\boldmath $\alpha$-asymptotics.}
To proceed, we need to make an assumption about the large-coupling behaviour of holomorphically perturbed defects. We already know that they flow to topological defects (at least away from anti-Stokes rays, see below). We assume that the leading contribution comes from a free energy density on the defect, or, equivalently, from a perturbation by the identity field. For dimensional reasons, if the perturbing field $\psi$ on a topological defect $T$ has dimension $(h,0)$, this gives, for $\lambda = r e^{i \theta}$,
\begin{equation}\label{eq:intro-asymp-ansatz}
    \mathcal{D}_T(\lambda \psi) ~\underset{r \to \infty}{\sim}~ \mathcal{D}^\infty(\theta) \exp\!\big( u(\theta) r^\alpha \big) \ , \quad \alpha = \frac{1}{1-h} \ .
\end{equation}
Crucially, both the topological defect operator $\mathcal{D}^\infty(\theta)$ at the endpoint of the flow, and the precise asymptotic $u(\theta) \in \mathbb{C}$ do depend on the direction $e^{i\theta}$ in the complex plane along which one approaches infinity (as well as on the initial topological defect $T$ which we do not make explicit in the notation). 

We say that the $\psi$-perturbed defects have \textit{$\alpha$-asymptotics} if \eqref{eq:intro-asymp-ansatz} holds for each topological defect $T$ outside of a finite set $S_T$ of angles $\theta$ (the \textit{anti-Stokes rays}), and if 
on an open interval between two angles in $S_T$,
the function $u(\theta)$ is of the form $(\text{const}) e^{i \alpha \theta}$ and $\mathcal{D}^\infty(\theta)$ is constant.
The precise formulation is given in Definition~\ref{def:alpha-asymptotics}.

\smallskip

\noindent
\textbf{Asymptotic semiring.}
To model the change of the asymptotic behaviour \eqref{eq:intro-asymp-ansatz} under addition and composition, we introduce the \textit{asymptotic semiring} $\mathcal{A}$.
Its elements are pairs of functions $[\mathcal{D}^\infty(\theta),a(\theta)]$ -- where $\mathcal{D}^\infty(\theta)$ is a topological defect operator of the CFT and $a(\theta) \in \mathbb{R}$ -- modulo the relation that two such pairs are considered equal if they differ in only finitely many angles $\theta$. Multiplication and addition in $\mathcal{A}$ are defined as 
\begin{equation}
 [\mathcal{D}^\infty_1(\theta),a_1(\theta)] \cdot [\mathcal{D}^\infty_2(\theta),a_2(\theta)] = [\mathcal{D}^\infty_1(\theta)\mathcal{D}^\infty_2(\theta),a_1(\theta)+a_2(\theta)]
\end{equation}
and
\begin{equation}    [\mathcal{D}^\infty_1(\theta),a_1(\theta)]+[\mathcal{D}^\infty_2(\theta),a_2(\theta)] = \begin{cases}
        [\mathcal{D}^\infty_1(\theta),a_1(\theta)] &; a_1(\theta)>a_2(\theta) \\
        [\mathcal{D}^\infty_1(\theta) + \mathcal{D}^\infty_2(\theta),a_1(\theta)] &; a_1(\theta)=a_2(\theta) \\
        [\mathcal{D}^\infty_2(\theta),a_2(\theta)] &; a_1(\theta)<a_2(\theta) 
    \end{cases}
\end{equation}
The detailed description is given in Definition~\ref{def:asymptotic-semiring}. Note that in the second entry, multiplication becomes addition and addition becomes a maximum. This is known as the max-plus semiring, see Example~\ref{ex:semirings}.

If we represent the asymptotic behaviour in \eqref{eq:intro-asymp-ansatz} by the element
\begin{equation}\label{eq:intro-semiringhom}
    [\mathcal{D}^\infty(\theta), \mathrm{Re}(u(\theta))] \ ,
\end{equation}
then, for example, the above addition captures the leading behaviour of the sum of perturbed defects $\mathcal{D}_1(\lambda \psi) + \mathcal{D}_2(\lambda \psi)$: The exponential with the larger real part will dominate at large~$r$.

The key construction of this paper is the following theorem, whose detailed formulation is quite technical (Theorem~\ref{thm:semiring-hom}), but for the sake of the introduction we present it as:

\begin{theorem}\label{thm:intro-main}
If the $\psi$-perturbed topological defects have $\alpha$-asymptotics, the assignment \eqref{eq:intro-semiringhom} defines a semiring homomorphism
\begin{equation}
    \Lambda : \mathrm{Gr}(\mathcal{C}_X)_{\ge 0} \to \mathcal{A} \ .
\end{equation}
\end{theorem}

Hence, by constraining or classifying such semiring homomorphisms, we can constrain or even determine RG flow endpoints of holomorphically perturbed topological defects.
Below we will illustrate the power of this approach in the example of minimal models.

\smallskip

\noindent
\textbf{Lee-Yang example, part 2.}
The numerical truncated conformal space analysis from \cite{phys-paper} gives us (in particular) the following two inputs: 1) in the normalisation of defect fields we use, for $\theta=0$ the asymptotic behaviour in \eqref{eq:intro-asymp-ansatz} satisfies $u(0) \in \mathbb{R}_{>0}$, and 2) there is a single anti-Stokes ray along the negative real axis. We have $\alpha = \frac{5}{6}$ and by our assumption of $\alpha$-asymptotics, together with 2), we get that the image of $\mathcal{D}_\tau(\lambda)$ in the asymptotic semiring $\mathcal{A}$ is $[\mathcal{D}^\infty,a \cos(\frac{5}{6} \theta)]$ for some $a>0$ and $|\theta|<\pi$ (the anti-Stokes ray is at $\theta = \pm \pi$). In $\mathcal{A}$, the functional relation \eqref{eq:LY-fun-rel_intro} becomes 
\begin{equation}\label{eq:LY-asym-fun-rel_intro}
\big[\mathcal{D}^\infty \mathcal{D}^\infty , a \cos\big(\tfrac{5}{6} [\theta + \tfrac{2\pi}{5}]\big)+a \cos\big(\tfrac{5}{6} [\theta - \tfrac{2\pi}{5}]\big) \big]
    =
    \mathrm{id} + 
\big[\mathcal{D}^\infty ,a \cos\big(\tfrac{5}{6} \theta\big)\big]
\ ,
\end{equation}
where $[\varphi]$ is the representative of $\varphi$ modulo $2 \pi$ in the interval $(-\pi,\pi]$.
Evaluating the relation for $|\theta| < \frac{3 \pi}{5}$ (so that the cosine on the right is positive) gives $\mathcal{D}^\infty \mathcal{D}^\infty = \mathcal{D}^\infty$, while outside of that range we get $\mathcal{D}^\infty \mathcal{D}^\infty = \mathrm{id}$ (see Section~\ref{sec:2q-solution} for more details). Altogether, this implies $\mathcal{D}^\infty = \mathrm{id}$, fixing the flow endpoint of $\mathcal{D}_\tau(r e^{i \theta})$ for $|\theta|<\pi$, while for $|\theta|=\pi$ there is no well-defined endpoint of the form \eqref{eq:intro-asymp-ansatz}.
This behaviour is confirmed explicitly by the numerical analysis in \cite{phys-paper}.

\smallskip

\noindent
\textbf{General minimal models.}
Consider the diagonal Virasoro minimal model $M(p,q)$ with $p,q$ coprime and $4\le 2p<q$. The Lee-Yang model $M(2,5)$ is the smallest example in this series.

The simple topological defects $D_{(r,s)}$ in $M(p,q)$ (with defect operators $\mathcal{D}_{(r,s)}$) are indexed by Kac labels $(r,s)$ with $1 \le r < p$, $1 \le s <q$, modulo $(r,s) \sim (p-r,q-s)$ \cite{Petkova:2000ip}. The defects $D_{(1,s)}$ close under fusion with truncated $sl(2)$-fusion rules, as do the defects $D_{(r,1)}$. 
We consider the perturbation by the holomorphic defect field $\psi$ given by the primary field in representation $(1,3)$. Its weight is
\begin{equation}\label{eq:intro-h13-weight}
    h_{(1,3)} = -1+\frac{2p}q \ .
\end{equation}
The condition \eqref{eq:intro-h-constraint} would only require $q>\frac43p$. The assumption $q>2p$ made above is equivalent to $h_{(1,3)}<0$. The explicit  solution we describe below relies on the condition $q>2p$.

The holomorphic defect field $\psi$ occurs on the defect $D_{(1,s)}$, $s=2,3,\dots,q-2$ with multiplicity $1$ and does not occur on $D_{(1,1)} = \Id$ (the identity defect) or on $D_{(1,q-1)}$. For a suitable normalisation of $\psi$ (see Section \ref{sec:mm-funrel} for details), we denote the defect operator $\mathcal{D}_{(1,s)}$ perturbed by $\lambda\psi$ by
\begin{equation}
    \mathcal{D}_{(1,s)}(\lambda) \ .
\end{equation}
The fusion of these perturbed defects was computed in \cite{Runkel:2007wd} and we recall the computation in the present conventions in Section \ref{sec:mm-funrel}. The result is the Hirota system
\begin{equation}\label{eq:Mpq-fun-rel-intro}
    \mathcal{D}_{(1,s)}(\zeta\lambda)\, 
    \mathcal{D}_{(1,s)}(\zeta^{-1}\lambda)
    \,=\, 
    \Id
    +    
    \mathcal{D}_{(1,s-1)}(\lambda)\, 
    \mathcal{D}_{(1,s+1)}(\lambda)
 \ , \quad 
    \text{where}
    ~~\zeta = e^{i\pi p/q} \ .
\end{equation}
Here, $s=2,\dots,q-2$ and we set $\mathcal{D}_{(1,1)}(\lambda) = \mathcal{D}_{(1,1)} = \mathrm{id}$ and $\mathcal{D}_{(1,q-1)}(\lambda) = \mathcal{D}_{(1,q-1)}$. 
The perturbed defect operators mutually commute,  $\mathcal{D}_{(1,s)}(\lambda)\mathcal{D}_{(1,s')}(\lambda') = \mathcal{D}_{(1,s')}(\lambda')\mathcal{D}_{(1,s)}(\lambda)$ for all $s,s'$ and $\lambda, \lambda'$,
and we have the reflection symmetry $\mathcal{D}_{(1,q-1)} \, \mathcal{D}_{(1,s)}(\lambda) \,=\, \mathcal{D}_{(1,q-s)}((-1)^p \lambda)$. 
Note that for $p=2$, $q=5$ this agrees with the relation \eqref{eq:LY-fun-rel_intro} in the Lee-Yang model, where $\tau = (1,2) \sim (1,3)$.

We assume $\alpha$-asymptotics for the $\psi$-perturbed defects in $M(p,q)$ and hence by Theorem \ref{thm:intro-main} we get a semiring homomorphism $\Lambda$ to the asymptotic semiring $\mathcal{A}$ of $M(p,q)$. We denote the image of $\mathcal{D}_{(1,s)}(\lambda)$ under $\Lambda$ by
\begin{equation}\label{eq:Xs-intro}
    \mathcal{X}_s \equiv [\mathcal{D}^\infty_s(\theta),a_s(\theta)] ~ \in ~ \mathcal{A} \ .
\end{equation}
As $\Lambda$ is a homomorphism, the functional relations \eqref{eq:Mpq-fun-rel-intro} imply the identity
\begin{equation}\label{eq:Hirota-asym-intro}
(\mathsf{S}_\gamma \mathcal{X}_s) \, (\mathsf{S}_{-\gamma} \mathcal{X}_s)
\,=\,
1 + \mathcal{X}_{s-1} \mathcal{X}_{s+1} 
    ~~ ,
    ~~ 2 \le s \le q-2 \ ,
\end{equation}
in the asymptotic semiring. Here, $\gamma = \pi p/q$ is the phase of $\zeta$ in \eqref{eq:Mpq-fun-rel-intro}, and $\mathsf{S}_\gamma$ shifts the argument by $\gamma$, so that $\mathsf{S}_\gamma \mathcal{X}_s = [\mathcal{D}^\infty_s(\theta+\gamma),a_s(\theta+\gamma)]$.

In Proposition~\ref{prop:pq-solution} we give an explicit solution to \eqref{eq:Hirota-asym-intro}. The detailed description is slightly lengthy and we defer it to Section~\ref{sec:pq-solution}. Here we just note two features of this solution:
\begin{itemize}
    \item Each $\cX_s$, $s=2,\dots,q-2$ has either one or two anti-Stokes rays. 

\item The topological defects $\mathcal{D}_s^\infty(\theta)$ giving the flow endpoints of the $\mathcal{D}_{(1,s)}(\lambda)$, $\lambda = r e^{i\theta}$, take all values $\mathcal{D}_{(r,1)}$, $r=1,\dots,p-1$ as we vary $s$ and $\theta$.
\end{itemize}
If one demands that \textit{all} $\cX_s$ have a single anti-Stokes ray along the negative real axis and that $\cX_2$ satisfies a positivity condition, we can show that necessarily $p=2$ and that our solution is unique (Proposition~\ref{prop:one-solution}). It may be possible to show such a uniqueness result also for the general solution in Proposition~\ref{prop:pq-solution}, but we have not pursued this question here.
Numerical spot-checks using the truncated conformal space approach confirm that the solution is realised in $M(p,q)$, see \cite{phys-paper}.

\begin{figure}[tb]
    \centering
    \includegraphics[width=0.92\linewidth]{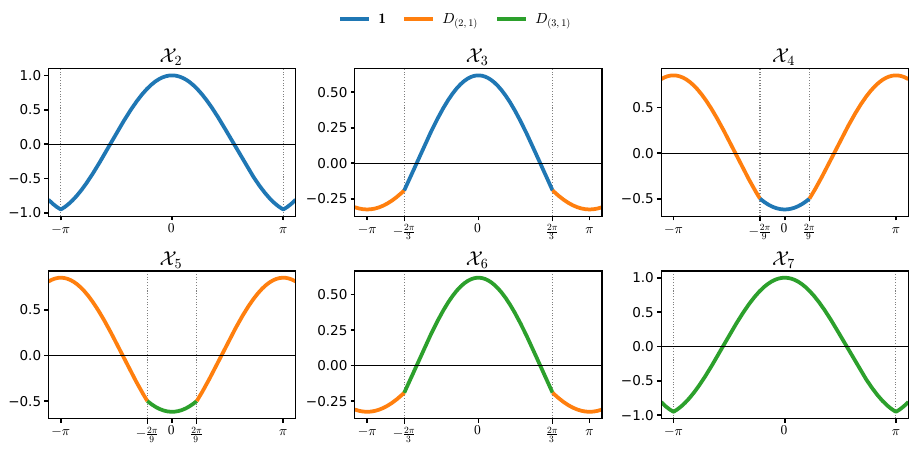}
    \caption{The functions $a_s(\theta)$ and the value of $\mathcal{D}_s^\infty(\theta)$ of the solution $\cX_s$ in $M(4,9)$ plotted against $\theta$.
The vertical lines in the plots indicate the angles $\theta_s^*$ of the anti-Stokes rays. Note that $\cX_2$ and $\cX_7$ only have a single anti-Stokes ray at $\theta = \pm \pi$, while $\cX_3,\dots,\cX_6$ have two anti-Stokes rays.}
    \label{fig:M49-intro}
\end{figure}

Consider the example $M(4,9)$. In Figure~\ref{fig:M49-intro} we plot the functions $a_s(\theta)$ from \eqref{eq:Xs-intro} and indicate the value of $\mathcal{D}_s^\infty(\theta)$. This illustrates how the flow endpoint $\mathcal{D}_s^\infty(\theta)$ depends on the angle $\theta$. A more detailed version of that figure which also shows the value of the two sides of the Hirota equation is given in Figure~\ref{fig:pq-M49} in Section~\ref{sec:pq-solution}.

\begin{figure}
    \centering
    \includegraphics[width=\linewidth]{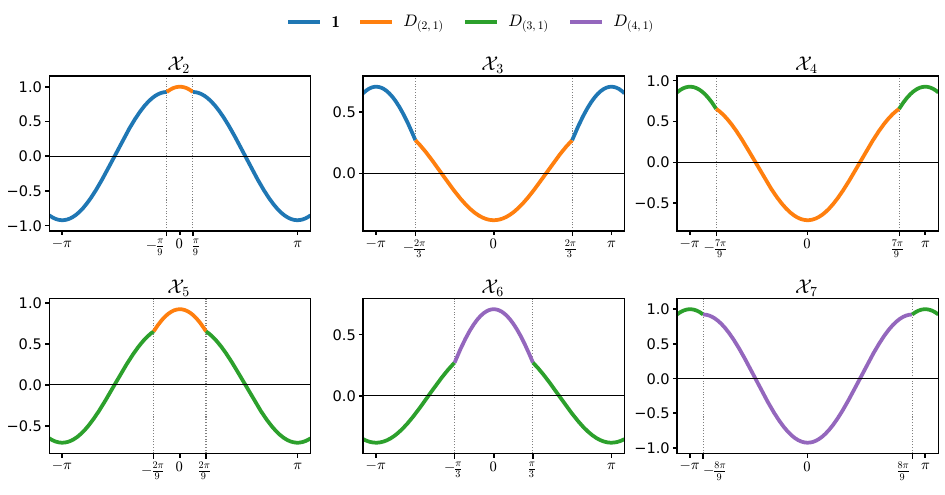}\caption{The functions $a_s(\theta)$ and the value of $\mathcal{D}_s^\infty(\theta)$ of the solution $\cX_s$ in $M(5,9)$ plotted against $\theta$. Compared to Figure~\ref{fig:M49-intro}, now $\mathcal{X}_2$, $\mathcal{X}_7$ also feature two anti-Stokes rays.  
    }
    \label{fig:M59-intro}
\end{figure}

We also have a solution in the regime $\frac{4}{3} p < q < 2p$ which amounts to $\frac12>h_{(1,3)}>0$, cf.\ \eqref{eq:intro-h13-weight}. We do not give the full details here, but we state the solution and outline a proof in Remarks~\ref{rem:pq-Tsys} and~\ref{rem:solutionless2p}, and we illustrate the asymptotic behaviour of the perturbed defects in $M(5,9)$ in Figure~\ref{fig:M59-intro} (see Figure~\ref{fig:M59solandhirota} for the two sides of the Hirota equation).\footnote{In Figure~\ref{fig:M59-intro} we report the solution of the functional relation for which $\mathcal{X}_2$ is positive at $\theta =0$; as discussed in Remark \ref{rem:solutionless2p}, this is obtained by $\mathcal{X}_s(\theta)\to \mathcal{X}_s(\theta+\pi)$ in \eqref{eq:pq-positive-count}.}
This should be contrasted with Figure~\ref{fig:M49-intro} which treats $M(4,9)$. One fundamental difference between the solutions in the two regimes is that for $q>2p$ the defects $\mathcal{D}_{(1,2)}(\lambda)$ 
and $\mathcal{D}_{(1,q-2)}(\lambda)$ have a single anti-Stokes ray, while for $\frac{4}{3} p < q < 2p$ they have two.
We stress that finding such a solution does not imply that it is necessarily realised by minimal model flows
(but we believe it is, and this has been checked in the case of $M(5,9)$ in \cite{phys-paper}).

\medskip

To summarise, we have identified an algebraic structure which allows us to describe defect flows of holomorphically perturbed defects in terms of a suitable semiring homomorphism. 
It is important to note that these flows are \textit{not} described by a semiring homomorphism between just the semirings of topological defects. Indeed, in the minimal model example, there is \textit{no semiring homomorphism at all} from the semiring formed by the defect labels $D_{(1,s)}$ to the semiring formed by the $D_{(r,1)}$ (see Remark~\ref{rem:no-semiring-hom}).
What makes the semiring homomorphism work is the angle dependence and the max-plus semiring structure of $\mathcal{A}$. It is a little bit magical to watch how the shifts on the left hand side of \eqref{eq:Hirota-asym-intro} and the dominance condition for the addition on the right hand side conspire to make the fusion rules for $D_{(r,1)}$ fit into the Hirota system for the $D_{(1,s)}$.

Similar (but different) algebraic structures have been considered in the description of defect/bulk flows in \cite{Gaiotto:2012np,Nakayama:2024msv, Fukusumi:2025clr,Antinucci:2025fjp}. 

\paragraph{Outline.}
This paper is organised as follows. In Section~\ref{sec:TDI-cat} we review the description of the monoidal category of holomorphically perturbed topological defects, and we give an equivalent formulation in terms of quiver modules which is helpful in explicit computations.
\\
Section~\ref{sec:semiring} contains the main new contribution of the present paper, the description of defect flows in terms of a semiring homomorphism, as well as the definition of the asymptotic semiring which serves as the target of the homomorphism.
\\
In Section~\ref{sec:minmod-13} we apply this formalism by considering topological defects in Virasoro minimal models, perturbed by the holomorphic $(1,3)$-field, giving, as a result, an explicit solution to the corresponding functional relation in the asymptotic semiring.
\\
We conclude in Section~\ref{sec:outlook} with an outlook on further questions which seem a natural continuation of the present work.

\section{The category of holomorphically perturbed defects}\label{sec:TDI-cat}

\subsection{Chiral TFT presentation of defect perturbations}\label{sec:chiral-TFT-defects}

To work with holomorphically perturbed defects, we use the representation of correlators of 2d CFT via chiral 3d topological field theory \cite{Fuchs:2002cm}. We will not review this in detail, but instead explain the representation of a holomorphic defect field in chiral TFT (see \cite{Runkel:2007wd,Ambrosino:2025pjj} for more on this): 
\begin{equation}\label{eq:defect-field-in-chiral-TFT}
\includegraphics[valign = c, scale=0.75]{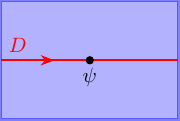} \quad=\quad 
\includegraphics[valign = c, scale=0.75]{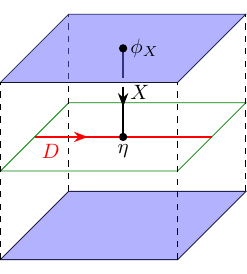} \quad .
\end{equation}
The left hand side shows a patch of the CFT world sheet $\Sigma$ with the topological line defect $D$ and the defect field $\psi$ inserted on it. On the right hand side we see the chiral TFT presentation of the same patch. The three-manifold is $\Sigma \times [-1,1]$, with the boundary component $\Sigma \times \{1\}$ supporting the holomorphic fields, and $\Sigma \times \{-1\}$ the antiholomorphic ones. 
The green plane at $\Sigma \times \{0\}$ indicates a topological surface defect, though in this paper we will only consider the case of the transparent surface defect, corresponding to diagonal CFTs. The topological defect $D$ is represented in chiral TFT by the line defect $D$ placed in the plane $\Sigma \times \{0\}$. 
The label $X$ of the vertical line defect corresponds to the representation of the chiral algebra that $\psi$ transforms in, and the point where the $X$-line meets the holomorphic boundary is marked by the corresponding element $\phi_X \in X$. 
The point where the $X$-line meets $D$ is marked by a defect junction $\eta$ of the chiral TFT.
In our approach we will fix $\phi_X$ and $X$ once and for all, and we will parametrise the holomorphically perturbed topological defects by the line defect $D$ and the junction $\eta : X \otimes D \to D$ in chiral TFT. We will write such a pair as
\begin{equation}
    D(\eta) \ .
\end{equation}

Next we briefly describe the perturbed defect operator for $D(\eta)$ on a cylinder of circumference $L$, for which we will use the two notations
\begin{equation}\label{eq:perturbed-defect-notation}
    \mathcal{D}[D(\eta)] \equiv \mathcal{D}_D(\eta) \ ,
\end{equation}
see also \cite[Sec.\,3.3]{Manolopoulos:2009np} and \cite[Sec.\,2]{phys-paper}. It is given by the usual perturbative expansion. To have an explicit expansion parameter, we use $\lambda\eta$ for some $\lambda \in \mathbb{C}$,
\begin{equation}\label{eq:defect-operator-as-pert-series}
    \mathcal{D}_D(\lambda\eta) = \mathcal{D}_D \exp\!\Big(\lambda 
    L^{h-1} 
    \hspace{-4pt}
    \int_0^L 
    \hspace{-4pt}
    \psi(x) dx\Big) = 
\mathcal{D}_D\sum_{n=0}^\infty \lambda^n 
    L^{n(h-1)}
    \int_{0<x_1<\cdots<L} 
    \hspace{-3.5em}
    \psi(x_1) \cdots \psi(x_n) \,dx_1 \cdots dx_n \ ,
\end{equation}
where the integral on the right hand side is over amplitudes on a cylinder of circumference $L$ and height $R$ with a circular $D$ defect and insertions of defect fields $\psi$ at $x_1,\dots,x_n$:
\begin{equation}
\includegraphics[valign=c,width=0.6\linewidth]{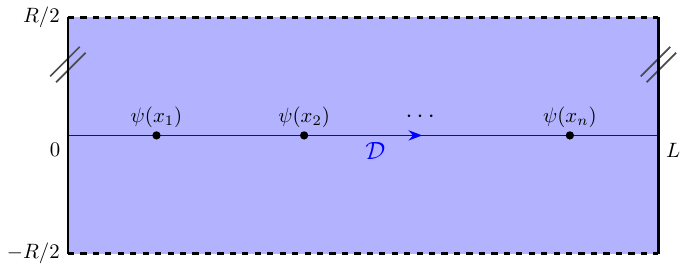}
\end{equation}
The factor $L^{h-1}$ is included in \eqref{eq:defect-operator-as-pert-series} to make $\lambda$ dimensionless.
For finite height $R$ this defines the operator
\begin{equation}\label{eq:defect-op-finite-height-cyl}
    e^{- \frac{R}2  H} \,\mathcal{D}_D(\lambda \eta) \,e^{-\frac{R}2 H} : \mathcal{H} \to \mathcal{H} \ ,
\end{equation}
where $H$ is the Hamiltonian. We will assume:
\begin{quote}
    \textbf{Assumption.} The multiple integrals in \eqref{eq:defect-operator-as-pert-series} converge, and the sum over $n$ has a non-zero radius of convergence in $\lambda$. 
\end{quote}
This restricts which defects $D$ and perturbing fields $\psi$ the categorical methods developed in the following can be applied to. If the integrals require regularisation, this can affect the resulting functional equations, see the discussion in \cite[Sec.\,4]{Ambrosino:2025pjj} (based on \cite{Gaiotto:2020fdr}) for an example.

Finally, to get the operator $\mathcal{D}_D(\lambda \eta)$, one takes the limit $R \to 0$. 
As explained in \cite[Sec.\,2.2]{Runkel:2007wd}, the operators in \eqref{eq:defect-op-finite-height-cyl} commute with $H$, we can take the limit in each finite-dimensional eigenspace of $H$, and there the limit is just given by composing with $e^{R H}$.
In the following, we no longer include the parameter $\lambda$ and take $D(\eta)$ to be defined by analytic continuation along $r \eta$, $r \in [0,1]$ if necessary.

\medskip

The fusion of two holomorphically perturbed defects $D(\eta)$ and $E(\zeta)$ is again given by a holomorphically perturbed defect. The underlying topological defect is $D \otimes E$, the fusion of the topological defects $D$ and $E$, and the perturbing field $\psi$ is the sum of the perturbing fields on $D$ and on $E$. The chiral TFT presentation is
\begin{equation}\label{eq:tensor-prod-in-chiral-TFT}
\includegraphics[valign =c,scale=0.75]{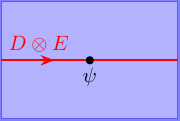} ~~ = ~~ \includegraphics[valign =c,scale=0.75]{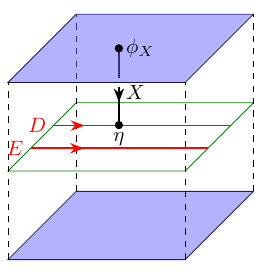} ~ + ~ \includegraphics[valign =c,scale=0.75]{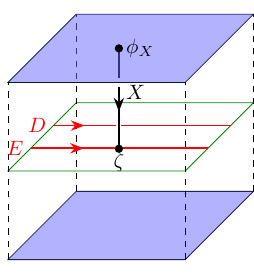} \quad .
\end{equation}

As $\phi_X \in X$ is kept fixed, we can describe the different defect perturbations and their fusion purely in terms of the braided monoidal category of line defects of the chiral TFT. In the next subsection we review the relevant mathematical structure in more generality, without assuming e.g.\ non-degeneracy of the braiding (which holds for line defects in the chiral TFT).

\subsection{Category of perturbed defects and functional relations}\label{sec:cat-of-pert-def}

In this section we fix an additive braided monoidal category $\mathcal{C}$.  

\begin{definition}[{\!\cite[Def.\,2.1]{Manolopoulos:2009np}}] \label{def:CX}
    Let $X \in \mathcal{C}$. The category $\mathcal{C}_X$ has
    \begin{itemize}
        \item objects: pairs $D(\eta) = (D,\eta)$, where $D \in \mathcal{C}$ and $\eta : X \otimes D \to D$ is a morphism in $\mathcal{C}$.
        \item morphisms $D(\eta) \to E(\zeta)$: a morphism $f : D \to E$ in $\mathcal{C}$ such that $f \circ \eta = \zeta \circ (\mathrm{id}_X \otimes f)$.
    \end{itemize}
\end{definition}

We assume for simplicity of notation that $\mathcal{C}$ is strictly monoidal (i.e.\ it has trivial associators and unitors). Then the category $\mathcal{C}_X$ is also strictly monoidal with tensor product on objects defined as \cite[Sec.\,2.2]{Manolopoulos:2009np}
\begin{equation}
  D(\eta) \otimes E(\zeta) ~=~ (D \otimes E)\big( T(\eta,\zeta) \big) \ .
\end{equation}
Here
\begin{equation}\label{eq:Deta-Ezeta-T}
    T(\eta,\zeta) = \eta \otimes \mathrm{id}_E + (\mathrm{id}_D \otimes \zeta) \circ (c_{X,D} \otimes \mathrm{id}_E)    
    : X \otimes D \otimes E \to D \otimes E \ ,
\end{equation}
and $c_{X,D} : X \otimes D \to D \otimes X$ denotes the braiding of $\mathcal{C}$.
In string diagram notation (read from bottom to top in this paper), this becomes
\begin{equation}
    T(\eta,\zeta) = \hspace{-.2cm}
 \includegraphics[scale=0.9, valign = c]{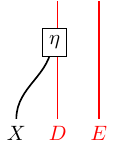} \hspace{-0.05cm} +
 \includegraphics[scale=0.9, valign = c]{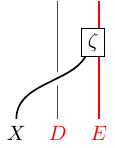}
 \quad .
\end{equation}
On morphisms $f : D(\eta) \to D'(\eta')$, $g : E(\zeta) \to E'(\zeta')$, the tensor product $f \otimes g$ is simply that of the underlying category $\mathcal{C}$. The tensor unit is given by $\mathbf{1} \equiv \mathbf{1}(0)$.

The monoidal category $\mathcal{C}_X$ has also been studied from the perspective of quiver representations in \cite{schaumann2023fusionquivers}.

The next lemma explains in which sense $\mathcal{C}_X$ is an extension of $\mathcal{C}$:

\begin{lemma}\label{lem:C-CX-embed}
There is a fully faithful monoidal functor $\iota: \mathcal{C} \to \mathcal{C}_X$ given by $D \mapsto D(0)$.
\end{lemma}

\begin{proof}
This is immediate from the definitions. Indeed, from Definition~\ref{def:CX} we see that morphisms $D(0) \to E(0)$ are precisely the morphisms $D \to E$ in $\mathcal{C}$ as the additional conditions become trivial, hence the functor is fully faithful. For monoidality note that $T(0,0)=0$ and so $D(0) \otimes E(0) = (D\otimes E)(0)$.
\end{proof}

\begin{remark}
In the application to CFT, the monoidal category $\mathcal{C}_X$ describes the translation invariant defects obtained from $\mathcal{C}$-labelled defects, perturbed by a  holomorphic field $\phi_X$ in the representation $X \in \mathcal{C}$ of the chiral algebra, as well as the behaviour of these defects under fusion. Indeed, once we fix $\phi_X \in X$, an object $D(\eta) \in \mathcal{C}_X$ gives the data $X$, $\phi_X$, and $\eta$ used in \eqref{eq:defect-field-in-chiral-TFT} to describe a holomorphic field on the defect $D$, and $T(\eta,\zeta)$ is precisely the perturbing field on the fused defect as described in chiral TFT in \eqref{eq:tensor-prod-in-chiral-TFT}. 
This results in the following compatibility between the tensor product in $\mathcal{C}_X$ and the composition of perturbed defect operators \cite[Thm.\,3.2]{Manolopoulos:2009np}:
\begin{equation}
    \mathcal{D}[D(\eta)] \,
    \mathcal{D}[E(\zeta)]
    \,=\,
    \mathcal{D}[D(\eta) \otimes E(\zeta)] \ ,
\end{equation}
where we used the first notation in \eqref{eq:perturbed-defect-notation}.
The embedding in Lemma~\ref{lem:C-CX-embed} amounts to the statement that topological defects are a full subcategory of translation invariant defects.
\end{remark}

\begin{proposition}\label{prop:CX-properties}
Suppose $\mathcal{C}$ is in addition abelian and rigid. Then $\mathcal{C}_X$ is an abelian rigid monoidal category with exact tensor product. 
\end{proposition}

This is shown in \cite[Thms.\,2.8,\,2.11]{Manolopoulos:2009np}, together with the observation that in a rigid category, the tensor product is automatically exact in each argument (since $V \otimes -$ is both a left and a right adjoint, cf.\,\cite[Rem.\,2.10.9]{etingof2015tensor}).

An abelian monoidal category with exact tensor product has a well-defined Grothendieck ring, and so under the assumptions of the above proposition, $\mathcal{C}_X$ has a well-defined Grothen\-dieck ring $\mathrm{Gr}(\mathcal{C}_X)$. For an object $D(\eta) \in \mathcal{C}_X$ we write $[D(\eta)] \in \mathrm{Gr}(\mathcal{C}_X)$ for its element in the Grothendieck ring. By definition, a short exact sequence
\begin{equation}\label{eq:exact-seq-in-CX}
    0 \to A(\alpha) \to B(\beta) \to C(\gamma) \to 0
    \quad \text{in} ~~\mathcal{C}_X
\end{equation}
gives rise to the identity
\begin{equation}
    [B(\beta)] = [A(\alpha)] + [C(\gamma)] 
    \quad \text{in} ~~\mathrm{Gr}(\mathcal{C}_X) \ .
\end{equation}

\begin{remark}\label{rem:pert-def-op-factor-Gr-ring}
In the application to CFT, recall that $\mathcal{D}_D(\eta)$ denotes the defect operator corresponding to $D(\eta) \in \mathcal{C}_X$ (and recall that we assume the iterated integrals to converge). The defect operators have the important property that they depend on $D(\eta)$ only through the class $[D(\eta)]$ in the Grothendieck ring \cite[Cor.\,3.3]{Manolopoulos:2009np}. Thus, short exact sequences in $\mathcal{C}_X$ result in relations among defect operators:
\begin{equation}
    \eqref{eq:exact-seq-in-CX}
    \quad \Rightarrow \quad
    \mathcal{D}[B(\beta)] = \mathcal{D}[A(\alpha)] + \mathcal{D}[C(\gamma)] \ .
\end{equation}
We will recover this implication in terms of trace-equivalence later on in Remark~\ref{rem:trace-equiv-gives-direct-sum}.
\end{remark}

In the setting we are interested in, $\mathcal{C}$ is in addition $\mathbb{C}$-linear. In this case, for $\lambda \in \mathbb{C}$ we define the functor $\mathcal{U}_\lambda : \mathcal{C}_X \to \mathcal{C}_X$ as
\begin{equation}
    \mathcal{U}_\lambda(D(\eta)) = D(\lambda\eta)
\end{equation}
on objects and as the identity on morphisms. Since $T(\lambda\eta,\lambda\zeta) = \lambda T(\eta,\zeta)$, the functor $\mathcal{U}_\lambda$ is strictly monoidal. It furthermore satisfies 
\begin{equation}
    \mathcal{U}_\lambda \circ \mathcal{U}_\mu = \mathcal{U}_{\lambda\mu} \ .
\end{equation}
In particular, $\mathcal{U}_\lambda$ is invertible iff $\lambda\neq 0$. This shows:

\begin{proposition}\label{prop:Cx-action-on-CX}
Let $\mathcal{C}$ be in addition $\mathbb{C}$-linear. Then $\mathbb{C}^\times$ acts on $\mathcal{C}_X$ via strict tensor autoequivalences as $\lambda \mapsto \mathcal{U}_\lambda$.
\end{proposition}

This induces an action of $\mathbb{C}^\times$ on $\mathrm{Gr}(\mathcal{C}_X)$ as ring automorphisms:
\begin{equation}\label{eq:Cx-action-GrCX}
    \mathsf{U} : \mathbb{C}^\times \to \mathrm{Aut}(\mathrm{Gr}(\mathcal{C}_X))
    ~~,\quad
    \mathsf{U}_\lambda([D(\eta)]) := [\mathcal{U}_\lambda( D(\eta))] = [D(\lambda\eta)] \ .
\end{equation}

\subsection{Computing with perturbed defects via quivers}\label{sec:quiver}

Suppose now that $\mathcal{C}$ is a braided fusion category, that is, in addition to being braided monoidal it is rigid and semisimple with a finite number of simple objects. Write $\mathcal{I}$ for (a choice of representatives of the isomorphism classes of) the simple objects in $\mathcal{C}$. Then every object in $\mathcal{C}$ is, up to isomorphism, given by a collection of finite dimensional $\mathbb{C}$-vector spaces $(V_a)_{a \in \mathcal{I}}$: for all $D \in \mathcal{C}$ we have
\begin{equation}\label{eq:D-via-a-Va}
    D \cong \bigoplus_{a \in \mathcal{I}} a \otimes V_a
    \quad \text{where} \quad
    V_a = \mathrm{Hom}_{\mathcal{C}}(a,D) \ ,
\end{equation}
where $a \otimes V_a$ stands for $\dim_\mathbb{C}(V_a)$ many copies of $a$. More formally, $\mathcal{C}$ is a right (and also a left) module category over $\mathrm{Vec}_\mathbb{C}$ and we identify $V_a \equiv \mathbf{1} \triangleleft V_a$ (or we can directly write $D \cong \bigoplus_{a \in \mathcal{I}} a \triangleleft V_a$). 
In fact, as a $\mathbb{C}$-linear category, $\mathcal{C}$ is equivalent to $\mathrm{Vec}_\mathbb{C}^\mathcal{I}$, the category of finite-dimensional $\mathcal{I}$-graded $\mathbb{C}$-vector spaces.

\medskip

From here on, we will for simplicity assume that
\begin{equation}\label{eq:assumption-C}
    \text{$X$ is simple}
    \quad \text{and} \quad
    \text{$N_{ab}^{~c} \in \{0,1\}$ for all $a,b,c \in \mathcal{I}$ ,} 
\end{equation}
that is, $\mathcal{C}$ has no fusion multiplicities. These assumptions are not necessary, but are sufficient for our application and make the exposition less notation-heavy.

For each $a,b,c \in \mathcal{I}$ such that $N_{ab}^{~c} \neq 0$ pick a non-zero morphism $\lambda_{ab}^c : a \otimes b \to c$. By assumption, this morphism is a basis of $\mathrm{Hom}_\mathcal{C}(a \otimes b,c)$. We use the string diagram notation
\begin{equation}\label{eq:lambda-basis-in-Hom}
    \lambda_{ab}^c = \includegraphics[valign = c, scale =0.75]{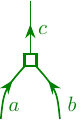} \quad .
\end{equation}
In terms of this, our conventions for the fusing and braiding matrices are
\begin{equation}
   \includegraphics[valign = c, scale =0.75]{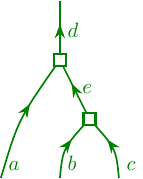} \hspace{-.5em}=~ \sum_{f} \mathrm{F}^{(abc)d}_{ef}    \includegraphics[valign = c, scale =0.75]{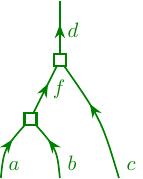}
    ~~,\quad
     \includegraphics[valign = c, scale =0.75]{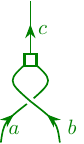} \hspace{-.5em}=~ \mathrm{R}^{(ab)c}  \includegraphics[valign = c, scale =0.75]{R0.pdf}
    \ .
\end{equation}
The inverse relations are
\begin{equation}
    \includegraphics[valign = c, scale =0.75]{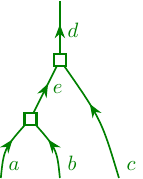} \hspace{-.5em}=~ \sum_{f} \mathrm{G}^{(abc)d}_{ef}  \includegraphics[valign = c, scale =0.75]{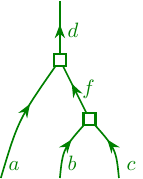}
    ~~,\quad
     \includegraphics[valign = c, scale =0.75]{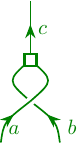} \hspace{-.5em}=~ \frac{1}{\mathrm{R}^{(ba)c}} ~ \includegraphics[valign = c, scale =0.75]{R0.pdf} 
    \ .
\end{equation}

With these preparations, we can describe $\mathcal{C}_X$ as modules for a certain quiver $Q_X$. This is the perspective taken in \cite{schaumann2023fusionquivers}, whose notation we follow here.

\subsubsection*{\boldmath  The quiver $Q_X$ and its modules}

A \textit{quiver} $Q$ is a tuple $(Q^0,Q^1,s,t)$, where $Q^0$ is the set of vertices, $Q^1$ is the set of arrows, and $s,t : Q^1 \to Q^0$ denote the source and target maps. 
A \textit{module of a quiver $Q$} is a pair $(V,f)$ where $V = (V_a)_{a \in Q^0}$ is a collection of vector spaces indexed by vertices $a \in Q^0$ of $Q$, and $f = (f_{\mu} : V_a \to V_b)_{\mu : a \to b}$ is a collection of linear maps indexed by arrows $\mu \in Q^1$. 
A \textit{morphism of $Q$-modules} $h : (V,f) \to (W,g)$ is a collection $h=(h_a : V_a \to W_a)_{a \in Q^0}$ of linear maps such that for each arrow $\mu : a \to b$ the following diagram commutes:
\begin{equation}\label{eq:quiver-module-morphism}
    \begin{tikzcd}
        V_a \arrow[r,"f_\mu"] \arrow[d,"h_a"] & V_b \arrow[d,"h_b"]
        \\
        W_a \arrow[r,"g_\mu"] & W_b
    \end{tikzcd}
\end{equation}
We write $\mathrm{mod}(Q)$ for the category of finite-dimensional $Q$-modules.

\medskip

We now define the quiver $Q_X$ which describes $\mathcal{C}_X$. It has vertices labelled by the index set $Q^0=\mathcal{I}$ of the simple objects of $\mathcal{C}$. There are $N_{Xa}^{~b}$ arrows $a \to b$, i.e.\ by \eqref{eq:assumption-C} there is either one or no arrow $a \to b$. 
The set of arrows is therefore 
\begin{equation}\label{eq:QX-quiver-arrows}
Q^1 = \{ (b,a) | b,a \in \mathcal{I}, N_{Xa}^{~b} \neq 0 \} \ ,
\end{equation}
with $s,t : Q^1 \to Q^0$ given by $s(b,a)=a$ and $t(b,a)=b$.
In a $Q_X$-module $(V,f)$ we write $f_{ba} : V_a \to V_b$ for the linear map assigned to the unique arrow $a \to b$ (if it exists), else we set $f_{ba}=0$. 

The modules of $Q_X$ are an equivalent description of $\mathcal{C}_X$:

\begin{lemma}\label{lem:QX-CX-equiv}
There is an equivalence $\Gamma : \mathrm{mod}(Q_X) \to \mathcal{C}_X$ of $\mathbb{C}$-linear categories which sends an object $(V,f)$ to $D(\eta)$ with $D = \bigoplus_{a \in \mathcal{I}} a \otimes V_a$ and
\begin{equation}
    \eta : X \otimes D = \bigoplus_{a \in \mathcal{I}} X \otimes a \otimes V_a \xrightarrow{\sum_{a,b} \lambda_{Xa}^b \otimes f_{ba}} \bigoplus_{b \in \mathcal{I}} b \otimes V_b = D \ ,
\end{equation}
and a morphism $h : (V,f) \to (W,g)$ is sent to 
\begin{equation}
    \Gamma(h) = \Big[ \bigoplus_{a \in \mathcal{I}} a \otimes V_a 
    \xrightarrow{\sum_a \mathrm{id}_a \otimes h_a}
    \bigoplus_{a \in \mathcal{I}} a \otimes W_a 
    \Big]
    \ .
\end{equation}
\end{lemma}

In the sum $\sum_{a,b} \lambda_{Xa}^b \otimes f_{ba}$ in the definition of $\eta$ it is understood that $\lambda_{Xa}^b \otimes f_{ba}$ is non-zero only on the summand  $X \otimes a \otimes V_a$ which it maps to $b \otimes V_b$. An analogous statement holds for $\Gamma(h)$.
In terms of string diagrams, the definition of $\Gamma$ reads
\begin{equation}\label{eq:modQ-CX-functor-diag}
    \Gamma(V,f) \,=\, \Bigg(\, \bigoplus_{a \in \mathcal{I}} a \otimes V_a \ , \
    \sum_{a,b \in \mathcal{I}} \ \,
    \includegraphics[valign = c]{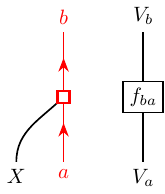}
    \, \Bigg)
   \, ,  \qquad
    \Gamma(h) \,=\, \sum_{a \in \mathcal{I}} \ \,
    \includegraphics[valign = c]{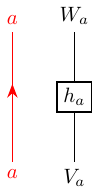}
\end{equation}

\begin{proof}[Proof of Lemma~\ref{lem:QX-CX-equiv}]
We will show that $\Gamma$ is fully faithful and essentially surjective. Faith\-ful\-ness is clear. For fullness, a morphism $h : \Gamma(V,f) \to \Gamma(W,g)$ in $\mathcal{C}$ (not $\mathcal{C}_X$ yet) is necessarily of the form $\bigoplus_a \mathrm{id}_a \otimes h_a$, with $h_a : V_a \to W_a$. It has to satisfy the condition in Definition~\ref{def:CX}, which, together with the explicit expression \eqref{eq:modQ-CX-functor-diag}, is easily seen to be equivalent to the condition \eqref{eq:quiver-module-morphism} to be a morphism in $\mathrm{mod}(Q_X)$. To see that $\Gamma$ is essentially surjective, pick any $D(\eta) \in \mathcal{C}_X$, together with an isomorphism $\phi : D \to \bigoplus_a a \otimes V_a$, and use that to pull back $\eta$ to $u = \phi \circ \eta \circ (\mathrm{id}_X \otimes \phi^{-1})$. Then $u$ is necessarily of the form $\Gamma(V,f)$ shown on the left in \eqref{eq:modQ-CX-functor-diag}.
\end{proof}

\begin{remark}
By the above lemma, the data in a $Q_X$-module $(V,f)$ has a clear interpretation in terms of perturbed topological defects: The vector space $V_a$, $a \in \mathcal{I}$, gives the multiplicity space of the elementary topological defect $a$ in the overall topological defect $D$, and $f_{ba}$ describes how the component $X \otimes a \to b$ of the defect field acts on the multiplicity spaces. By abuse of notation, we write
\begin{equation}
    \mathcal{D}[V,f] \equiv \mathcal{D}[\Gamma(V,f)]
\end{equation}
for the defect operator corresponding to a quiver module $(V,f)$, cf.\ \eqref{eq:perturbed-defect-notation}.
\end{remark}

\begin{example}\label{ex:LY-Qtau-modules}
The Lee-Yang CFT is the Virasoro minimal model $M(2,5)$ of central charge $c=-22/5$. There are two irreducible representations of the Virasoro vertex operator algebra, namely $\mathbf{1}$ of lowest $L_0$-weight $h_{\mathbf{1}}=0$ and $\tau$ of lowest weight $h_\tau = -\frac15$, and so $\mathcal{I} = \{ \mathbf{1}, \tau \}$. 
The fusion rule is $\tau \otimes \tau = \mathbf{1} \oplus \tau$. For the perturbing field we take the primary field in the representation $X = \tau$. The quiver $Q_\tau$ is then given by:
        \begin{equation}\label{eq:LY-quiver}
            Q_\tau =
\raisebox{-.5\height}{\begin{tikzpicture}[>=latex]
\coordinate (A) at (0,0);
\coordinate (B) at (2,0);

\begin{scope}[very thick,decoration={
    markings,
    mark=at position 0.55 with {\arrow{>}}}
    ] 
\draw[cyan, very thick,postaction={decorate}] (A)  to[bend right] (B);
\draw[cyan, very thick,postaction={decorate}] (B)  to[bend right] (A);
\end{scope}
\begin{scope}[very thick,decoration={
    markings,
    mark=at position 0.8 with {\arrow{>}}}
    ] 
    \node (E2) at (1.9,0) {};
\draw[cyan, very thick,postaction={decorate}] (E2) to[out=-30, in=30, looseness=26] (E2);%
\end{scope}
\fill[black] (A)circle (2pt);
\fill[black] (B)circle (2pt);
\node at (A) [below=2pt] {$\mathbf{1}$} ;
\node at (B) [below=4pt] {$\tau$} ;
\end{tikzpicture}}
        \end{equation}
For example, the arrow from $\tau$ to itself arises from $N_{Xa}^{~b}=1$ for $a=b=\tau$ (and anyway $X=\tau$).
A $Q_\tau$-module $(V,f)$ consists of two vector spaces $V_{\mathbf{1}}$ and $V_\tau$, and of three linear maps 
$f_{\mathbf{1}\tau} : V_\tau \to V_\mathbf{1}$, 
        $f_{\tau\mathbf{1}} : V_\mathbf{1} \to V_\tau$, 
        $f_{\tau\tau} : V_\tau \to V_\tau$.
For example:
\begin{enumerate}
    \item 
Consider $(V,f)$ with $V_{\mathbf{1}} = 0$, $V_{\tau} = \mathbb{C}$. Of the three arrows in the quiver $Q_\tau$ in \eqref{eq:LY-quiver}, only the arrow $\tau \to \tau$ can have a non-zero linear map, and we set $f_{\tau\tau} = \lambda \in \mathbb{C}$ (understood as a linear map from $V_\tau = \mathbb{C}$ to itself). We abbreviate $(V,f)=(\tau,\lambda)$. We can ask when $(\tau,\lambda)$ is isomorphic to $(\tau,\mu)$. The defining condition \eqref{eq:quiver-module-morphism} amounts to the existence of $\phi_\tau\in \mathbb{C}^\times$ such that $\phi_\tau \mu = \lambda \phi_\tau$. Altogether, $(\tau,\lambda) \cong (\tau,\mu)$ iff $\lambda = \mu$. As $V$ is one-dimensional, $(V,\lambda)$ is simple for all $\lambda \in \mathbb C$. This shows that $\mathrm{mod}(Q_\tau) \cong \mathcal{C}_\tau$ has continuously many pairwise non-isomorphic simple objects.

\item
Next take
$V_{\mathbf{1}} = V_\tau = \mathbb{C}$. Then the three components of $f$ are just numbers, $f_{\mathbf{1}\tau}, f_{\tau\mathbf{1}}, f_{\tau\tau} \in \mathbb{C}$. We can now ask if $(V,f)$ is isomorphic to $(V,g)$. The defining condition \eqref{eq:quiver-module-morphism} amounts to the existence of $\phi_{\mathbf{1}}, \phi_\tau\in \mathbb{C}^\times$ such that:
\begin{align}
    a=\mathbf{1}, \, b = \tau ~~:~~& \phi_\tau f_{\tau\mathbf{1}} = g_{\tau\mathbf{1}} \phi_{\mathbf{1}}
\nonumber \\
    a=\tau, \, b =\mathbf{1} ~~:~~& \phi_{\mathbf{1}} f_{\mathbf{1}\tau} = g_{\mathbf{1}\tau} \phi_{\tau}
 \\
    a=\tau, \, b = \tau ~~:~~& \phi_\tau f_{\tau\tau} = g_{\tau\tau} \phi_{\tau}
\nonumber 
\end{align}
Equivalently, there has to be a $\xi \in \mathbb{C}^\times$ such that
\begin{equation}\label{eq:LY-isom}
     g_{\tau\tau} = f_{\tau\tau}
     ~~,\quad
     g_{\tau\mathbf{1}} = \xi \, f_{\tau\mathbf{1}} 
     ~~,\quad
     g_{\mathbf{1}\tau} = \xi^{-1} \, f_{\mathbf{1}\tau} \ .
\end{equation}
To see if $(V,f)$ is simple, we have to check if there is a non-zero morphism $\phi : (\mathbf{1},0) \to (V,f)$ or $\psi : (\tau,\lambda) \to (V,f)$. The first case amounts to finding $\phi_{\mathbf{1}} \in \mathbb{C}^\times$, and the second case to finding $\psi_\tau \in \mathbb{C}^\times$, such that
\begin{align}
    a=\mathbf{1}, \, b = \tau ~~:~~& 0 = f_{\tau\mathbf{1}} \phi_{\mathbf{1}}
    && 0 = 0
\nonumber \\
    a=\tau, \, b =\mathbf{1} ~~:~~& 0=0
&& 0 = f_{\mathbf{1}\tau} \psi_{\tau} \\
    a=\tau, \, b = \tau ~~:~~& 0=0
&& \psi_\tau \lambda = f_{\tau\tau} \psi_{\tau}\nonumber 
\end{align}
Thus $(V,f)$ is simple iff $f_{\tau\mathbf{1}} \neq 0$ and $f_{\mathbf{1}\tau} \neq 0$. 
We can then use \eqref{eq:LY-isom} to set $f_{\mathbf{1}\tau} =1$ and conclude that $(f_{\tau\tau},f_{\tau\mathbf{1}}) \in \mathbb{C} \times \mathbb{C}^\times$ parametrise pairwise non-isomorphic simple objects.
\end{enumerate}
        \end{example}

\subsubsection*{\boldmath Tensor product of $Q_X$-modules}

We will now define a tensor product $\circledast$ on $\mathrm{mod}(Q_X)$ and will then show that the equivalence in Lemma~\ref{lem:QX-CX-equiv} becomes a monoidal equivalence. This tensor product is, in slightly different language, also described in \cite{Manolopoulos:2009np,schaumann2023fusionquivers}.

Given $(V,g), (W,h) \in \mathrm{mod}(Q_X)$, we define
\begin{equation}\label{eq:VgxWh-Z}
    (Z,f) = (V,g) \circledast (W,h)  
\end{equation}
as follows. Firstly $Z = (Z_b)_{b \in \mathcal{I}}$ is given by (recall our restriction \eqref{eq:assumption-C})
\begin{equation}\label{eq:fused-quiver-vsp}
    Z_b ~=\hspace{-0.6em} 
    \bigoplus_{c,d \in \mathcal{I}, N_{cd}^{~b} \neq 0} \hspace{-0.6em} V_c \otimes W_d \ .
\end{equation}
Given an arrow $b \to a$ in $Q_X$, the corresponding linear map $f_{ab}$ can be expressed componentwise on the direct summands of $Z_b$ and $Z_{a}$ as
\begin{equation}\label{eq:VgWh-tensor-f}
    f_{ab} : Z_b  \xrightarrow{\sum_{c,d,c',d'} (f_{ab})_{cd}^{c'd'}} Z_{a}
\quad \text{with} \quad
    (f_{ab})_{cd}^{c'd'} : V_c \otimes W_d \to  V_{c'} \otimes W_{d'} 
\end{equation}
given by\footnote{\label{fn:FG-rel}
To make the connection to \cite[Sec.\,4]{phys-paper}, one needs to substitute $G^{(Xcd)a}_{c'b} = F^{(dcX)a}_{c'b}$, etc., which holds in minimal model CFTs.
}
\begin{equation}\label{eq:tensor-Q-mod-on-obj}
    (f_{ab})_{cd}^{c'd'}
    \,=\, \delta_{d,d'} \, G^{(Xcd)a}_{c'b} \, g_{c'c} \otimes \mathrm{id}_{W_d} ~+~ \delta_{c,c'} \, \frac{R^{(d'c)a}}{R^{(dc)b}} \, G^{(Xdc)a}_{d'b} \, \mathrm{id}_{V_c} \otimes h_{d'd} \ .
\end{equation}
This expression already appears in \cite[App.\,A.3]{Manolopoulos:2009np}, and versions where $V_c, W_d$ are one-di\-men\-si\-o\-nal can be found in \cite[Sec.\,2.4]{Runkel:2007wd} and in \cite[Sec.\,2.4.2]{Ambrosino:2025myh}.
On morphisms $p : (V,g) \to (V',g')$ and $q : (W,h) \to (W',h')$, the tensor product is defined componentwise according to \eqref{eq:fused-quiver-vsp},
\begin{equation}
    (p \circledast q)_b ~=\hspace{-0.6em} 
    \bigoplus_{c,d \in \mathcal{I}, N_{cd}^{~b} \neq 0} \hspace{-0.6em} p_c \otimes q_d 
\quad \text{where} \quad
p_c \otimes q_d : V_c \otimes W_d \to V'_c \otimes W'_d \ .
\end{equation}

This tensor product can easily be implemented on a computer. A Mathematica file with this algorithm in the case of minimal models is included in the ancillary files on the arXiv page. Importantly, the above tensor product agrees with that on $\mathcal{C}_X$ (an analogous result is shown in \cite[Thm.\,A.10]{Manolopoulos:2009np}).

\begin{proposition}\label{prop:modQX-CX-monoidal}
$\mathrm{mod}(Q_X)$ with tensor product $\circledast$ is monoidally equivalent to $\mathcal{C}_X$.
\end{proposition}

\begin{proof}[Proof (sketch).]
We will describe the coherence isomorphisms 
\begin{equation}
    \Gamma^2_{(V,g),(W,h)} : \Gamma(V,g) \otimes \Gamma(W,h) \to \Gamma((V,g) \circledast (W,h))
\end{equation} 
explicitly. Naturality and the hexagon identity for $\Gamma^2$ are satisfied, and $\Gamma$ is compatible with the tensor units, but we skip the details.
We have
\be
\begin{split}
    \Gamma(V,g) \otimes \Gamma(W,h) &= \Big(\bigoplus_{c,d \in \mathcal{I}} c \otimes V_c \otimes d \otimes W_d ,
    T \Big)
    \ ,
\\  \Gamma((V,g) \circledast (W,h)) &= \Big(\bigoplus_{b \in \mathcal{I}} b \otimes Z_b , F \Big)
    \quad , \quad
    Z_b = \bigoplus_{c,d \in \mathcal{I}, N_{cd}^{~b} \neq 0} V_c \otimes W_d \ .
    \end{split}
\end{equation}
Here, $T$ is obtained from combining the explicit form of the functor $\Gamma$ in \eqref{eq:modQ-CX-functor-diag} and the tensor product of $\mathcal{C}_X$ in \eqref{eq:Deta-Ezeta-T}, and $F$ is obtained in the same way from \eqref{eq:VgWh-tensor-f}. In string diagrams:
\begin{equation}
\begin{aligned}
    T ~&=~ \sum_{c,d,c',d'} \ \delta_{d,d'} \ \
    \includegraphics[valign = c, ]{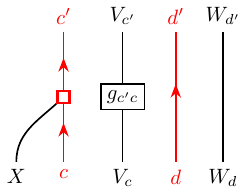}
    \ \ + \ \ \delta_{c,c'} \ \
    \includegraphics[valign = c]{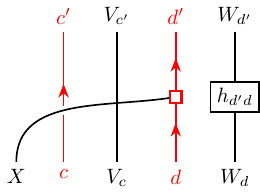}
    \\[3ex]
    F ~&=~ \sum_{a,b \in \mathcal{I}} \ \sum_{c,d,c',d'} \ \
    \includegraphics[valign = c, ]{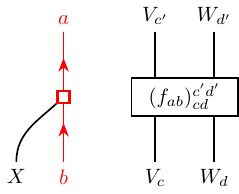}
\end{aligned}
\end{equation}
Recall from the text below \eqref{eq:D-via-a-Va} that $V_c$ and $W_d$ are just multiples of the tensor unit and we do not need to distinguish under- from overcrossings.
The isomorphism $\Gamma^2$ and its inverse are
\begin{equation}
    \Gamma^2_{(V,g),(W,h)} ~=~ \sum_{b,c,d} \
    \includegraphics[valign = c]{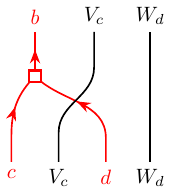}
    \qquad
    \big( \Gamma^2_{(V,g),(W,h)} \big)^{-1} ~=~ \sum_{b,c,d} \
    \includegraphics[valign = c]{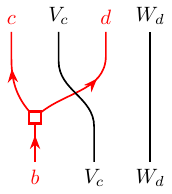}
\end{equation}
Here, in the target of $\Gamma^2$ and the source of $(\Gamma^2)^{-1}$, $b \otimes V_c \otimes W_d$ is understood as a subobject of $b \otimes Z_b$ with $Z_b$ as given in \eqref{eq:fused-quiver-vsp}.

\begin{figure}[tbp]
    \centering
    $\begin{aligned}
      & \delta_{d,d'} \ \includegraphics[valign = c]{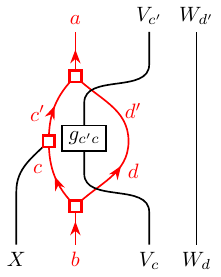}
      \ + \ \delta_{c,c'} \ \includegraphics[valign = c]{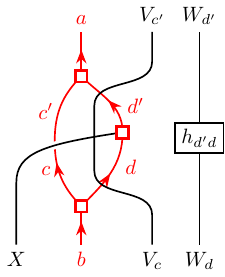}
      \\[2.5ex]
      ={}& \delta_{d,d'} \ \includegraphics[valign = c]{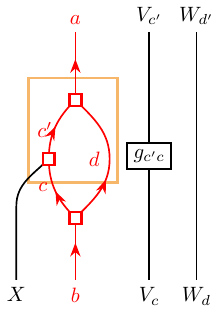}
      \ + \ \delta_{c,c'} \ \includegraphics[valign = c]{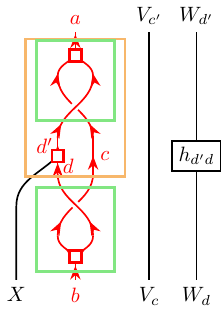}
      \\[2.5ex]
      ={}& \delta_{d,d'} \, \Gbox{\mathrm{G}^{(Xcd)a}_{c'b}} \
      \includegraphics[valign = c]{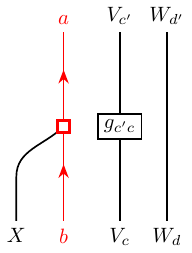}
      \ + \ \delta_{c,c'} \, \Rbox{\frac{\mathrm{R}^{(d'c)a}}{\mathrm{R}^{(dc)b}}}
      \Gbox{\mathrm{G}^{(Xdc)a}_{d'b}} \
      \includegraphics[valign = c]{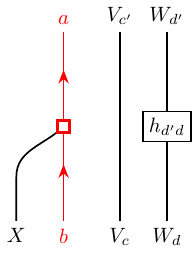}
    \end{aligned}$
    \caption{String diagram computation showing \eqref{eq:prop-modQX-CX-monoidal-aux1} componentwise.}
    \label{fix:prop-modQX-CX-monoidal-aux2}
\end{figure}

We will now check that $\Gamma^2$ is indeed a morphism in $\mathcal{C}_X$ by verifying the identity
\begin{equation}\label{eq:prop-modQX-CX-monoidal-aux1}
    \Gamma^2_{(V,g),(W,h)} \circ T 
    = F \circ \big(\mathrm{id}_X \otimes \Gamma^2_{(V,g),(W,h)}\big) \ .
\end{equation}
Composing both sides with $\mathrm{id}_X \otimes (\Gamma^2_{(V,g),(W,h)})^{-1}$ from the right and starting from the left hand side, the computation showing the above identity componentwise is given in Figure~\ref{fix:prop-modQX-CX-monoidal-aux2}, with the last expression being equal to $F$ with components $(f_{ab})_{cd}^{c'd'}$ as given in \eqref{eq:tensor-Q-mod-on-obj}.
\end{proof}

\begin{example}\label{ex:LY-tensor}
Continuing the Lee-Yang example, consider the $Q_\tau$-module $(\tau,\lambda)$ from part 1 in Example~\ref{ex:LY-Qtau-modules}. We will compute the tensor product $(Z,h) = (\tau,\lambda) \circledast (\tau,\mu)$ for $\lambda,\mu \in \mathbb{C}$, reproducing the result of \cite[Sec.\,4.3]{Manolopoulos:2009np}. By \eqref{eq:fused-quiver-vsp}, the multiplicity spaces of the underlying topological defect are
(recall that for $(\tau,\lambda)$ we have $V_{\mathbf{1}} = 0$, $V_{\tau} = \mathbb{C}$)
\be\begin{split}
    Z_{\mathbf{1}} &\,=\, V_{\mathbf{1}} \otimes V_{\mathbf{1}} \,\oplus\, V_{\tau} \otimes V_{\tau}
    \,=\, V_{\tau} \otimes V_{\tau} \,=\, \mathbb{C} \ ,
 \\    
    Z_{\tau} &\,=\, V_{\mathbf{1}} \otimes V_{\tau} 
    \,\oplus\, V_{\tau} \otimes V_{\mathbf{1}}
    \,\oplus\, V_{\tau} \otimes V_{\tau}
    \,=\, V_{\tau} \otimes V_{\tau} \,=\, \mathbb{C} \ .
\end{split}
\end{equation}
The components of $f$ are (we use the relation between $F$- and $G$-symbols mentioned in Footnote~\ref{fn:FG-rel})
\begin{equation}
    (f_{ab})_{\tau\tau}^{\tau\tau}
    \,=\, F^{(\tau\tau\tau)a}_{\tau b} \Big( \lambda + \frac{R^{(\tau\tau)a}}{R^{(\tau\tau)b}} \, \mu \Big) \ .
\end{equation}
We recall the conventions we use for minimal models in Section~\ref{sec:mm-funrel}, but let us state the values of the above constants for the Lee-Yang CFT already here,\footnote{\label{fn:MR-uses-conjugate-R}
These values agree with \cite[App.\,A.4]{Manolopoulos:2009np}, except that there the complex conjugate convention for $R$ is used. That is, \cite{Manolopoulos:2009np} uses $R^{(ab)c} = e^{-\pi i (h_c-h_a-h_b)}$, while here we use $R^{(ab)c} = e^{\pi i (h_c-h_a-h_b)}$.}
\begin{align}
    &R^{(\tau\tau)\mathbf{1}} = e^{2\pi i/5}
    ~,~~
    R^{(\tau\tau)\tau} = e^{\pi i/5} 
    ~,
    \nonumber \\
    &F^{(\tau\tau\tau)\mathbf{1}}_{\tau\tau}
    = 1
    ~,~~
    F^{(\tau\tau\tau)\tau}_{\tau \mathbf{1}}
    = \tfrac{1}{wd}
    ~,~~
    F^{(\tau\tau\tau)\tau}_{\tau \tau}
    = -\tfrac{1}{d}
    ~~,~~~\text{where}~
d = \tfrac{1-\sqrt{5}}2
\ .
\end{align}
The constant $w$ is a normalisation, and with the canonical normalisation of minimal model conformal blocks, it is given by
\begin{equation}
      w = \frac{\Gamma\big(\tfrac15\big)\Gamma\big(\tfrac65\big)}{\Gamma\big(\tfrac35\big)\Gamma\big(\tfrac45\big)} = 2.431.. ~,
\end{equation}
but one may as well set $w$ to $1$. The constant $d$ is the quantum dimension of $\tau$, $d = \dim(\tau)$. Note that as opposed to the unitary Fibonacci fusion category, for the non-unitary Lee-Yang model we have $d<0$. 
Substituting the above values, we get
\begin{equation}
    (f_{\mathbf{1}\tau})_{\tau\tau}^{\tau\tau} = 
    \lambda + e^{\pi i/5}  \mu 
    ~,~~
    (f_{\tau\mathbf{1}})_{\tau\tau}^{\tau\tau} 
    = \frac{1}{wd} \big( \lambda +  e^{-\pi i/5} \mu \big) 
    ~,~~
    (f_{\tau\tau})_{\tau\tau}^{\tau\tau} 
    = -\frac{1}{d} ( \lambda +  \mu )  
    \ .
\end{equation}
Comparing to Example~\ref{ex:LY-Qtau-modules}\,(2), we find that $(\tau,\lambda) \circledast (\tau,\mu)$ is simple unless $\mu = - e^{\pm \pi i /5} \lambda = e^{\mp 4 \pi i /5} \lambda$. For these two values of $\mu$ we get short exact sequences as follows:
\begin{align}\label{eq:LY-quiver-product}
    \mu &= e^{4 \pi i /5} \lambda
    ~:
    & (\tau,e^{2 \pi i /5} \lambda) &\to (\tau,\lambda) \circledast (\tau,e^{4 \pi i /5} \lambda)
    \to (\mathbf{1},0)
    \nonumber\\
    \mu &= e^{-4 \pi i /5} \lambda
    ~:
    &(\mathbf{1},0) &\to (\tau,\lambda) \circledast (\tau,e^{-4 \pi i /5} \lambda)
    \to (\tau,e^{-2 \pi i /5} \lambda)
\end{align}
Indeed, in the first line $(f_{\mathbf{1}\tau})_{\tau\tau}^{\tau\tau} = 0$, and so $(\tau,\nu)$ is a submodule of $(\tau,\lambda) \circledast (\tau,\mu)$ for 
\begin{equation}
    \nu = (f_{\tau\tau})_{\tau\tau}^{\tau\tau} 
    = -\frac{1}{d} \big( 1 +  e^{4 \pi i /5} \big) \lambda 
    = e^{2 \pi i /5} \lambda \ .
\end{equation}
In the second line, $(f_{\tau\mathbf{1}})_{\tau\tau}^{\tau\tau} = 0$, and $(\mathbf{1},0)$ is a submodule.
These exact sequences agree with \cite[Sec.\,4.3]{Manolopoulos:2009np} (recall Footnote~\ref{fn:MR-uses-conjugate-R}).
\end{example}

\subsubsection*{Trace relations}

Given a module $(V,f)$ of the quiver $Q_X$ and an oriented cycle (possibly of length zero) in $Q_X$, we can consider the trace of the corresponding linear maps. If for two modules $(V,f)$ and $(W,g)$ all such traces agree, we say that they are \textit{trace-equivalent} and write 
\begin{equation}
    (V,f) \sim_{\tr} (W,g) \ .
\end{equation}
In more detail, let $a_0,a_1,\dots,a_N$ be a sequence of vertices of the quiver $Q_X$ such that $a_0 = a_N$ and such that for all $n=0,\dots,N-1$ there is an arrow $a_n \to  a_{n+1}$. This defines an oriented cycle in $Q_X$. Then $(V,f)$ and $(W,g)$ are trace-equivalent if for all such oriented cycles we have that
\begin{equation}\label{eq:Vf-Wg-same-traces}
    \tr_{V_{a_0}}\!\big( \, f_{a_N a_{N-1}} \circ \cdots \circ f_{a_2 a_1} \circ f_{a_1 a_0} \, \big)
    ~=~
    \tr_{W_{a_0}} \! \big(\, g_{a_N a_{N-1}} \circ \cdots \circ g_{a_2 a_1} \circ g_{a_1 a_0} \,\big) \ .
\end{equation}
If the cycle is of length zero at a vertex $a_0$, by convention we take the above formula to mean
\begin{equation}
    \tr_{V_{a_0}}(\mathrm{id})
    ~=~
    \tr_{W_{a_0}}(\mathrm{id})
    \qquad \Leftrightarrow \qquad 
    \dim_\mathbb{C}(V_{a_0}) = \dim_\mathbb{C}(W_{a_0}) \ .
\end{equation}
Thus, trace equivalence implies that all component vector spaces are isomorphic ($V_a \cong W_a$ for all $a \in \mathcal{I}$). If we pick a family of isomorphisms $\phi_a : V_a \to W_a$, $a \in \mathcal{I}$, we can rewrite trace equivalence solely in terms of $V$:
\begin{equation}\label{eq:trace-equiv-same-space}
    (V,f) \sim_{\tr} (W,g) 
    \qquad \Leftrightarrow \qquad 
    (V,f) \sim_{\tr} (V,g') 
    ~~\text{for}~~
    g'_{ba} = \phi_b^{-1} \circ g_{ba} \circ \phi_a
    \ .
\end{equation}
In other words, when working with trace equivalent quiver modules, up to isomorphism we may assume that the underlying vector spaces are the same.

The traces in \eqref{eq:Vf-Wg-same-traces} appear in the context of geometric invariant theory in the Le Bruyn--Procesi Theorem \cite[Thm.\,1]{le1990semisimple} which states that the ring of polynomial invariants of quiver-modules with fixed dimension vector is generated by the above traces.

\begin{remark}\label{rem:trace-equiv-gives-direct-sum}
In the application to defect operators in CFT, we note that if two $Q_X$-modules are trace equivalent, they have the same defect operator:
\begin{equation}
    (V,f) \sim_{\tr} (W,g)
    \quad \Rightarrow \quad
    \mathcal{D}[V,f] = \mathcal{D}[W,g] \ .
\end{equation}
To see this, let $D(\eta) = \Gamma(V,f)$ and recall the definition of the perturbed defect operator in \eqref{eq:defect-operator-as-pert-series}. 
Consider $n$ defect fields $\psi(x_1) \cdots \psi(x_n)$ inserted on the circular defect $D$ as in the integrand of \eqref{eq:defect-operator-as-pert-series}.
From the explicit form of $\Gamma$ in \eqref{eq:modQ-CX-functor-diag}, we can expand $D$ into simple topological defects and multiplicity spaces, and each such integrand becomes a sum over a product of defect changing fields $a_i \to a_{i+1}$ between simple defects, which is independent of $(V,f)$, and a trace of the product of the corresponding maps $f_{a_{i+1}a_i}$ between the multiplicity spaces.
\end{remark}

Note that when $(V,f)$ and $(W,g)$ are isomorphic, they evidently have the same traces, 
\begin{equation}
    (V,f) \cong (W,g) \quad \Rightarrow \quad (V,f) \sim_{\tr} (W,g) \ .
\end{equation}
The converse does not hold as we will see in the example of the Lee-Yang CFT below. 

A more interesting trace relation arises from an invariant submodule. Namely given $(V,f)$, suppose we find vector subspaces $U_a \subset V_a$, $a \in \mathcal{I}$ such that $f$ preserves the $U_a$ in the sense that 
\begin{equation}\label{eq:quiver-sub-module}
    f_{ba}(U_a) \subset U_b \quad \text{for each arrow} ~~a \to b 
    \ .
\end{equation}
We denote the restriction of $f$ to $U$ by $f|_U$ and obtain the new quiver module $(U,f|_U)$. 
Because of \eqref{eq:quiver-sub-module} we also obtain well-defined maps between the quotient vector spaces $V_a/U_a$, namely, for each arrow $a \to b$ we get an induced linear map $\overline f_{ba} : V_a/U_a \to V_b/U_b$. Let us denote the resulting quiver module by $(V/U, \overline f)$. 

It is not hard to see that whenever we have a submodule $(U,f|_U)  \subset (V,f)$, all maps inside the traces \eqref{eq:Vf-Wg-same-traces} are block-triangular, and that the traces for the two diagonal blocks coincide with the traces for $(U,f|_U)$ and $(V/U, \overline f)$:
\begin{equation}
    (U,f|_U) \, \subset \, (V,f)
    \quad \Rightarrow \quad
    (V,f) \sim_{\tr} (U,f|_U) \oplus (V/U, \overline f) \ .
\end{equation}
We summarise the above discussion in the following lemma:

\begin{lemma}
If $(A,a) \to (B,b) \to (C,c)$ is a short exact sequence in $\mathrm{mod}(Q_X)$ then $(B,b) \sim_{\tr} (A,a) \oplus (C,c)$.
\end{lemma}

\begin{remark}\label{rem:trace-is-Gr}
In fact, one can show via geometric invariant theory that trace equivalence is precisely equality in the Grothendieck ring,\footnote{
    We thank Alastair King for explaining this point to us.
}
\begin{equation}
    (V,f) \sim_{\tr} (W,g)
    \quad \Leftrightarrow \quad
    [V,f] = [W,g] \text{ in }
    \mathrm{Gr}(\mathrm{mod}(Q_X)) \ .
\end{equation}
The easy direction $\Leftarrow$ amounts to the above lemma. For the difficult direction $\Rightarrow$, one uses that by \cite[Thm.\,1]{le1990semisimple} the traces \eqref{eq:Vf-Wg-same-traces} generate the ring of invariants for quiver modules with a fixed dimension vector, and hence that trace equivalence is GIT equivalence. Then in \cite[Prop.\,3.2]{king1994moduli} (see also \cite[(12.6)]{Artin:1969}), GIT equivalence is shown to be the same as having equivalent Jordan-Hölder series.
Note that in \cite[Prop.\,3.2]{king1994moduli} one has to set $\theta=0$, so that all modules are semistable and the stable ones are precisely the simple modules.
\end{remark}

\begin{example}\label{ex:LY-trace}
As an example of trace-equivalence, consider the Lee-Yang CFT and specifically the quiver module from Example~\ref{ex:LY-Qtau-modules}\,(2). Namely $(V,f)$ with $V_{\mathbf{1}} = V_\tau = \mathbb{C}$ and three components $f_{\mathbf{1}\tau}, f_{\tau\mathbf{1}}, f_{\tau\tau} \in \mathbb{C}$ for $f$. It is enough to evaluate the traces on a generating set of closed loops, for which we take $\tau \to \tau$ and $\tau \to \mathbf{1} \to \tau$. Given another quiver module $(V,g)$ this gives
\begin{equation}\label{eq:LY-trace-eq-example}
    (V,f) \sim_{\tr} (V,g)
    \quad \Leftrightarrow \quad
    f_{\tau\tau} = g_{\tau\tau} \text{ and }
    f_{\tau\mathbf{1}} f_{\mathbf{1}\tau} 
    = g_{\tau\mathbf{1}} g_{\mathbf{1}\tau} \ .
\end{equation}
This is weaker than being isomorphic. 
For example if $f_{\tau\mathbf{1}} = 0$, $f_{\mathbf{1}\tau} \neq 0$ and 
$g_{\tau\mathbf{1}} = 0$, $g_{\mathbf{1}\tau} = 0$, then by \eqref{eq:LY-isom} we have $(V,f) \ncong (V,g)$, but still $(V,f) \sim_{\tr} (V,g)$.
As in this case $(V,g)$ has no non-zero maps between $V_\mathbf{1}$ and $V_\tau$, it decomposes into a direct sum $(V,g) = \mathbf{1} \oplus (\tau,g_{\tau\tau})$. Together with \eqref{eq:LY-trace-eq-example} this shows:
\begin{equation}
    f_{\tau\mathbf{1}} = 0
    \quad \Rightarrow \quad (V,f) \sim_{\tr} \mathbf{1} \oplus (\tau,f_{\tau\tau})\ .
\end{equation}
The same conclusion holds if $f_{\mathbf{1}\tau} = 0$. Comparing to Example~\ref{ex:LY-Qtau-modules}\,(2), we see that these are precisely the cases where $(V,f)$ is not simple and becomes the middle term in an exact sequence, as has to be the case by Remark~\ref{rem:trace-is-Gr}.
\end{example}

Trace relations respect the tensor product in $\mathrm{mod}(Q_X)$:

\begin{lemma}
Let $(V,g), (V',g'), (W,h), (W',h')$ be $Q_X$-modules with $(V,g) \sim_{\tr} (V',g')$ and $(W,h) \sim_{\tr} (W',h')$. Then also 
$(V,g) \circledast (W,h)
~\sim_{\tr}~
(V',g') \circledast (W',h')$.
\end{lemma}

\begin{proof}
By \eqref{eq:trace-equiv-same-space} we may assume that $V=V'$ and $W=W'$. Write
$(Z,f) = (V,g) \circledast (W,h)$ and
$(Z,f') = (V,g') \circledast (W,h')$ as in \eqref{eq:VgxWh-Z}. We need to show that $(Z,f) \sim_{\tr} (Z,f')$.
Let $a_0,\dots,a_N$ be an oriented cycle and consider the trace
\begin{equation}
    T = \tr_{Z_{a_0}}\!\big( \, f_{a_N a_{N-1}} \circ \cdots \circ f_{a_2 a_1} \circ f_{a_1 a_0} \, \big) \ .
\end{equation}
By \eqref{eq:VgWh-tensor-f},
each term $f_{a_{i+1},a_i}$ consists of components $(f_{a_{i+1}a_i})_{c_id_i}^{c_{i+1}d_{i+1}}$. The above trace can be expanded into a sum over these components,
\begin{equation}
    T = 
    \sum_{c_0,c_1,\dots,c_{N-1}} 
    \sum_{d_0,d_1,\dots,d_{N-1}} 
    \tr_{V_{c_0} \otimes W_{d_0}}\!\big( \, 
    \cdots (f_{a_{i+1}a_i})_{c_id_i}^{c_{i+1}d_{i+1}} \cdots \, \big) \ .
\end{equation}
Each $(f_{a_{i+1}a_i})_{c_id_i}^{c_{i+1}d_{i+1}}$ is a sum of two terms as in \eqref{eq:tensor-Q-mod-on-obj}, one involving $\delta_{d_i,d_{i+1}} \, g_{c_{i+1}c_i} \otimes \mathrm{id}_{W_{d_i}}$ and the other $\delta_{c_i,c_{i+1}} \, \mathrm{id}_{V_{c_i}} \otimes h_{d_{i+1}d_i}$.
Expanding this sum in turn, each resulting summand takes the form of a prefactor times a product of a trace over $h$'s for some oriented cycle with a trace over $g$'s for a potentially different cycle. By assumption, in these traces we can replace $h$ by $h'$ and $g$ by $g'$. Undoing the above expansions results in $(Z,f')$.
\end{proof}

\begin{remark}
For the application to perturbed defects, we can now summarise the tools we have developed. Recall that we are studying the properties of topological defects $D$, perturbed by a holomorphic defect field described by a fixed vector $\phi_X$ in a representation $X$ of the VOA, and a three-fold junction $\eta$ from $X$ and $D$ to $D$ in the 3d\,TFT as in \eqref{eq:defect-field-in-chiral-TFT}. 
The quiver $Q_X$ is obtained from the fusion of simple line defects in the 3d\,TFT with $X$, see \eqref{eq:QX-quiver-arrows}. 
Via the equivalence $\Gamma$ in Proposition~\ref{prop:modQX-CX-monoidal}, the data $D(\eta)$ of a defect perturbation can be given in terms of quiver modules $(V,f) \in \mathrm{mod}(Q_X)$.
\begin{itemize}
    \item 
Under the convergence assumption in Section~\ref{sec:chiral-TFT-defects}, to each $(V,f) \in \mathrm{mod}(Q_X)$ we can assign a defect operator $\mathcal{D}[V,f] : \mathcal{H} \to \mathcal{H}$ on the state space of the CFT which commutes with $L_0$ and $\overline L_0$. 

\item 
If $(V,f) \sim_\mathrm{tr} (W,g)$, then $\mathcal{D}[V,f] = \mathcal{D}[W,g]$. Trace-equivalence is described in \eqref{eq:Vf-Wg-same-traces}.

\item 
The tensor product of quiver modules corresponds to the composition of defect operators: for $(V,f), (W,g) \in \mathrm{mod}(Q_X)$ we have
\be    \mathcal{D}[V,f]\,\mathcal{D}[W,g] 
    \,=\,
    \mathcal{D}\big[(V,f) \circledast  (W,g)\big] \ .
\ee
The tensor product $(V,f) \circledast  (W,g)$ is made explicit in \eqref{eq:tensor-Q-mod-on-obj}.
\end{itemize}
These three points are what we use to determine the functional relations between perturbed defect operators. \end{remark}

\begin{example}\label{ex:LY-fun-rel}
Let us illustrate the procedure outlined in the above remark in the case of the Lee-Yang CFT. Its simple topological defects are labelled by $\mathcal{I} = \{\mathbf{1},\tau\}$, and we take $X = \tau$ for the representation of the perturbing field. The quiver $Q_X$ is given in Example~\ref{ex:LY-Qtau-modules}.
Consider the quiver module $(\tau,\lambda)$ from Example~\ref{ex:LY-Qtau-modules}\,(1). In Example~\ref{ex:LY-tensor} we computed the tensor product, and, using the language of trace-equivalence, we found
\begin{equation}\label{eq:LY-fun-rel-quiver}
    (\tau,\zeta \lambda) \circledast (\tau,\zeta^{-1} \lambda)
    \sim_{\tr} 
    \mathbf{1} \oplus (\tau,\lambda) 
    \qquad \text{where} \quad \zeta = e^{2 \pi i / 5} \ .
\end{equation}
This implies the following functional relation for corresponding perturbed defect operators:
\begin{equation}\label{eq:LY-fun-rel}
    \mathcal{D}_\tau(\zeta \lambda) \, \mathcal{D}_\tau(\zeta^{-1} \lambda) \,=\,
    \mathrm{id} + \mathcal{D}_\tau(\lambda)  \ .
\end{equation}
\end{example}

\section{Functional relations in the asymptotic semiring}\label{sec:semiring}

This section contains the main new contributions of the present paper: the asymptotic semiring constructed from the fusion rule semiring of topological defects (Definition~\ref{def:asymptotic-semiring}), the condition of $\alpha$-asymptotics which is an exponential growth condition on perturbed defects (Definition~\ref{def:alpha-asymptotics}), and, finally, Theorem~\ref{thm:semiring-hom}, which states that under the assumption of $\alpha$-asymptotics, the defect RG-flow defines a semiring homomorphism from the fusion rule semiring of translation invariant defects to the asymptotic semiring. The actual definitions and statements in this section are phrased in greater generality, but defect flows are the main use case we have in mind.

\subsection{The asymptotic semiring}

We review the definition of a semiring and then go through the examples and constructions we will need.

\begin{definition}\label{def:semiring}
A \textit{semiring} is a tuple $(S,+,0,\cdot,1)$ where $(S,+,0)$ is a commutative monoid, and $\cdot : S\times S \to S$ is an associative product with unit $1$, which is distributive,
\[
    (a+b)\cdot c = a \cdot c + b \cdot c
    ~~,\quad
    a \cdot (b+c) = a \cdot b + a \cdot c \ ,
\]
and which satisfies $a \cdot 0 = 0 = 0 \cdot a$.
\end{definition}

A semiring is also called a \textit{rig} (``ring without negatives''). A semiring where each element has an additive inverse (so that $(S,+,0)$ is an abelian group) is a ring. In a ring, $a \cdot 0 = 0$ follows from $a \cdot 0 = a \cdot (0+0)$, but that argument uses the negative of $a \cdot 0$.
For more on semirings, see e.g.\ \cite{golan2013semirings}.

\begin{example}\label{ex:semirings}
\begin{enumerate}
    \item 
\textit{The semiring $\mathbb{R}_\mathrm{max}$:} As a set $\mathbb{R}_\mathrm{max} := \mathbb{R} \cup \{- \infty \}$, that is, the real numbers together with minus infinity. Define the two operations $\oplus,\odot$ for $a,b \in \mathbb{R}_\mathrm{max}$ as
\begin{equation}
    a \oplus b = \max(a,b)
    ~~,\quad 
    a \odot b = a+b \ .
\end{equation}
Then $\mathbb{R}_\mathrm{max} = (\mathbb{R} \cup \{-\infty\}, \oplus, -\infty, \odot, 0)$ is a semiring. Let us check distributivity as an example: For $a,b,c \in \mathbb{R}_\mathrm{max}$,
\begin{equation}
    (a \oplus b) \odot c
    = \max(a,b)+c
    ~~,\quad
    (a \odot c) \oplus (b \odot c) = \max(a+c,b+c) \ .
\end{equation}
This is clearly equal if $a,b,c\in\mathbb{R}$, and one can check the cases where one or more of $a,b,c$ are $-\infty$.
For some applications of $\mathbb{R}_\mathrm{max}$, see e.g.\ \cite{quadrat_icm1994}.

\item \textit{Fusion rule semirings:} A \textit{unital $\mathbb{Z}_+$-ring} \cite[Sec.\,3.1]{etingof2015tensor} is a ring $F$ with $\mathbb{Z}$-basis $\{b_i\}_{i \in \mathcal{I}}$ for some set $\mathcal{I}$, 
\begin{equation}\label{eq:based-ring-direct-sum}
    F = \bigoplus_{i \in \mathcal{I}} \mathbb{Z} \,b_i \ ,
\end{equation}
such that the unit $1$ is one of the basis elements, and the structure constants are non-negative integers:
\begin{equation}
    b_i \cdot b_j = \sum_{k \in \mathcal{I}} N_{ij}^{~k} \, b_k
    \qquad \text{where} ~~ N_{ij}^{~k} \in \mathbb{Z}_{\ge 0} \ .
\end{equation}
Note that for fixed $i,j \in \mathcal{I}$, $N_{ij}^{~k} \neq 0$ only for finitely many $k$, as $b_i \cdot b_j$ has to lie in the direct sum \eqref{eq:based-ring-direct-sum}.

Next suppose there is a bijection $(-)^* : \mathcal{I} \to \mathcal{I}$
which induces an anti-automorphism on $F$ and for which $b_{i^*} \cdot b_i = d_{i} 1 + \dots$ with $d_i>0$ and ``\dots'' a $\mathbb{Z}$-linear combination of basis elements other than $1$, and similar for $b_{i} \cdot b_{i^*}$.

By $F_{\ge0}$ we denote the semiring which is obtained from $F$ by only allowing non-negative integral linear combinations of basis elements,
\begin{equation}
    F_{\ge 0} = \bigoplus_{i \in \mathcal{I}} \mathbb{Z}_{\ge 0} \, b_i \ .
\end{equation}
Note that in $F_{\ge 0}$, $xy=0$ implies $x=0$ or $y=0$. Indeed, suppose $xy=0$ but $x,y \neq 0$. Then also $x^* x y=0$, but $x^*x = n 1 + r$, where $n>0$ and $r$ does not contain the basis element $1$. Hence $x^*x y = n y + r y \neq 0$ as already the first summand is non-zero, and all structure constants are non-negative, which is a contradiction.

As our main examples come from fusion rules in suitable rigid monoidal categories as in the next example, we refer to $F$ as above with the bijection $(-)^*$ also as a \textit{fusion rule}, and to the corresponding semiring $F_{\ge 0}$ as a \textit{fusion rule semiring}.\footnote{ 
We avoid the more obvious name ``fusion semiring'' as the term ``fusion ring'' is already used in
\cite[Sec.\,3.1]{etingof2015tensor} to denote a based ring with a finite indexing set $\mathcal{I}$ for its basis, in accordance with a ``fusion category'' having a finite set of isomorphism classes of simple objects. 
Furthermore, in a based ring $(-)^*$ is required to be an involution with an extra property, rather than just a bijection.
In our setting, we need to relax both conditions. The indexing set $\mathcal{I}$ of a basis can be infinite in our examples, and we do not necessarily have $i^{**} = i$. Both effects can be seen in the Lee-Yang example: By Example~\ref{ex:LY-Qtau-modules} there are already continuously many simple objects of the form $(\tau,\lambda)$, and by \cite[Sec.\,4.3]{Manolopoulos:2009np} (see also \eqref{eq:LY-quiver-product}) only the 5-fold dual of $(\tau,\lambda)$ is again isomorphic to $(\tau,\lambda)$, not the 2-fold dual (there is a typo in \cite{Manolopoulos:2009np} which states 10-fold dual). Hence we chose the slightly different name to avoid a direct clash of definitions.}

\item \textit{Grothendieck semirings:} Building on the previous example, let $\mathcal{M}$ be an abelian $\mathbb{C}$-linear rigid monoidal category with simple tensor unit and where all objects have finite composition length. The Grothendieck ring $\mathrm{Gr}(\mathcal{M})$ is the free abelian group generated by the (possibly infinitely many) isomorphism classes $X_i$ of simple objects in $\mathcal{M}$. An object $X \in \mathcal{M}$ corresponds to the element $[X] \in \mathrm{Gr}(\mathcal{M})$ with $[X] = \sum_i [X:X_i] X_i$, where $[X:X_i]$ is the multiplicity of $X_i$ in the composition series of $X$, see \cite[Def.\,1.5.8]{etingof2015tensor}. The product on $\mathrm{Gr}(\mathcal{M})$ is given by $[X] \cdot [Y] = [X \otimes Y]$, see \cite[Sec.\,4.5]{etingof2015tensor}. The \textit{Grothendieck semiring} is the fusion rule semiring defined as
\begin{equation}
    \mathrm{Gr}(\mathcal{M})_{\ge 0} := \bigoplus_i \mathbb{Z}_{\ge 0} X_i \ .
\end{equation}
If $X \neq 0$ in $\mathcal{M}$, then also $[X] \neq 0$ in $\mathrm{Gr}(\mathcal{M})_{\ge 0}$ as $X$ has a non-zero number of composition factors. If $X \otimes Y =0$ in $\mathcal{M}$, then already $X =0$ or $Y=0$. For example, if $X \neq 0$, rigidity of $\mathcal{M}$ gives the evaluation map $X^* \otimes X \to \mathbf{1}$, which is non-zero and hence surjective as $\mathbf{1}$ is simple. The tensor product of $\mathcal{M}$ is exact (as $\mathcal{M}$ is rigid) and we get a surjective morphism $X^* \otimes X \otimes Y \to Y$. Hence $X \otimes Y = 0$ with $X\neq 0$ implies $Y=0$. Also, clearly, if $X \oplus Y=0$, then $X=Y=0$. As an aside, if the tensor unit is not simple, we can have situations where $X \otimes Y=0$ for $X,Y \neq 0$, e.g.\ take finite-dimensional modules over the $\mathbb{C}$-algebra $\mathbb{C} \oplus \mathbb{C}$ and take the tensor product of a module where the first copy of $\mathbb{C}$ acts trivially with one where the second copy of $\mathbb{C}$ acts trivially.
\end{enumerate}
\end{example}

Let $S$ be a semiring. On $S \times \mathbb{R}_\mathrm{max}$ we define a new addition and multiplication, different from using pointwise operations:
for $(a,x), (b,y) \in S \times \mathbb{R}_\mathrm{max}$ we set
\begin{equation}\label{eq:SxRmax_operations}
    (a,x)+(b,y) = \begin{cases}
        (a,x) &; x>y \\
        (a+b,x) &; x=y \\
        (b,y) &; x<y 
    \end{cases}
    \quad , \qquad
    (a,x) \cdot (b,y) = (ab,x+y) \ .
\end{equation}
The zero for addition is $(0,-\infty)$, and the unit for multiplication is $(1,0)$. This is \textit{not} a semiring as it is not distributive (try $\big((1,1)+(1,0)\big)\cdot(1,-\infty)$). Before we repair this, let us motivate why we want to look at these operations in the first place.

\begin{remark}\label{rem:Dmax-and-exp}
In our application to renormalisation group flows of holomorphically perturbed defects, we want to describe the behaviour of terms $D \exp(a e^{i \alpha \theta} r^\alpha)$ for $D$ a defect operator, $a \in \mathbb{C}$ fixed, $\theta$ an angle, and $r$ large. For the product we have
\begin{equation}
    D_1 \exp(a_1 e^{i \alpha \theta} r^\alpha)\, D_2  \exp(a_2 e^{i \alpha \theta} r^\alpha)
    = D_1 D_2  \exp((a_1+a_2) e^{i \alpha \theta} r^\alpha) 
\end{equation}
For the sum it depends on which term is dominant. Write $R_j = \mathrm{Re}(a_j e^{i \alpha \theta})$, $j=1,2$. For $R_1=R_2$ there is no dominant behaviour, while for $R_1\neq R_2$ we have
\begin{equation}
    D_1 \exp(a_1 e^{i \alpha \theta} r^\alpha)\,+\, D_2 \exp(a_2 e^{i \alpha \theta} r^\alpha)
    \underset{r \to \infty}{\sim} 
    \begin{cases}
        D_1 \exp(a_1 e^{i \alpha \theta} r^\alpha) &; R_1>R_2 \\
        D_2 \exp(a_2 e^{i \alpha \theta} r^\alpha) &; R_1<R_2
    \end{cases}
\end{equation}
If we write $\mathbb{D}_\infty$ for the fusion rule semiring of topological defects, the above behaviour of terms $D \exp(a e^{i \alpha \theta} r^\alpha)$ under addition and composition looks almost like that of $(D, \mathrm{Re}(a e^{i \alpha \theta})) \in \mathbb{D}_\infty \times \mathbb{R}_\mathrm{max}$ from \eqref{eq:SxRmax_operations}, except for two points:
\begin{itemize}
    \item If $D=0$, then also $D \exp(a e^{i \alpha \theta} r^\alpha) = 0$, independent of $a$, while $(D, \mathrm{Re}(a e^{i \alpha \theta})) \neq 0$ in $\mathbb{D}_\infty \times \mathbb{R}_\mathrm{max}$.
    \item If $R_1=R_2$ in the notation above, then there may be no dominant term. For example, take $a_1=i$, $a_2=-i$, $\theta=0$. The sum $D_1 e^{i r^\alpha} + D_2 e^{-i r^\alpha}$ does not have a leading  term of the form $D e^{b r^\alpha}$, while 
    $(D_1,\mathrm{Re}(i r^\alpha)) + (D_2,\mathrm{Re}(-i r^\alpha)) = (D_1,0) + (D_2,0) = (D_1+D_2,0)$.
\end{itemize}
\end{remark}

To fix the issue in the first point, we need two more notions \cite[p.\,4]{golan2013semirings}:

\begin{definition}
A semiring $S$ is called
\begin{itemize}
    \item \textit{zero-sum-free} if for all $s,t\in S$ we have that $s+t=0$ implies $s=t=0$;
    \item \textit{entire} if it has no zero divisors, that is, if for all $s,t \in S$ we have that $st=0$ implies $s=0$ or $t=0$.
\end{itemize}
\end{definition}

All three semirings in Example~\ref{ex:semirings} are zero-sum-free and entire. A ring with $0 \neq 1$ is never zero-sum-free as $1+(-1)=0$. As the property of being zero-sum-free is required by some of the constructions below, we need to work with semirings also in Examples~\ref{ex:semirings}\,(2,3), even though there are canonical ring completions in these cases.

\begin{lemma}\label{lem:Dmax-def}
Let $S$ be a zero-sum-free and entire semiring. Then
\begin{equation}
    S_\mathrm{max} ~:=~ \big\{(0,-\infty)\big\} \,\cup\, \big\{ (a,x) \,\big|\, a \in S, a \neq 0, x \in \mathbb{R} \big\} ~\subset~ S \times \mathbb{R}_\mathrm{max} \ ,
\end{equation}
with operations induced from $S \times \mathbb{R}_\mathrm{max}$ as in \eqref{eq:SxRmax_operations},
is a zero-sum-free and entire semiring. Furthermore, the map $\pi_2 : S_\mathrm{max} \to \mathbb{R}_\mathrm{max}$, $(a,x) \mapsto x$ is a semiring homomorphism.
\end{lemma}

\begin{proof}
The addition in \eqref{eq:SxRmax_operations} closes on $S_\mathrm{max}$ because $S$ is zero-sum-free. The multiplication closes because $S$ is entire.
That $S_\mathrm{max}$ is again zero-sum-free and entire is immediate. Next we check the defining properties of a semiring as in Definition~\ref{def:semiring}.

Associativity of the multiplication is clear. Associativity of addition requires distinguishing cases. We need to show
\begin{equation}
    \big( (a,x) + (b,y) \big) + (c,z)
    ~=~
    (a,x) + \big( (b,y) + (c,z) \big) \ .
\end{equation}
If one of $x,y,z$ is strictly larger than the other two, then both sides are just equal to that term. If $x=y=z$, both sides are also equal. It remains to consider the three cases where two of $x,y,z$ are equal and larger than the third. Let us do $x=y>z$ as an example:
\begin{align}
    \big( (a,x) + (b,y) \big) + (c,z) &= (a+b,x) + (c,z) = (a+b,x)
     \ ,
     \nonumber \\
    (a,x) + \big( (b,y) + (c,z) \big) 
    &= (a,x) + (b,y) = (a+b,x)
    \ .
\end{align}
The other cases are similar. For distributivity we need to show
\begin{equation}
    \big( (a,x)+(b,y) \big) \cdot (c,z)
    ~=~
    (a,x)\cdot (c,z)+(b,y)\cdot (c,z) \ .
\end{equation}
We distinguish the cases $z \in \mathbb{R}$ and $z = -\infty$. Suppose first $z \in \mathbb{R}$.
In each case of $x>y$, $x=y$, or $x<y$, one can check that both sides are equal. For example, for $x>y$:
\begin{align}
    \big( (a,x)+(b,y) \big) \cdot (c,z)
    &= (a,x) \cdot (c,z) = (ac,x+z)  \ ,
    \nonumber \\
    (a,x)\cdot (c,z)+(b,y)\cdot (c,z)
    &= (ac,x+z)+(bc,y+z) = (ac,x+z) \ .
\end{align}
Note that in the second line we use that $x>y$ implies $x+z>y+z$, which is where we need the assumption $z \in \mathbb{R}$.
In case $z=-\infty$, by construction we have $c=0$, and both lines above are equal to $(0,-\infty)$.
This uses $(s,x) \cdot (0,-\infty) = (0,-\infty) = (0,-\infty) \cdot (s,x)$, which also establishes the last requirement in Definition~\ref{def:semiring}.

Finally, that the projection onto $\mathbb{R}_\mathrm{max}$ is a semiring homomorphism is clear since on the second entry the operations \eqref{eq:SxRmax_operations} are just those of $\mathbb{R}_\mathrm{max}$.
\end{proof}

This solves the first issue in Remark~\ref{rem:Dmax-and-exp}, because now the only element in $S_\mathrm{max}$ of the form $(0,x)$ is $(0,-\infty)$, which is the zero element in $S_\mathrm{max}$.
To solve the second issue, we will look at $\theta$-dependent functions with values in $S_\mathrm{max}$, up to an equivalence relation. Let
\begin{equation}
    \mathbb{S} = \mathbb{R} / 2\pi\mathbb{Z}
\end{equation}
be the unit circle, parametrised as $e^{i\theta}$, $\theta \in \mathbb{S}$. The functions
$\mathrm{Fun}(\mathbb{S},S_\mathrm{max})$
are again a semiring with pointwise addition and multiplication (still zero-sum-free but no longer entire). The idea is to ``avoid'' the problematic points where there is no dominant asymptotic behaviour in the example in Remark~\ref{rem:Dmax-and-exp} by declaring two functions to be equal if they agree outside an arbitrary finite set. We can now state the first main definition in this section:

\begin{definition}\label{def:asymptotic-semiring}
Let $S$ be a zero-sum-free and entire semiring. The \textit{asymptotic semiring} is
\begin{equation}
    \mathcal{A}(S) := \mathrm{Fun}(\mathbb{S},S_\mathrm{max}) / \sim \ ,
\end{equation}
where $f \sim g$ iff $f(\theta)=g(\theta)$ for all but finitely many $\theta \in \mathbb{S}$.
For $[f] \in \mathcal{A}(S)$ with $f(\theta) = (a(\theta),x(\theta))$ we will use the notation 
\begin{equation}
[f] = [a(\theta),x(\theta)] \ ,    
\end{equation}
and we will refer to $a(\theta)$ as the \textit{asymptotic defect} and $x(\theta)$ as the \textit{asymptotic exponent}.
\end{definition}

\begin{remark}\label{rem:asympt-semiring-prop}
\begin{enumerate}
\item The map $j : S \to \mathcal{A}(S)$ given by $j(0) = [0,-\infty]$ and, for $s \neq 0$ by $j(s) = [s,0]$, where $[s,0]$ is independent of $\theta$, is a semiring homomorphism. 

    \item The semiring homomorphism $\pi_2$ in Lemma~\ref{lem:Dmax-def} induces a semiring ho\-mo\-mor\-phism
\begin{equation}
    \mathcal{A}(S) \to 
    \mathrm{Fun}(\mathbb{S},\mathbb{R}_\mathrm{max}) / \sim
    ~~,\quad [f] \mapsto [\pi_2 \circ f] \ .
\end{equation}

\item 
Any bijection $h : \mathbb{S} \to \mathbb{S}$ induces a semiring automorphism of $\mathcal{A}(S)$ by pullback along $h$, 
\begin{equation}
    h^*([a(\theta),x(\theta)]) = [a(h(\theta)),x(h(\theta))] \ .
\end{equation}
Any function $s : \mathbb{S} \to \mathbb{R}_{>0}$ gives a semiring automorphism of $\mathcal{A}(S)$ via
\begin{equation}
    [a(\theta),x(\theta)] \mapsto [a(\theta),s(\theta)\,x(\theta)] \ .
\end{equation}

\item Combining the two automorphisms in the previous point, we get a family of $\mathbb{C}^\times$-actions on $\mathcal{A}(S)$ which will be important below. Namely, fix $\alpha \in \mathbb{R}$ and define
\begin{equation}\label{eq:Cx-action-V-on-asymp}
    \mathsf{V}^\alpha : \mathbb{C}^\times \to \mathrm{Aut}(\mathcal{A}(S))
\end{equation}
as, for $\lambda = r e^{i \varphi}$, $r>0$, 
\begin{equation}
    \mathsf{V}^\alpha_\lambda([a(\theta),x(\theta)])
    =[a(\theta+\varphi),r^\alpha x(\theta+\varphi)] \ .
\end{equation}
\end{enumerate}
\end{remark}

\subsection{Homomorphism from perturbed defects to the asymptotic semiring}

Next we want to generate elements in the asymptotic semiring via some kind of limiting process, mimicking the large coupling constant limit of a holomorphically perturbed defect. 
The setting we will use to talk about limits is to start from a (possibly infinite-dimensional, not necessarily complete) complex inner product space $\mathcal{H}$. We equip the endomorphisms $\mathrm{End}(\mathcal{H})$
with the weak operator topology. That is, a sequence $b_n$ in $\mathrm{End}(\mathcal{H})$ converges to $b$ if $\langle u , b_n v \rangle \to \langle u , b v \rangle$ for all $u,v \in \mathcal{H}$. 

\begin{remark}\label{rem:H-state-space-CFT}
In our application to defect flows in 2d\,CFT, $\mathcal{H}$ will be the state space of the CFT, graded by scaling dimension $\Delta = h + \bar h$ of the fields,
\begin{equation}
    \mathcal{H} = \bigoplus_\Delta \mathcal{H}_\Delta \ .
\end{equation}
The direct sum is over the spectrum of scaling dimensions. We assume that each $\mathcal{H}_\Delta$ is finite dimensional, while $\mathcal{H}$ itself is in general of course infinite dimensional. On $\mathcal{H}$ we pick an inner product $\langle-,-\rangle$ such that different $\mathcal{H}_\Delta$ are orthogonal. Since the $\mathcal{H}_\Delta$ are finite-dimensional and mutually orthogonal, the weak topology on $\mathrm{End}(\mathcal{H})$ (and hence the notion of convergence) is actually independent of the choice of inner product on $\mathcal{H}$. 
\end{remark}

With these preparations, here is the second main definition of this section.

\begin{definition}\label{def:alpha-asymptotics}
Let $\mathcal{H}$ be a complex inner product space, and let $\alpha \in \mathbb{R}_{>0}$.
Let $\mathbb{D}$ be a non-zero semiring with $\mathbb{C}^\times$-action $\mathsf{U} : \mathbb{C}^\times \to \mathrm{Aut}(\mathbb{D})$, $\lambda \mapsto \mathsf{U}_\lambda$. Let $\mathcal{D} : \mathbb{D} \to \mathrm{End}(\mathcal{H})$ be a semiring homomorphism.
We say that \textit{$\mathbb{D}$ has $\alpha$-asymptotics on $\mathcal{H}$ relative to $\mathsf{U}$ and $\mathcal{D}$} if the following holds:
For all $b \in \mathbb{D}$, $b \neq 0$, 
\begin{itemize}
    \item there is a finite subset $S_b\subset \mathbb{S}$, and
    \item for each $\theta_0 \in \mathbb{S} \setminus S_b$ there are elements $u(b,\theta_0) \in \mathbb{C}$ and $D(b,\theta_0) \in \mathrm{End}(\mathcal{H})$, $D(b,\theta_0)\neq 0$,
\end{itemize}
such that, for each $I \subset \mathbb{S} \setminus S_b$ a connected component (i.e.\ an open interval with boundary points being two consecutive points in $S_b$), and for each $\theta_0, \theta \in I$,
\begin{equation}\label{eq:limit-condition}
\lim_{r \to \infty}
\exp\!\Big( -u(b,\theta_0) e^{i \alpha(\theta-\theta_0)}r^\alpha \Big) \, \mathcal{D}\big(\mathsf{U}_{r e^{i \theta}}(b)\big)
~=~
D(b,\theta_0) \ .
\end{equation}
\end{definition}

This definition is somewhat technical, so let us first relate it to the expected behaviour of perturbed defects, and then see how all the assumptions are used in Theorem~\ref{thm:semiring-hom} to obtain a semiring homomorphism from $\mathbb{D}$ into a suitable asymptotic semiring.

\begin{remark}\label{rem:hol-pert-to-top-in-CFT}
For $E(\eta) \in \mathcal{C}_X$, 
consider the holomorphically perturbed defect $\mathcal{D}[E,\eta]$ and let $h = h_{\phi_X}$ be the $L_0$-weight of the  perturbing field. Set 
\begin{equation}
    \alpha = \frac{1}{1-h} \ .
\end{equation}
Write $\lambda = r e^{i \varphi}$. As argued in \cite[Sec.\,5]{phys-paper} and recalled in \eqref{eq:intro-asymp-ansatz}, we expect that the asymptotic behaviour for large $r$ and generic $\varphi$ is
\begin{equation}\label{eq:pert-def-asymp-assume}
    \mathcal{D}\big( \mathsf{U}_\lambda [E,\eta]\big)
    =
    \mathcal{D}_E(r e^{i \varphi} \eta)
    \underset{r \to \infty}{\sim}
    \mathcal{D}^\infty(\varphi) 
    \exp\!\big( a(\varphi) e^{i \alpha\varphi} r^\alpha \big) 
    \ .
\end{equation}
By ``generic $\varphi$'' we mean that for certain angles, so-called \textit{anti-Stokes rays}, the dominant behaviour changes, so that on the ray itself there is no well-defined dominant term. The set $S_b \subset \mathbb{S}$ in the above definition is meant to be the set of angles of the anti-Stokes rays. 
In the wedge between two neighbouring anti-Stokes rays (a \textit{Stokes sector}), the limiting defect $\mathcal{D}^\infty$ (called $\mathcal{D}_\mathrm{IR}$ in \cite{phys-paper}) is topological and is expected to be constant (independent of $\varphi$). We will see in Corollary~\ref{cor:angle-dep-of-u-D} below how this behaviour is recovered from the above definition.
To make contact with \eqref{eq:limit-condition}, take $\mathbb{D} = \mathrm{Gr}(\mathcal{C}_X)_{\ge 0}$, and
\begin{equation}
    b = [E,\eta] ~,~~
    \theta = \theta_0=\varphi ~,~~
    u(b,\theta_0) = a(\varphi) e^{i \alpha \varphi}
    ~,~~
    D(b,\theta_0) = \mathcal{D}^\infty(\varphi) \ .
\end{equation}
Then the limit \eqref{eq:limit-condition} is what we mean more precisely by the asymptotic behaviour \eqref{eq:pert-def-asymp-assume}.
\end{remark}

Next we turn to consequences of Definition~\ref{def:alpha-asymptotics}.

\begin{lemma} \label{lem:u-D-unique}
\textit{(Uniqueness)}
Suppose $b \neq 0$ and $u(b,\theta_0) \in \mathbb{C}$, $D(b,\theta_0) \neq 0$ satisfy \eqref{eq:limit-condition} for some $\theta_0=\theta$. Suppose also $\tilde u$ and $\tilde D \neq 0$ satisfy \eqref{eq:limit-condition} for this $\theta_0$. Then already
\begin{equation}
	\tilde u = u(b,\theta_0)
	\quad \text{and} \quad
	\tilde D = D(b,\theta_0) \ .
\end{equation}
\end{lemma}

\begin{proof}
Write $u=u(b,\theta_0)$ and $D=D(b,\theta_0)$.
Suppose $\mathrm{Re}(\tilde u)>\mathrm{Re}(u)$. 
Then
\begin{equation}
    D 
= \lim_{r \to \infty}
\exp\!\big( -u r^\alpha \big) \, \mathcal{D}\big(\mathsf{U}_{r e^{i \theta_0}}(b)\big)
= \lim_{r \to \infty} \exp\!\big( (\tilde u-u) r^\alpha \big)
\underbrace{\exp\!\big( -\tilde u r^\alpha\big) \, \mathcal{D}\big(\mathsf{U}_{r e^{i \theta_0}}(b)\big)}_{\to \tilde D \text{ for } r\to \infty} \ .
\end{equation}
The right hand side cannot converge as the first exponential is divergent, contradicting the assumption that the left hand side converges. 
The case $\mathrm{Re}(\tilde u)<\mathrm{Re}(u)$ leads to an analogous contradiction.
Finally, suppose $\mathrm{Re}(\tilde u)=\mathrm{Re}(u)$ but $\tilde u \neq u$. Then in the above equation we have $\tilde u - u = i( \mathrm{Im}(\tilde u) - \mathrm{Im}(u)) \neq 0$. Hence again the term $\exp\!\big( (\tilde u-u) r^\alpha \big)$ does not converge for $r \to \infty$, while the rest of the right hand side converges to $\tilde D$. 

Thus we must have $\tilde u = u$. 
\end{proof}

The above lemma implies that $D(b,\theta)$ is constant on connected components of $\mathbb{S} \setminus S_b$:

\begin{corollary}\label{cor:angle-dep-of-u-D}
Suppose $\mathbb{D}$ has $\alpha$-asymptotics.
Let $b \neq 0$ and $I \subset \mathbb{S} \setminus S_b$ be connected. For $\theta,\theta' \in I$, we have 
\begin{equation}
  u(b,\theta') = e^{i \alpha(\theta'-\theta)} u(b,\theta)
  \quad \text{and} \quad
  D(b,\theta') = D(b,\theta) \ .
\end{equation}
\end{corollary}

\begin{proof}
Evaluating \eqref{eq:limit-condition} after replacing $\theta_0$ and $\theta$ with $\theta'$ yields
\begin{equation}
\lim_{r \to \infty}
\exp\!\big( -u(b,\theta') r^\alpha \big) \, \mathcal{D}\big(\mathsf{U}_{r e^{i \theta'}}(b)\big)
~=~D(b,\theta') \ .
\end{equation}
Next, replacing $\theta_0$ with $\theta$, and $\theta$ with $\theta'$ in \eqref{eq:limit-condition} gives
\begin{equation}
\lim_{r \to \infty}
\exp\!\big( -u(b,\theta) e^{i \alpha(\theta'-\theta)} r^\alpha \big) \, \mathcal{D}\big(\mathsf{U}_{r e^{i \theta'}}(b)\big)
~=~
D(b,\theta) 
\ .
\end{equation}
The claim now follows from Lemma~\ref{lem:u-D-unique}.
\end{proof}

Suppose that there is a zero-sum-free and entire semiring $\mathbb{D}_\infty \subset \mathrm{End}(\mathcal{H})$ which contains the elements $D(b,\theta_0)$ for all $b \in \mathbb{D}$, $b \neq 0$, and $\theta_0 \in \mathbb{S} \setminus S_b$. In our application, $\mathbb{D}_\infty$ consists of the topological defect operators. 
Together with the uniqueness lemma, we can now define a map from $\mathbb{D}$ to the asymptotic semiring for $\mathbb{D}_\infty$. 
Under the extra assumption of a direct-sum decomposition being preserved, we show that this map is a semiring homomorphism.
This is the main result of this section.
Recall $(\mathbb{D}_\infty)_\mathrm{max}$ from Lemma~\ref{lem:Dmax-def} and $\mathcal{A}(\mathbb{D}_\infty)$ from Definition~\ref{def:asymptotic-semiring}.

\begin{theorem}\label{thm:semiring-hom}
Let $\mathcal{H}$ be a complex inner product space, $\mathbb{D}$ a non-zero zero-sum-free and entire
semiring with $\mathbb{C}^\times$-action $\mathsf{U}$, let $\mathcal{D} : \mathbb{D} \to \mathrm{End}(\mathcal{H})$ be a semiring homomorphism, and
let $\mathbb{D}_\infty \subset \mathrm{End}(\mathcal{H})$ be a sub-semiring such that
\begin{enumerate}
    \item $\mathbb{D}$ has $\alpha$-asymptotics on $\mathcal{H}$ relative to $\mathsf{U}$ and $\mathcal{D}$ in the sense of Definition~\ref{def:alpha-asymptotics},
    \item $\mathbb{D}_\infty$ is zero-sum-free and entire, and for all $b \in \mathbb{D}$, $b \neq 0$, and $\theta \in \mathbb{S} \setminus S_b$ we have $D(b,\theta) \in \mathbb{D}_\infty$.

\item There is an orthogonal direct sum decomposition 
$\mathcal{H} = \bigoplus_\Delta \mathcal{H}_\Delta$ with all $\mathcal{H}_\Delta$ finite-dimensional, such that for all $b \in \mathbb{D}$, $\mathcal{D}(b)$ preserves $\mathcal{H}_\Delta$.
\end{enumerate}
For $b \in \mathbb{D}$ let $f_b : \mathbb{S} \to \mathbb{D}_\infty \times \mathbb{R}_\mathrm{max}$ be the function
\begin{equation}\label{eq:fb-def}
	f_b(\theta) = 
	\begin{cases}
		(0,-\infty) &; b=0 
		\\
		\big(D(b,\theta),\, \mathrm{Re} \, u(b,\theta)\big) &; b \neq 0 , \theta \notin S_b
		\\
		(1,0) &; b \neq 0  , \theta \in S_b   
	\end{cases} 	
\end{equation}
where $u(b,\theta)$, $D(b,\theta)$ are given by \eqref{eq:limit-condition}.
Then the image of $f_b$ lies in $(\mathbb{D}_\infty)_\mathrm{max}$ and the map
\begin{equation}\label{eq:Lambda-def}
	\Lambda : \mathbb{D} \to \mathcal{A}(\mathbb{D}_\infty)
    ~,~~
	\Lambda(b) = [f_b]
\end{equation}
is a semiring homomorphism.
\end{theorem}

The value of $f_b(\theta)$ for $b \neq 0$ and $\theta \in S_b$ in \eqref{eq:fb-def} is arbitrary as long as it is an element of $(\mathbb{D}_\infty)_\mathrm{max}$, in any case the class $[f_b]$ depends on $f_b$ only up to its value at finitely many points.
Condition 3 can be weakened but not omitted as we need some way to control that the limit of the composition of operators in the image of $\mathcal{D}$ is equal to the composition of the limits. With the present assumption this follows as on each finite-dimensional $\mathcal{H}_\Delta$, the operator norm topology and the weak operator topology coincide. By Remark~\ref{rem:H-state-space-CFT}, this is also the case relevant to the application to perturbed defects.

Before proving the theorem we need an auxiliary lemma.

\begin{lemma}\label{lem:M-finite}
   For $b,c \in \mathbb{D}$ the set 
\begin{equation}
M_{b,c} = \big\{ \theta \in \mathbb{S} \setminus (S_b \cup S_c) \,\big|\, 
  \mathrm{Re} \, u(b,\theta) = \mathrm{Re} \, u(c,\theta) \,,~ 
  \mathrm{Im} \, u(b,\theta) \neq \mathrm{Im}  \, u(c,\theta)
  \big\}    
\end{equation}
is finite. 
\end{lemma}

\begin{proof}
By assumption, the set $S_b \cup S_c$ is finite. Let $I$ be one of the finitely many connected components of $\mathbb{S} \setminus (S_b \cup S_c)$. We will show that $M \cap I$ is finite. 
Suppose for $\theta_0 \in I$ we have 
$\mathrm{Re} \, u(b,\theta_0) = \mathrm{Re} \, u(c,\theta_0)$. By Corollary~\ref{cor:angle-dep-of-u-D}, for $\theta \in I$,
\begin{equation}\label{eq:lem-M-finite-aux}
  u(b,\theta) = e^{i \alpha(\theta-\theta_0)} u(b,\theta_0)
  ~~,\quad
  u(c,\theta) = e^{i \alpha(\theta-\theta_0)} u(c,\theta_0)
  \ .
\end{equation}
If also $\mathrm{Im} \, u(b,\theta_0) = \mathrm{Im} \, u(c,\theta_0)$, then $u(b,\theta_0) = u(c,\theta_0)$ and \eqref{eq:lem-M-finite-aux} implies $u(b,\theta) = u(c,\theta)$ for all $\theta \in I$ so that $M \cap I = \emptyset$.
If $\mathrm{Im} \, u(b,\theta_0) \neq \mathrm{Im} \, u(c,\theta_0)$, then $\mathrm{Re}(e^{i \alpha(\theta-\theta_0)} u(b,\theta_0)) = \mathrm{Re}(e^{i \alpha(\theta-\theta_0)} u(c,\theta_0))$ only has finitely many solutions $\theta \in I$.
\end{proof}

\begin{proof}[Proof of Theorem~\ref{thm:semiring-hom}]
To see that the image of $f_b$ lies in $(\mathbb{D}_\infty)_\mathrm{max} \subset \mathbb{D}_\infty \times \mathbb{R}_\mathrm{max}$, it remains to check that for $b \neq 0$ and $\theta \notin S_b$ we have $D(b,\theta)\neq 0$. But this is ensured by $\alpha$-asymptotics (Definition~\ref{def:alpha-asymptotics}).
	
Next we check that $\Lambda$ is a semiring homomorphism. 

\medskip

\noindent
\textit{$\Lambda(0) = 0$}: This holds by construction.

\medskip

\noindent
\textit{$\Lambda(1) = 1$}: As $\mathsf{U}_\lambda$ and $\mathcal{D}$ are semiring homomorphisms, we have $\mathcal{D}(\mathsf{U}_\lambda(1))=\mathrm{id}_\mathcal{H}$. As 
$\mathbb{D} \neq \{0\}$ by assumption, we have 
$1 \neq 0$ in $\mathbb{D}$, and so for $\theta_0 \notin S_1$,  \eqref{eq:limit-condition} gives
\begin{equation}
	\lim_{r \to \infty}
	\exp\!\big( -u(1,\theta_0) r^\alpha \big) 
	\,=\,
	D(1,\theta_0) \,\neq\, 0 \ .
\end{equation}
This can only hold for $u(1,\theta_0)=0$, and hence $D(1,\theta_0)=1$. Altogether, $\Lambda(1) = [f_1]$ with $f_1(\theta) = [1,0]$, which is the unit in $\mathcal{A}(\mathbb{D}_\infty)$.

\medskip

\noindent
\textit{$\Lambda(ab) = \Lambda(a)\Lambda(b)$ for all $a,b \in \mathbb{D}$}: If one of $a,b$ is zero, there is nothing to check. Suppose therefore that $a,b \neq 0$. As $\mathbb{D}$ is entire, also $ab \neq 0$. 
Write $T = S_{ab} \cup S_a \cup S_b$. Then
\begin{equation}
	f_x(\theta) = \big(D(x,\theta),u(x,\theta)\big)
	\quad \text{for} \quad 
    \theta \in \mathbb{S} \setminus T ~~\text{and}~~ x \in \{a,b,ab\} \ .
\end{equation}
We will check that $f_{ab}(\theta) = f_a(\theta) f_b(\theta)$ holds in $(\mathbb{D}_\infty)_\mathrm{max}$ for all $\theta \in \mathbb{S} \setminus T$. As elements in $\mathcal{A}(\mathbb{D}_\infty)$ are equal if the corresponding functions only differ on a finite number of points, this implies that $[f_{ab}] = [f_a][f_b]$ in $\mathcal{A}(\mathbb{D}_\infty)$, as required.
Explicitly, we need to check
\begin{equation}\label{eq:thm-semiring-hom-aux1}
	\big(D(ab,\theta),\,\mathrm{Re} \, u(ab,\theta)\big)
	=
	\big(D(a,\theta)D(b,\theta),\,
	\mathrm{Re} \, u(a,\theta) + 
	\mathrm{Re} \, u(b,\theta)\big) \ .
\end{equation}
We compute
\be\begin{split}
D(a,\theta)D(b,\theta) 
&=
\lim_{r \to \infty}
\exp\!\big( -u(a,\theta) r^\alpha \big) \, \mathcal{D}(\mathsf{U}_{r e^{i \theta}}(a))
\exp\!\big( -u(b,\theta) r^\alpha \big) \, \mathcal{D}(\mathsf{U}_{r e^{i \theta}}(b)) 
\\
&=
\lim_{r \to \infty}
\exp\!\big( -(u(a,\theta)+u(b,\theta)) r^\alpha \big) \, \mathcal{D}(\mathsf{U}_{r e^{i \theta}}(ab)) \ ,
\end{split}
\ee
and
\begin{equation}
	D(ab,\theta) 
	=
	\lim_{r \to \infty}
	\exp\!\big( -u(ab,\theta) r^\alpha \big) \, \mathcal{D}(\mathsf{U}_{r e^{i \theta}}(ab)) \ .
\end{equation}
Note that, as $a,b \neq 0$, by assumption also $D(a,\theta), D(b,\theta) \neq 0$, and as $\mathbb{D}_\infty$ is entire, furthermore the product $D(a,\theta) D(b,\theta)$ is nonzero. 
We can thus apply Lemma~\ref{lem:u-D-unique}, which shows \eqref{eq:thm-semiring-hom-aux1}.

\medskip

\noindent
\textit{$\Lambda(a+b) = \Lambda(a)+\Lambda(b)$ for all $a,b \in \mathbb{D}$}: We start in the same way as before. It is enough to consider $a,b \neq 0$. As $\mathbb{D}$ is zero-sum-free, also $a+b \neq 0$. 
Write $T = S_{a+b} \cup S_a \cup S_b \cup M_{a,b}$, where $M_{a,b}$ was defined in Lemma~\ref{lem:M-finite}. For $\theta \in \mathbb{S} \setminus T$ we have the three limits
\be\begin{split}\label{eq:thm-semiring-hom-aux2}
	D(a,\theta) 
	&=
	\lim_{r \to \infty}
	\exp\!\big( -u(a,\theta) r^\alpha \big) \, \mathcal{D}(\mathsf{U}_{r e^{i \theta}}(a))\ ,
	\\
	D(b,\theta) 
&=
\lim_{r \to \infty}
\exp\!\big( -u(b,\theta) r^\alpha \big) \, \mathcal{D}(\mathsf{U}_{r e^{i \theta}}(b)) \ ,
\\
	D(a+b,\theta) 
&=
\lim_{r \to \infty}
\exp\!\big( -u(a+b,\theta) r^\alpha \big) \, \mathcal{D}(\mathsf{U}_{r e^{i \theta}}(a+b)) \ .
\end{split}\ee
Write $R_x = \mathrm{Re} \, u(x,\theta)$, $x = a,b$. 
We need to distinguish three cases:
\begin{itemize}
	\item 
$R_a>R_b$: In this case we have
\begin{align}
&	\lim_{r \to \infty}
\exp\!\big( -u(a,\theta) r^\alpha \big) \, \mathcal{D}(\mathsf{U}_{r e^{i \theta}}(b))
	\nonumber 
\\
&=
	\lim_{r \to \infty}
\underbrace{\exp\!\big( (u(b,\theta)-u(a,\theta)) r^\alpha \big)}_{\to 0 \text{ for $r \to \infty$ as $R_a>R_b$}} \, \underbrace{\exp\!\big( -u(b,\theta) r^\alpha \big) \, \mathcal{D}(\mathsf{U}_{r e^{i \theta}}(b))}_{\to D(b,\theta) \text{ by construction}} = 0 \ .
\end{align}
With this we see that
\begin{equation}
\lim_{r \to \infty}
	\exp\!\big( -u(a,\theta) r^\alpha \big) \, \mathcal{D}(\mathsf{U}_{r e^{i \theta}}(a+b)) = D(a,\theta) \ .
\end{equation}
Comparing to \eqref{eq:thm-semiring-hom-aux2}, and using Lemma~\ref{lem:u-D-unique}, we conclude that $D(a+b,\theta) = D(a,\theta)$ and $u(a+b,\theta) = u(a,\theta)$, that is, $f_{a+b}(\theta) = f_a(\theta)$. 
On the other hand, the addition in $(\mathbb{D}_\infty)_\mathrm{max}$ is given by (using $R_a>R_b$) 
\begin{equation}
	\big(D(a,\theta),\mathrm{Re}\,u(a,\theta)\big)
	+
	\big(D(b,\theta),\mathrm{Re}\,u(b,\theta)\big)
	=
	\big(D(a,\theta),\mathrm{Re}\,u(a,\theta)\big) \ .
\end{equation}
Hence also $f_a(\theta)+f_b(\theta) = f_a(\theta)$ and altogether $f_a(\theta)+f_b(\theta) = f_{a+b}(\theta)$. 

\item $R_a<R_b$: Analogous to the previous case.

\item $R_a=R_b$:
As $T$ includes the set $M_{a,b}$ from Lemma~\ref{lem:M-finite} and $\theta \in \mathbb{S}\setminus T$, we know that also 
$\mathrm{Im} \, u(a,\theta) = \mathrm{Im} \, u(b,\theta)$, so that $u(a,\theta) = u(b,\theta)$. From this and the limits in \eqref{eq:thm-semiring-hom-aux2} we conclude 
\begin{equation}
	\lim_{r \to \infty}
	\exp\!\big( -u(a,\theta) r^\alpha \big) \, \mathcal{D}(\mathsf{U}_{r e^{i \theta}}(a+b)) = D(a,\theta)+D(b,\theta) \ .
\end{equation}
As $a,b \neq 0$, by assumption also $D(a,\theta), D(b,\theta) \neq 0$, and as $\mathbb{D}_\infty$ is zero-sum-free, furthermore $D(a,\theta) + D(b,\theta) \neq 0$. 
We can thus apply Lemma~\ref{lem:u-D-unique} to conclude that $D(a+b,\theta) = D(a,\theta)+D(b,\theta)$ and $u(a+b,\theta) = u(a,\theta)$. The addition in $(\mathbb{D}_\infty)_\mathrm{max}$ is given by (using $R_a=R_b$) 
\begin{equation}
	\big(D(a,\theta),\mathrm{Re}\,u(a,\theta)\big)
	+
	\big(D(b,\theta),\mathrm{Re}\,u(b,\theta)\big)
	=
	\big(D(a,\theta)+D(b,\theta),\mathrm{Re}\,u(a,\theta)\big) \ .
\end{equation}
It again follows that $f_a(\theta)+f_b(\theta) = f_{a+b}(\theta)$. 
\end{itemize}
So far we have shown $f_{a+b}(\theta) = f_a(\theta) + f_b(\theta)$ for all $\theta \in \mathbb{S} \setminus T$. 
As $T \subset \mathbb{S}$ is finite, it follows that $[f_a]+[f_b] = [f_{a+b}]$.
\end{proof}

\begin{proposition}\label{prop:Lambda-equivariant}
The semiring homomorphism $\Lambda : \mathbb{D} \to \mathcal{A}(\mathbb{D}_\infty)$ from Theorem~\ref{thm:semiring-hom} is equivariant with respect to the $\mathbb{C}^\times$ actions $\mathsf{U}$ on $\mathbb{D}$ and $\mathsf{V}^\alpha$ on $\mathcal{A}(\mathbb{D}_\infty)$ from \eqref{eq:Cx-action-V-on-asymp}: for all $\lambda \in \mathbb{C}^\times$ we have
\begin{equation}\label{eq:Lambda-equivariant}
    \Lambda \circ \mathsf{U}_\lambda  = \mathsf{V}^\alpha_\lambda \circ \Lambda \ .
\end{equation}
\end{proposition}

\begin{proof}
We verify \eqref{eq:Lambda-equivariant} on a given $b \in \mathbb{D}$. For $b=0$ the identity holds, so assume $b \neq 0$. 
Write $\lambda = s e^{i \varphi}$, $s>0$ and let $b' = \mathsf{U}_{s e^{i \varphi}}(b)$. Let $T = S_{b'} \cup (S_{b}-\varphi)$, where $S_b-\varphi = \{ \vartheta - \varphi \, | \,\vartheta \in S_b \}$, and take $\theta \in \mathbb{S} \setminus T$. Then $\theta + \varphi \notin S_b$ and we have
\begin{align}
	D(b,\theta+\varphi) 
	&=
	\lim_{r \to \infty}
	\exp\!\big( -u(b,\theta+\varphi) r^\alpha \big) \, \mathcal{D}\big(\mathsf{U}_{r e^{i (\theta+\varphi)}}(b)\big) \ .
\end{align}
Since $\theta \notin S_{b'}$ we get
\be\begin{split}
	D(b',\theta) 
	&=
	\lim_{r \to \infty}
	\exp\!\big( -u(b',\theta) r^\alpha \big) \, \mathcal{D}\big(\mathsf{U}_{r e^{i \theta}}(b') \big)
     \\
	&=
	\lim_{r \to \infty}
	\exp\!\big( -u(b',\theta) s^{-\alpha} \, (sr)^\alpha \big) \, \mathcal{D}\big(\mathsf{U}_{rs e^{i (\theta+\varphi)}}(b)\big)
\\
	&=
	\lim_{\tilde r \to \infty}
	\exp\!\big( -u(b',\theta) s^{-\alpha} \, {\tilde r}^\alpha \big) \, \mathcal{D}\big(\mathsf{U}_{\tilde r e^{i (\theta+\varphi)}}(b)\big)
    \ .
\end{split}\ee
Using Lemma~\ref{lem:u-D-unique}, this shows $D(b',\theta) = D(b,\theta+\varphi)$ and $u(b',\theta) s^{-\alpha} = u(b,\theta+\varphi)$. Equivalently, for $\theta \in \mathbb{S} \setminus T$,
\begin{equation}
    f_{b'}(\theta) 
    = \big( D(b',\theta),u(b',\theta) \big)
    = \big( D(b,\theta+\varphi), s^\alpha u(b,\theta+\varphi) \big) = (V_\lambda^\alpha(f_b))(\theta) \ .
\end{equation}
As $T$ is finite, this shows the proposition:
$\Lambda(\mathsf{U}_\lambda(b)) = [f_{b'}]  
    = [\mathsf{V}^\alpha_\lambda(f_b)]
    = \mathsf{V}^\alpha_\lambda(\Lambda(b))$.
\end{proof}

\begin{remark}\label{rem:semiring-hom-Gr}
In our application to 2d\,CFT, the topological defects under consideration form a ribbon fusion category $\mathcal{C}$ (the line defects of the chiral TFT, cf.\ Section~\ref{sec:chiral-TFT-defects}). 
The vector space $\mathcal{H}$ is the state space of the CFT on a circle as in Remark~\ref{rem:H-state-space-CFT}, $\mathbb{D} = \mathrm{Gr}(\mathcal{C}_X)_{\ge 0}$ is the Grothendieck semiring (cf.\ Example~\ref{ex:semirings}\,(3)) of the category of holo\-morphically perturbed defects, and $\mathcal{D}$ assigns the corresponding defect operators (Remark~\ref{rem:pert-def-op-factor-Gr-ring}). $\mathbb{D}_\infty$ is the semiring of topological defect operators.

For a rational CFT, the map assigning a topological defect operator in $\mathbb{D}_\infty$ to an element in the Grothendieck semiring $\mathrm{Gr}(\mathcal{C})_{\ge 0}$ is a bijection (this is equivalent to non-degeneracy of the $S$-matrix). 
Combining with Theorem~\ref{thm:semiring-hom} and Proposition~\ref{prop:Lambda-equivariant} we get a $\mathbb{C}^\times$-equivariant semiring homomorphism
\begin{equation}
    \Lambda : \mathrm{Gr}(\mathcal{C}_X)_{\ge 0}
    \to \mathcal{A}(\mathrm{Gr}(\mathcal{C})_{\ge 0})
\end{equation}
which describes the renormalisation group flow of holomorphically perturbed defects. The constant $\alpha \in \mathbb{R}$ entering the $\mathbb{C}^\times$-action on $\mathcal{A}(\mathrm{Gr}(\mathcal{C})_{\ge 0})$ is determined by the conformal weight $h$ of the holomorphic defect field as $\alpha = 1/(1-h)$, see Remark~\ref{rem:hol-pert-to-top-in-CFT}.

Recall the semiring homomorphism $j : \mathrm{Gr}(\mathcal{C})_{\ge 0} \to \mathcal{A}(\mathrm{Gr}(\mathcal{C})_{\ge 0})$ from Remark~\ref{rem:asympt-semiring-prop}\,(1) and the monoidal functor $\iota: \mathcal{C} \to \mathcal{C}_X$ from Lemma~\ref{lem:C-CX-embed}. By abuse of notation, we will also denote the semiring homomorphism induced by $\iota$ on the Grothendieck semiring by $\iota : \mathrm{Gr}(\mathcal{C})_{\ge 0} \to \mathrm{Gr}(\mathcal{C}_X)_{\ge 0}$.
As the image of $\iota$ consists of unperturbed defects, the map $\mathcal{D}$ is the identity in this case, which translates into the following identity,
\begin{equation}\label{eq:Lambda-iota-j-compatible}
    \Lambda \circ \iota = j ~:~ \mathrm{Gr}(\mathcal{C})_{\ge 0} \to \mathcal{A}(\mathrm{Gr}(\mathcal{C})_{\ge 0}) \ .
\end{equation}

More generally, one would not take the surface defect in \eqref{eq:defect-field-in-chiral-TFT} to be trivial, see \cite[Sec.\,5]{Ambrosino:2025pjj} for the modifications of the formalism in that case. In particular one would then start from a (not necessarily braided) fusion category $\mathcal{F}$ of topological line defects on the surface defect and $X$ is an object in the Drinfeld centre of $\mathcal{F}$.
\end{remark}

Turning the logic of the above remark around, we are led to the following mathematical classification question:
\begin{quote}
Given a $\mathbb{C}$-linear abelian rigid braided monoidal category $\mathcal{C}$, an object $X \in \mathcal{C}$, and $\alpha \in \mathbb{R}$, classify semiring homomorphisms
\begin{equation}\label{eq:classify-hom-classify-rg}
    \mathrm{Gr}(\mathcal{C}_X)_{\ge 0}
    \to \mathcal{A}(\mathrm{Gr}(\mathcal{C})_{\ge 0}) \ ,
\end{equation}
which are $\mathbb{C}^\times$-equivariant with respect to the functorial action $\mathcal{U}$ on $\mathcal{C}_X$ and the action $\mathsf{V}^\alpha$ by semiring automorphisms on $\mathcal{A}(\mathrm{Gr}(\mathcal{C})_{\ge 0})$, and which satisfy \eqref{eq:Lambda-iota-j-compatible}.
\end{quote}
The $\mathbb{C}^\times$-action on $\mathcal{C}_X$ can be found in Proposition~\ref{prop:Cx-action-on-CX}, and $\mathsf{V}^\alpha$ is given in Remark~\ref{rem:asympt-semiring-prop}\,(4).
More generally, $\mathcal{C}$ need not be braided and then $X \in \mathcal{Z}(\mathcal{C})$.

\begin{remark}\label{rem:categorify}
Most ingredients in the mathematical classification question above are de\-ca\-te\-go\-rifications of a categorical notion. It seems plausible that one can define a suitable monoidal category $\mathcal{A}(\mathcal{C})$ and modify the CFT analysis so that renormalisation group flows of holomorphically perturbed defects should produce a monoidal functor $\mathcal{C}_X \to \mathcal{A}(\mathcal{C})$. Passing to the Grothendieck ring on both sides should then reduce this to \eqref{eq:classify-hom-classify-rg}. We will return to this question in future work.
\end{remark}

To conclude this subsection, let us make some not very systematic connections to tropical mathematics. 
The limit~\eqref{eq:limit-condition} defining an element in the asymptotic semiring is similar to ``Maslov dequantisation'', see \cite{litvinov2005maslov}, except that the latter only remembers the exponent and not the coefficient. 
The map $\pi_2$ in Lemma~\ref{lem:Dmax-def} has properties analogous to that of a valuation, which is a map from a field to $\mathbb{R}_{\text{min}}$, cf.\ \cite[Sec.\,2.1]{Maclagan-Sturmfels-book}. The closest analogue is the map from the field of Puiseux series to $\mathbb{R}_{\text{min}}$ which picks out the lowest exponent. The tropicalisation of algebraic varieties \cite[Sec.\,3]{Maclagan-Sturmfels-book} may have relations to analysing functional equations in the asymptotic semiring, and it would be interesting to explore this further. Tropical methods were used e.g.\ in \cite{Fomin:2001rc,Inoue_2013} to analyse periodicity of Y-systems.

\subsection{Two examples}

\subsubsection*{Two-dimensional topological field theory}

A very simple yet instructive example is the case that the 2d\,CFT in question is actually a 2d\,TFT, i.e.\ it is the unitary CFT at central charge $c=0$. This case allows one to verify all assumptions explicitly, yet it still illustrates how anti-Stokes rays arise in the limit.

Mathematically, the 2d\,TFT is a functor from a category of two-dimensional defect bordisms to the symmetric monoidal category $\mathrm{Vec}_\mathbb{C}$ of finite-dimensional vector spaces. We refer to \cite{Davydov:2011kb,Carqueville:2016nqk} for details. 

\medskip

We take the 2d\,TFT to have a one-dimensional state space $\mathcal{H}=\mathbb{C}$ on the circle. The topological line defects in the TFT are described by $\mathcal{C}=\mathrm{Vec}_\mathbb{C}$, and a finite-dimensional vector space $V$ is interpreted as $\dim_\mathbb{C}(V)$-many copies of the trivial line defect. As the state space is $\mathbb{C}$, the defect operators are just complex numbers. For example, the topological defect operator assigned to $V$ is
\begin{equation}
    \mathcal{D}_V = \dim_\mathbb{C}(V) \ .
\end{equation}

For the representation of the perturbing field we take $X = \mathbb{C}$ (and $\phi_X = 1 \in \mathbb{C}$), so that $\mathcal{C}_X$ is the category formed by pairs $(V,f)$ of a vector space and a linear endomorphism $f \in \mathrm{End}(V)$. This coincides with the description by quiver modules: the quiver $Q_X$ has a single vertex $\mathbf{1}$ and a single arrow $\mathbf{1} \to \mathbf{1}$. The defect operator for $(V,f)$ for a circle of circumference $L$ is, by \eqref{eq:defect-operator-as-pert-series},
\begin{equation}\label{eq:tft-example-D-map}
    \mathcal{D}[V,f] = \tr_V \exp(f) \ .
\end{equation}
Note that the $L$-dependence cancels: The factor $L^{h-1} = L^{-1}$ cancels against the factor of $L$ produced by the integration along the defect circumference. For $f=0$ this reproduces the unperturbed operator, $\mathcal{D}[V,0] = \tr_V \mathrm{id} = \mathcal{D}_V$. 

Next we check the $\alpha$-asymptotics condition from Definition~\ref{def:alpha-asymptotics} by giving the data entering the definition and then checking the condition.

\smallskip

\noindent
\textit{Data:} We have $\mathcal{H}=\mathbb{C}$, and 
as $h_X=0$ we set $\alpha=1$ (Remark~\ref{rem:hol-pert-to-top-in-CFT}).
We take $\mathbb{D} = \mathrm{Gr}(\mathrm{mod}(Q_X))_{\ge 0}$.
The $\mathbb{C}^\times$ action is $\mathsf{U}_\lambda[V,f] = [V,\lambda f]$.
The semiring homomorphism $\mathcal{D} : \mathbb{D} \to \mathrm{End}(\mathcal{H})$ is the assignment of defect operators in \eqref{eq:tft-example-D-map}.

\smallskip

\noindent
\textit{Condition:} Let $[V,f] \in \mathbb{D}$,  $[V,f] \neq 0$. 
Let $\mathrm{Spec}(f)$ be the set of eigenvalues of $f$ (without multiplicities). We take
\begin{equation}
	S_{[V,f]} 
	= \{\, \theta \in \mathbb{S} \,|\,
	\text{there is no unique $\lambda \in \mathrm{Spec}(f)$ such that 
	$\mathrm{Re}(e^{i \theta} \lambda)$ is maximal}\, \}
    \ .
\end{equation}
For a connected component $I \subset \mathbb{S} \setminus S_{[V,f]}$ we define $\lambda_I$ to be the unique eigenvalue of $f$ such that $\mathrm{Re}(e^{i \theta} \lambda_I)$ is maximal for all $\theta \in I$.
For $\theta_0 \in I$ we set
\begin{equation}
	u([V,f],\theta_0) = e^{i \theta_0} \lambda_I
	~~,\quad
	D([V,f],\theta_0) = \dim_\mathbb{C} \mathrm{Eig}(f,\lambda_I)
	\ ,
\end{equation}
where
$\mathrm{Eig}(f,\lambda_I) = \mathrm{ker}(f - \lambda_I \mathrm{id})^N$, for $N$ large enough,
is the generalised eigenspace for $\lambda_I$. By construction, $\dim_{\mathbb{C}} \mathrm{Eig}(f,\lambda_I)$ is the number of $\lambda_I$-entries on the diagonal in the Jordan normal form of $f$.

Let $I \subset \mathbb{S} \setminus S_{[V,f]}$ be a connected component and let $\theta_0, \theta \in I$. 
Checking the limit condition \eqref{eq:limit-condition} amounts to the following computation:
\begin{align}
&\lim_{r \to \infty}
\exp\!\big( -u([V,f],\theta_0) e^{i \alpha (\theta-\theta_0)}r^\alpha \big) \, \mathcal{D}\big(\mathsf{U}_{r e^{i \theta}}[V,f]\big)
\nonumber \\
&=
\lim_{r \to \infty}
\exp\!\big( - \lambda_I e^{i \theta}r \big) \,
\tr_V \exp(r e^{i \theta} f)
\nonumber \\
&\overset{(*)}=
\lim_{r \to \infty}
\exp\!\big( - \lambda_I e^{i \theta}r \big) \,
\tr_{\mathrm{Eig}(f,\lambda_I)} \exp(r e^{i \theta} \lambda_I)
= 
\dim_\mathbb{C} \mathrm{Eig}(f,\lambda_I) \ .
\end{align}
For $\overset{(*)}=$  note that the leading contribution in $\tr_V \exp(r e^{i \theta} f)$ will come from the generalised eigenspace for $\lambda_I$ as it is the unique eigenvalue with largest real part $\mathrm{Re}(e^{i \theta} \lambda_I)$. 

This concludes the verification of $\alpha$-asymptotics.
We can thus apply Theorem~\ref{thm:semiring-hom} with the choice $\mathbb{D}_\infty = \mathrm{Gr}(\mathrm{Vec}_\mathbb{C})_{\ge 0} = \mathbb{Z}_{\ge 0}$. This results in a semiring homomorphism 
\begin{equation}
	\Lambda : \mathrm{Gr}(\mathrm{mod}(Q_X))_{\ge 0} \to \mathcal{A}(\mathrm{Gr}(\mathrm{Vec}_\mathbb{C})_{\ge 0})
\end{equation}
which from the above analysis we can easily describe explicitly. For $[V,f] \neq 0$, $\Lambda[V,f]$ is the class of the function
\begin{equation}
	\mathbb{S} \setminus S_{[V,f]} \to 
	\mathbb{D}_\infty \times \mathbb{R}_\mathrm{max}
	~~,\quad 
	\theta \mapsto 
	\big(\dim_{\mathbb{C}} \mathrm{Eig}(f,\lambda_I) \,,\, \mathrm{Re}(
    e^{i\theta} \lambda_I ) \,\big) \ ,
\end{equation}
where as above, $\lambda_I$ is the eigenvalue of $f$ associated to the connected component $I \subset \mathbb{S} \setminus S_{[V,f]}$ containing $\theta$. At the angles in $S_{[V,f]}$ the dominant behaviour of $\tr_V \exp(r e^{i \theta} f)$ changes, so these are indeed the anti-Stokes rays.

\subsubsection*{Lee-Yang CFT}

The example of the Lee-Yang CFT serves as a warm-up for the general treatment in the next section. Here we already cannot give the full answer, and we have to rely on some assumptions. 
Our aim is to illustrate the constraints coming from the semiring homo\-mor\-phism \eqref{eq:classify-hom-classify-rg} using the results on the Lee-Yang CFT already developed in Examples~\ref{ex:LY-Qtau-modules}, \ref{ex:LY-tensor}, \ref{ex:LY-trace} and \ref{ex:LY-fun-rel}. 
Here, $\mathrm{Gr}(\mathcal{C})_{\ge 0}$ is given by $F_{\ge 0} = \mathbb{Z}_{\ge 0} \mathbf{1} \oplus \mathbb{Z}_{\ge 0} \tau$ with product $\tau\cdot \tau = \mathbf{1} + \tau$. 
By Proposition~\ref{prop:modQX-CX-monoidal}, in place of $\mathcal{C}_X$ we may use $\mathrm{mod}(Q_X)$. We would like to constrain the image of $[\tau,\lambda]$ in $\mathcal{A}(F_{\ge 0})$. 

In terms of Theorem~\ref{thm:semiring-hom}, we assume that the perturbed defects $[V,f]$ have $\alpha$-asymptotics with $\alpha = 1/(1-h_\tau) = \frac{5}{6}$, and that $\mathbb{D}_\infty \cong F_{\ge 0}$ via the bijection between $F_{\ge 0}$ and topological defect operators noted in Remark~\ref{rem:semiring-hom-Gr}.

Recall the $\mathbb{C}^\times$-action on $\mathrm{Gr}(\mathrm{mod}(Q_X))$ in \eqref{eq:Cx-action-GrCX}. Abbreviating $T = [\tau,1]$, we have $[\tau,\lambda] = \mathsf{U}_\lambda T$.
Our main tool is the functional relation \eqref{eq:LY-fun-rel-quiver} (which also holds in the Grothendieck ring), which we rewrite using the $\mathbb{C}^\times$-action as
\begin{equation}
    (\mathsf{U}_\zeta T) \cdot (\mathsf{U}_{\zeta^{-1}}T) = \mathbf{1} + T 
    \qquad \text{where} \quad \zeta = e^{2 \pi i / 5} \ .
\end{equation}
Write $\mathcal{X} = \Lambda(T)$ for the image of $T$ in the asymptotic semiring $\mathcal{A}(F_{\ge 0})$. As $\Lambda$ is $\mathbb{C}^\times$-equivariant, we have
\begin{equation}\label{eq:LY-asymp-fun-rel}
    (\mathsf{V}^\alpha_{\zeta} \mathcal{X}) \cdot 
    (\mathsf{V}^\alpha_{\zeta^{-1}} \mathcal{X})
    =
    \mathbf{1} + \mathcal{X} \ .
\end{equation}

We now add two extra inputs from the numerical analysis via the truncated conformal space approach (TCSA). The first one is that there is a single anti-Stokes ray along the negative real axis. By Corollary~\ref{cor:angle-dep-of-u-D} this forces $\cX$ to be given by, for $-\pi<\theta<\pi$,
\begin{equation}
    \cX = [D, \mathrm{Re}(a e^{i \alpha \theta})] 
\end{equation}
for some constant $a \in \mathbb{C}$ and $D \in F_{\ge 0}$ independent of $\theta$.
The second input is that $a \in \mathbb{R}_{> 0}$.
We will see in Proposition~\ref{prop:one-solution} below that with these assumptions the functional relation \eqref{eq:LY-asymp-fun-rel} forces $D = \mathbf{1}$. 

This predicts the renormalisation group flows for the perturbed defects $\mathcal{D}_\tau(e^{i \theta} r)$ for $r \to \infty$ and $-\pi<\theta<\pi$ to end at the identity defect $\mathbf{1}$, and it also predicts the leading free energy contribution to be $-a\exp(i \frac{5}{6} \theta)$. Both are confirmed beautifully by the TCSA analysis, see \cite[Sec.\,3]{phys-paper}.

\section{Chiral (1,3)-perturbation of defects in minimal models}\label{sec:minmod-13}

In this section we present a more non-trivial example for diagonal minimal model CFTs $M(p,q)$, namely the perturbation of topological defects with Kac labels $(1,s)$ by the holomorphic defect field $(1,3)$. We will show in Section~\ref{sec:2q-solution} that under the assumption that there is a single Stokes sector and an additional positivity condition, one is forced to have $p=2$ and that there is a unique solution to the functional relations in the asymptotic semiring. For general $p$, in Section~\ref{sec:pq-solution} we present one solution, but we do not discuss its uniqueness, though we believe it is the one realised by minimal model flows.

\subsection{Functional relations}\label{sec:mm-funrel}

\subsubsection*{Conventions}

The diagonal Virasoro minimal model $M(p,q)$, for $2 \le p < q$ coprime has central charge 
\begin{equation}
    c = 13 - 6 (t+t^{-1})
    \qquad \text{where} ~~ 
    t = \frac{p}{q} \ .
\end{equation}
The category $\mathcal{C}$ of representations of the corresponding Virasoro vertex operator algebra has simple representations
indexed by Kac labels
\begin{equation}\label{eq:Mpq-irreps}
    (r,s) 
    \qquad \text{where}~~
    r=1,\dots,p-1~,~~s=1,\dots,q-1 \ ,
\end{equation}
modulo the relation $(r,s) \sim (p-r,q-s)$. The lowest $L_0$-weight in each representation is
\begin{equation}
    h_{(r,s)} = \frac{(d_{r,s})^2-(d_{1,1})^2}{4t}
    \qquad \text{where}~~
    d_{r,s} = r - st \ .    
\end{equation}
The minimal model $R$-matrix is\footnote{
Note that this is conjugate to the convention used in \cite{Runkel:2007wd} -- this has to be taken into account when comparing equations.}
\begin{equation}
    R^{(ab)c} = e^{\pi i(h_c-h_a-h_b)} \ .
\end{equation}
The $F$-matrix entries we will need are listed explicitly in \cite[App.\,A.4]{Runkel:2007wd}.

\subsubsection*{Relations in the Grothendieck ring of quiver modules}

The perturbation we consider is by the primary field in $X = (1,3)$ of conformal weight $h_{(1,3)} = -1+2t$. The relevant functional relations have already been obtained in \cite{Runkel:2007wd}, but we rederive them here as an illustration of the quiver-based methods in Section~\ref{sec:quiver}. We will focus on the full subcategory $\mathcal{S} \subset \mathcal{C}$ given by the first row in the Kac-table, i.e.\ by direct sums of the simple objects $(1,s)$, $s=1,\dots,q-1$. 
In this subsector we have truncated $sl(2)$-fusion rules: $(1,1)$ is the tensor unit, $(1,q-1)$ is tensor-invertible with $(1,q-1) \otimes (1,s) = (1,q-s)$, and
\begin{align}
    (1,2) \otimes (1,s) &= (1,s-1) \oplus (1,s+1) 
    &~~\text{for}~~ 2 \le s \le q-2 \ ,
    \nonumber
    \\
    (1,3) \otimes (1,s) &= (1,s-2) \oplus (1,s) \oplus (1,s+2)
    &~~\text{for}~~ 3 \le s \le q-3 \ ,
    \label{eq:mm-13-fusion}
\end{align}
and so on.
The quiver $Q_X$ for $\mathcal{S}$ and $X = (1,3)$ is, for $p>2$ and $q$ odd,
\begin{equation}
        \includegraphics[valign = c, scale=1]{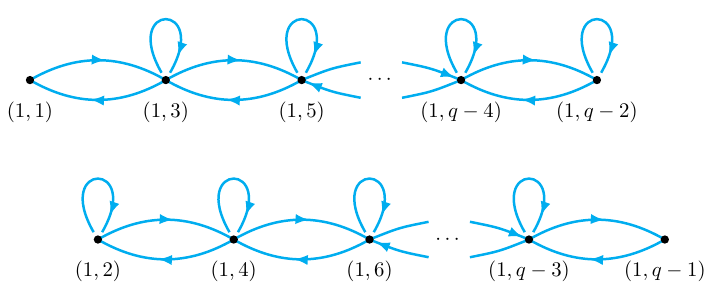}
\end{equation}
For $p=2$ we necessarily have $q$ odd and the two sub-quivers are identified because of the relation $(1,s) \sim (1,q-s)$.
For $q$ even, the even and odd sub-quivers end in opposite ways, the top row ends at $(1,q-1)$ and that vertex has no self-loop, and the bottom row ends at $(1,q-2)$, which has a self-loop.
Note that for the Lee-Yang model $M(2,5)$, this reproduces the quiver in Example~\ref{ex:LY-Qtau-modules}.
By Proposition~\ref{prop:modQX-CX-monoidal} we have $\mathcal{S}_X \cong \mathrm{mod}(Q_X)$.

The one-dimensional quiver modules are $(\mathbb{C}_{(1,s)},\lambda)$. For $p=2$ we take $(1,s)$ modulo the relation $(1,s) \sim (1,q-s)$. We are interested in the fusion 
\begin{equation}
    (Z,f) \,=\, (\mathbb{C}_{(1,2)},\lambda) \circledast (\mathbb{C}_{(1,s)},\mu) \ .
\end{equation}
From \eqref{eq:fused-quiver-vsp} 
and \eqref{eq:mm-13-fusion}
we read off
\begin{align}
    Z_{(1,s-1)} &= \begin{cases}
        \mathbb{C}_{(1,2)} \otimes_{\mathbb{C}} \mathbb{C}_{(1,s)} = \mathbb{C}
        &; s>1
        \\
        0 &; s=1
    \end{cases}
    \quad ,
    \nonumber 
    \\
    Z_{(1,s+1)} &= \begin{cases}
        \mathbb{C}_{(1,2)} \otimes_{\mathbb{C}} \mathbb{C}_{(1,s)} = \mathbb{C}
        &; s<q-1
        \\
        0 &; s=q-1
    \end{cases}
    \quad ,
\end{align}
with the other components $Z_b$ being zero.
For the components $(f_{ab})_{cd}^{c'd'}$ of $f$ in \eqref{eq:VgWh-tensor-f} we abbreviate, for $\epsilon,\nu =\pm 1$,
\begin{equation}
    f_{\epsilon\nu} = (f_{(1,s+\epsilon),(1,s+\nu)})_{(1,2),(1,s)}^{(1,2),(1,s)}    
\end{equation}
Then \eqref{eq:tensor-Q-mod-on-obj} reads, with $s \equiv (1,s)$ (recall the relation between $F$ and $G$ from Footnote~\ref{fn:FG-rel}),
\begin{equation}\label{eq:Mpq-D2-Ds-fusion-f}
    f_{\epsilon\nu}
    \,=\, F^{(s23)s+\epsilon}_{2,s+\nu} \, \lambda ~+~ \frac{R^{(s2)s+\epsilon}}{R^{(s2)s+\nu}} \, F^{(2s3)s+\epsilon}_{s,s+\nu} \, \mu 
    \quad 
    \text{for}
    ~~
    2 \le s \le q-2
    \ .
\end{equation}
For $s=1$ only $\epsilon=\nu=+$ are allowed, and for $s=q-1$ only $\epsilon=\nu=-$. This expression agrees with \cite{Runkel:2007wd}: $f_{\epsilon\nu}$ is equal to $\xi_{cd}$ in (2.26) there for $f=3$, $a=2$, $b=s$, $c=s+\nu$ and $d = s+\epsilon$ and all $\eta$'s set to $1$. 

We immediately see that for appropriate choices of $\mu$ in terms of $\lambda$ we can achieve $f_{+-}=0$ or $f_{-+}=0$ leading to exact sequences and hence to a relation in the Grothendieck ring $\mathrm{Gr}(\mathrm{mod}(Q_X))$. 
The computation of the functional relation in $\mathrm{Gr}(\mathrm{mod}(Q_X))$ is now the same as in \cite{Runkel:2007wd} and we only present the answers. It is convenient to introduce the constants (called $\eta^{(1,s)(1,s)}$ in \cite[Eqn.\,(3.4)]{Runkel:2007wd})
\begin{equation}
    \eta_s = \frac{\Gamma(2-3t)\Gamma(t)}{
    \Gamma\big(2-(s+1)t\big)\Gamma\big((s-1)t\big)} \ .
\end{equation}
Note that $\eta_2=1$.
Then, for $2 \le s \le q-2$ we have (this uses the $F$-matrix expressions in \cite[Eqn.\,(A.10)]{Runkel:2007wd}):
\begin{align}
    &f_{-+} = 0
    ~\Leftrightarrow~ \mu = \zeta^s \eta_s \lambda
    & \text{; then} &&
    f_{--} &= \zeta^{s+1} \eta_{s-1}  \lambda \ ,
    &
    f_{++} &=  \zeta^{s-1} \eta_{s+1} \lambda \ ,
    \nonumber
    \\
    &f_{+-} = 0
    ~\Leftrightarrow~ \mu = \zeta^{-s} \eta_s \lambda
    & \text{; then} &&
    f_{--} &= \zeta^{-(s+1)} \eta_{s-1} \lambda \ ,
    &
    f_{++} &= \zeta^{-(s-1)} \eta_{s+1} \lambda \ .
    \label{eq:f-mu-nu-zeros}
\end{align}
For $s=1$ the $f_{--}$ term is absent, and for $s=q-1$ the $f_{++}$ term is absent.
This suggests rescaling the parameter, that is, to work with the simple quiver modules $(\mathbb{C}_{(1,s)},\eta_s\mu)$ instead. For $1 \le s \le q-1$ we denote the corresponding class in $\mathrm{Gr}(\mathrm{mod}(Q_X))$ by 
\begin{equation}
    T_s(\mu) := [\mathbb{C}_{(1,s)},\eta_s\mu]
    \ .
\end{equation}
For $s=1$ and $s=q-1$ there is actually no $\mu$-dependence as the corresponding component $f_{(1,s),(1,s)}$ of the quiver module is zero (there is no loop from $(1,s)$ to itself in these cases). $T_1 \equiv T_1(\mu)$ is the unit in $\mathrm{Gr}(\mathrm{mod}(Q_X))$.

\begin{remark}
From the CFT point of view, working with $\eta_s \mu$ instead of $\mu$ amounts to changing the normalisation of the perturbing defect field. Namely, consider \eqref{eq:defect-field-in-chiral-TFT} with $D=(1,s)$ and $X = (1,3)$. Then in $(\mathbb{C}_{(1,s)},\mu)$ we are using the convention that the defect junction $\eta$ in \eqref{eq:defect-field-in-chiral-TFT} is $\mu \, \lambda_{(1,3),(1,s)}^{(1,s)}$. Here, $\lambda_{ab}^c$ is the standard basis element from \eqref{eq:lambda-basis-in-Hom} which leads to the unit-normalised minimal model $F$-matrices as in \cite{Runkel:1998he}. For $T_s(\lambda)$ as defined above, the junction is instead given by $\eta_s \mu \, \lambda_{(1,3),(1,s)}^{(1,s)}$.
\end{remark}

The functional relation we are after can now be easily read off from \eqref{eq:f-mu-nu-zeros} and is,
for $\sigma = \pm 1$ and $2 \le s \le q-2$,
\begin{equation}\label{eq:Mpq-fun-rel}
    T_2(\lambda)\, T_s(\zeta^{\sigma s} \lambda)  \,=\, T_{s-1}(\zeta^{\sigma(s+1)}\lambda) + T_{s+1}(\zeta^{\sigma(s-1)}\lambda) \ , \quad 
    \text{where}
    ~~\zeta = e^{i\pi t} 
    ~,~
    t = p/q
    \ .
\end{equation}
As in \cite[Sec.\,3.1]{Runkel:2007wd} one furthermore checks that 
\begin{equation}\label{eq:T-reflection-Mpq}
    T_{q-1} \, T_s(\lambda) \,=\, T_{q-s}((-1)^p \lambda) \ .
\end{equation}
All the $T_s$ mutually commute:

\begin{lemma}
For $1 \le s,s' \le q-1$ and $\lambda,\lambda' \in \mathbb{C}$ we have $T_s(\lambda) T_{s'}(\lambda') = T_{s'}(\lambda') T_s(\lambda)$.
\end{lemma}

\begin{proof}
This is shown in \cite[Sec.\,2.8]{Runkel:2007wd}, and we recall the proof in the current setting. By \eqref{eq:Mpq-fun-rel}, every $T_s(\lambda)$ can be written as a polynomial in $T_2(\mu)$ for different parameters $\mu$. It is therefore enough to show that $T_2(\lambda)T_2(\mu) = T_2(\mu)T_2(\lambda)$.

Write $(Z(\lambda,\mu),f(\lambda,\mu)) = (\mathbb{C}_{(1,2)},\lambda) \circledast (\mathbb{C}_{(1,2)},\mu)$. From \eqref{eq:Mpq-D2-Ds-fusion-f} we see
\begin{equation}
    f(\lambda,\mu)_{\epsilon\nu}
    \,=\, F^{(223)2+\epsilon}_{2,2+\nu} \left( \lambda ~+~ \frac{R^{(22)2+\epsilon}}{R^{(22)2+\nu}} \, \mu \right)
    \ .
\end{equation}
Note that $f(\lambda,\mu)_{\epsilon\epsilon} = f(\mu,\lambda)_{\epsilon\epsilon}$ and
$f(\lambda,\mu)_{+-}\, f(\lambda,\mu)_{-+} =
    f(\mu,\lambda)_{+-} \, f(\mu,\lambda)_{-+}$.
Using this, one checks that setting
\begin{equation}
    \phi_{(1,1)} = f(\mu,\lambda)_{-+}
    ~~,\quad
    \phi_{(1,3)} = f(\lambda,\mu)_{-+} \ ,
\end{equation}
defines a morphism $\phi : (Z(\lambda,\mu),f(\lambda,\mu)) \to (Z(\mu,\lambda),f(\mu,\lambda))$, that is, the squares in \eqref{eq:quiver-module-morphism} commute.

We now have to distinguish two cases. If both $\phi_{(1,1)}$ and $\phi_{(1,3)}$ are non-zero, $\phi$ is an isomorphism and we are done. If one of the two is zero, this is precisely the case where we get the two functional equations (evaluate \eqref{eq:Mpq-fun-rel} with $\lambda = \zeta^{-\sigma} \rho$, $\sigma=\pm1$) 
\begin{equation}
        T_2(\zeta^{-\sigma}\rho)\, T_2(\zeta^{\sigma} \rho)  \,=\, 1 + T_{3}(\rho) \ , 
\end{equation}
which shows $T_2(\lambda)T_2(\mu) = T_2(\mu)T_2(\lambda)$ also in this case.
\end{proof}

Using commutativity one can show that the relations for $\sigma=+$ and $\sigma=-$ in \eqref{eq:Mpq-fun-rel} are actually equivalent.
In terms of the $\mathbb{C}^\times$-action from \eqref{eq:Cx-action-GrCX} we can rewrite the functional relation \eqref{eq:Mpq-fun-rel} as follows. Set
\begin{equation}
    T_s := T_s(1) \quad \text{for} ~~ 2 \le s \le q-2 \ ,
\end{equation}
and recall that $T_1$ and $T_{q-1}$ have no parameter dependence in the first place. Then \eqref{eq:Mpq-fun-rel} becomes (for $\sigma=+$) 
\begin{equation}\label{eq:Mpq-T-system}
    T_2 \, (\mathsf{U}_{\zeta^s} T_s)
    \,=\, 
    \mathsf{U}_{\zeta^{s+1}} T_{s-1}
    + \mathsf{U}_{\zeta^{s-1}} T_{s+1} 
    \qquad
    \text{for}
    ~~ 2 \le s \le q-2 \ .
\end{equation}
Note that $\zeta^q = (-1)^p$, so that we can rewrite \eqref{eq:T-reflection-Mpq} as
\begin{equation}
    T_{q-1} \, T_s \,=\, \mathsf{U}_{\zeta^q} T_{q-s} 
        \qquad
    \text{for}
    ~~ 1 \le s \le q-1 \ .
\end{equation}
A standard recursive argument shows that the T-system  \eqref{eq:Mpq-T-system} implies the Hirota system
\begin{equation}\label{eq:Mpq-Hirota}
(\mathsf{U}_\zeta T_s) \, (\mathsf{U}_{\zeta^{-1}} T_s)
\,=\,
1 + T_{s-1} T_{s+1} \ .
\end{equation}
Note that this argument needs subtraction, so even though the Hirota system is an identity in the semiring, the proof uses the underlying Grothendieck ring $\mathrm{Gr}(\mathrm{mod}(Q_X))$.

\subsubsection*{Relations in the asymptotic semiring}

The category $\mathcal{C}$ of representations of the Virasoro VOA for the minimal model $M(p,q)$ is a modular fusion category and it has $(p-1)(q-1)/2$ simple objects $\mathcal{I}$ as described in \eqref{eq:Mpq-irreps}. The tensor product of $\mathcal{C}$  describes the fusion of topological line defects in $M(p,q)$. The relevant asymptotic semiring is therefore
\begin{equation}
    \mathcal{A}( \mathrm{Gr}(\mathcal{C})_{\ge 0} ) \ .
\end{equation}

We would like to (but will not be able to) answer the question asked around \eqref{eq:classify-hom-classify-rg}, i.e.\ what are all $\mathbb{C}^\times$-equivariant semiring homomorphisms
\begin{equation}\label{eq:semiring-hom-mpq}
    \Lambda : \mathrm{Gr}(\mathcal{C}_X)_{\ge 0}
    \to \mathcal{A}(\mathrm{Gr}(\mathcal{C})_{\ge 0}) 
\end{equation}
compatible with \eqref{eq:Lambda-iota-j-compatible}.
In fact, we cannot even write down a single such homomorphism. Instead, our approach is as follows:
\begin{enumerate}
    \item We believe the holomorphically perturbed defects for the $(1,3)$-perturbation have $\alpha$-asymptotics as in Definition~\ref{def:alpha-asymptotics} and Remark~\ref{rem:hol-pert-to-top-in-CFT}, provided the integrals \eqref{eq:defect-operator-as-pert-series} in the perturbative series are finite without counter terms. This requires $h_{(1,3)}<\frac12$, or, equivalently, 
    $q > \frac43p$. 
    \item Assuming $\alpha$-asymptotics, by Theorem~\ref{thm:semiring-hom} and Remark~\ref{rem:semiring-hom-Gr} we obtain a semiring homomorphism $\Lambda$ as in \eqref{eq:semiring-hom-mpq}.

    \item Supposing hence the existence of $\Lambda$, if we identify a functional relation among elements in 
    $\mathrm{Gr}(\mathcal{C}_X)_{\ge 0}$, by applying $\Lambda$ the corresponding relation has to hold in 
    $\mathcal{A}(\mathrm{Gr}(\mathcal{C})_{\ge 0})$. We can then use these relations to constrain the possible images of such elements under $\Lambda$. 
\end{enumerate}
As the images under $\Lambda$ are the endpoints of RG-flows, the output of step 3 constrains the possible RG-flow endpoints of the perturbed defects.

\medskip

Let us apply this approach to the example at hand. Write
\begin{equation}\label{eq:Xs-def}
    \mathcal{X}_s \equiv [D^\infty_s(\theta),a_s(\theta)] := \Lambda(T_s) \ .
\end{equation}
Denote the elements of $\mathrm{Gr}(\mathcal{C})_{\ge 0}$ (corresponding to the unperturbed topological defect operators) by $D_{(r,s)}$, $(r,s) \in \mathcal{I}$. For example, $\mathcal{X}_1 = [D_{(1,1)},0]$.
We assume $\alpha$-asymptotics with
\begin{equation}\label{eq:Mpq-alpha}
    \alpha = \frac{1}{1-h_{(1,3)}} = \frac{1}{2(1-t)} = \frac{q}{2(q-p)} \ .
\end{equation}
It will be convenient to further introduce the constants
\begin{equation}\label{eq:mpq-gamma-beta-delta}
\gamma=\pi t = \frac{\pi p}{q}
~~,\quad
\beta=\alpha\gamma=\frac{p\pi}{2(q-p)}
\ .
\end{equation}
The $\mathbb{C}^\times$-equivariance from Proposition~\ref{prop:Lambda-equivariant} states $\Lambda \circ \mathsf{U}_\lambda  = \mathsf{V}^\alpha_\lambda \circ \Lambda$.
As $\lambda$ is always a phase in the functional equations, it is convenient to instead just use the argument shift automorphism
\begin{equation}
    \mathsf{S}_\varphi[D(\theta),a(\theta)] = [D(\theta+\varphi),a(\theta+\varphi)] \ .
\end{equation}

Let us collect the results of the above discussion in a proposition.

\begin{proposition}\emergencystretch=1em
Suppose there exists a 
$\mathbb{C}^\times$-equivariant semiring homomorphism
$\Lambda: \mathrm{Gr}(\mathcal{C}_X)_{\ge 0}
    \to \mathcal{A}(\mathrm{Gr}(\mathcal{C})_{\ge 0})$
compatible with \eqref{eq:Lambda-iota-j-compatible}.    
Then the $\mathcal{X}_s$ in \eqref{eq:Xs-def} satisfy the equations
\begin{subequations}\label{eq:Mpq-asymp-funrel}
\begin{align}
&\mathcal{X}_1 = [D_{(1,1)},0] = 1 ~~\text{and}~~ \mathcal{X}_{q-1} = [D_{(1,q-1)},0] \ ,
\label{eq:Mpq-asymp-funrel_bndcond}
\\
    &\mathcal{X}_2 \, (\mathsf{S}_{s\gamma} \mathcal{X}_s)
    \,=\, 
\mathsf{S}_{(s+1)\gamma} \mathcal{X}_{s-1}
    + \mathsf{S}_{(s-1)\gamma} \mathcal{X}_{s+1} 
    &\text{for}
    ~~ 2 \le s \le q-2 \ ,
\label{eq:Mpq-asymp-funrel_Tsys}
\\
    &\mathcal{X}_{q-1} \, \mathcal{X}_s \,=\, \mathsf{S}_{\pi p} \mathcal{X}_{q-s} 
    &\text{for}
    ~~ 1 \le s \le q-1 \ ,
\label{eq:Mpq-asymp-funrel_reflect}
\\
&(\mathsf{S}_\gamma \mathcal{X}_s) \, (\mathsf{S}_{-\gamma} \mathcal{X}_s)
\,=\,
1 + \mathcal{X}_{s-1} \mathcal{X}_{s+1} 
    &\text{for}
    ~~ 2 \le s \le q-2 \ .
\label{eq:Mpq-asymp-funrel_Hirota}
\end{align}
\end{subequations}
\end{proposition}

\subsection{Solution in asymptotic semiring of \texorpdfstring{$M(2,q)$}{M(2,q)}}
\label{sec:2q-solution}

The aim of this section and the next is to give solutions to the system of equations in \eqref{eq:Mpq-asymp-funrel}. In this section we add the assumption that there is a single Stokes sector, which will force $p=2$, but which will allow us to also make a uniqueness statement for the solution (Proposition~\ref{prop:one-solution} below). In the next subsection we work with general $M(p,q)$ (for $q>\frac43 p$), and we give an explicit solution, but we do not investigate its uniqueness properties.

\medskip

We will need a function which brings an angle $\theta \in \mathbb{R}$ into the standard range $(-\pi,\pi]$:
\begin{equation}
    [~] : \mathbb{R} \to (-\pi,\pi]
    \quad , \quad
    [\theta] \in  (-\pi,\pi] \text{ unique such that }
    \theta - [\theta] \equiv 0 \mod 2 \pi \ .
\end{equation}
Recall the constants $\alpha$, $\beta$, $\gamma$ from \eqref{eq:Mpq-alpha} and \eqref{eq:mpq-gamma-beta-delta},
and write
\begin{equation}\label{eq:one-c-def}
    c(\theta) \,=\, \cos\bigl(\alpha\princ{\theta}\bigr) \ .
\end{equation}

\begin{proposition}
\label{prop:one-solution}
Suppose that for $s=2,\dots,q-2$ we have
$\cX_s=\big[D_s^\infty,\,\mathrm{Re}(a_s\,e^{i\alpha\princ{\theta}})\big]$ for some
$D_s^\infty \in \mathrm{Gr}(\mathcal{C})_{\ge 0}$ and $a_s \in \mathbb{C}$, 
that is, each $\cX_s$ has a single anti-Stokes ray at $\theta = \pi$. Suppose furthermore
\begin{equation}\label{eq:M2q-assumptions-for-uniqueness}
a_2>0 \ , 
 \quad
\gamma<\pi/2 
 \ ,
\end{equation}
and that the $\cX_s$ solve the Hirota system \eqref{eq:Mpq-asymp-funrel_Hirota} with boundary conditions \eqref{eq:Mpq-asymp-funrel_bndcond}. Then necessarily $p=2$ and
\begin{equation}
\label{eq:one-solution}
\cX_s
=
\left[
\Id,\;
a_s\,c(\theta)
\right]
\ , \qquad
a_s
=
a_2\,\frac{\sin\!\big((s-1)\beta\big)}{\sin\beta}
\ ,
\qquad
s=1,\dots,q-1.
\end{equation}

\end{proposition}

We stress, in particular, that the $a_s$ are uniquely fixed by the assumptions, and that all $D_s^\infty$ are forced to be $\mathbf{1}$.
Before we prove the proposition, we will make a quick remark on the assumptions and then show a few auxiliary statements.

\begin{remark}
Note that $\gamma < \frac{\pi}2$ amounts to $q>2p$,   or, equivalently $h_{(1,3)}<0$. 
For $p=2$ (as is implied by the assumptions in the proposition) this amounts to $q \ge 5$, which is not in fact a restriction. In Remark~\ref{rem:solutionless2p} in the next section, we also comment on the regime $2p>q>\frac43 p$.
Next, if we assume $\alpha$-asymptotics from the start, the hypothesis on the $\cX_s$ says that every perturbed defect operator $\mathcal{D}_{(1,s)}(\lambda)$ has a single anti-Stokes ray along the negative real axis; note that this is an assumption about the whole tower and not only about $\mathcal{D}_{(1,2)}(\lambda)$. For $s=2$ the conditions in \eqref{eq:M2q-assumptions-for-uniqueness} in addition demand that on the positive real axis, $\mathcal{D}_{(1,2)}(\lambda)$
has leading asymptotic behaviour $\mathcal{D}_{(1,2)}(r) \sim D_2^{\infty} \exp(a_2 r^\alpha)$ with $a_2>0$ and $D_2^{\infty}$ being some topological defect operator, cf.\ \eqref{eq:pert-def-asymp-assume}.
The specific functional form for $c(\theta)$ now follows because for $T_s$ the set $S_{T_s} \subset \mathbb{S}$ where the asymptotics is not well-defined (cf.\ Definition~\ref{def:alpha-asymptotics}) is just $S_{T_s} = \{ \pi \}$ by the first assumption. Corollary~\ref{cor:angle-dep-of-u-D} then implies $u(T_s,\theta) = a_s e^{i \alpha \theta}$ for some $a_s \in \mathbb{C}$ and $-\pi<\theta<\pi$. 
This gives $\cX_s = \Lambda(T_s) = [D_s^\infty,\mathrm{Re}(u(T_s,\theta))]  = [D_s^\infty,\,\mathrm{Re}(a_s\,e^{i\alpha\princ{\theta}})]$ as in Proposition~\ref{prop:one-solution}, cf.\ the definition of $\Lambda$ in \eqref{eq:fb-def}. 
We wrote the proposition in this way so that it is a self-contained statement about the asymptotic semiring $\mathcal{A}(\mathrm{Gr}(\mathcal{C})_{\ge 0})$ of the minimal model $M(p,q)$.
In any case, the key message is that if the negative real axis is the only anti-Stokes ray and $a_2>0$, then $p=2$ and all flow endpoints are fixed uniquely to be given by the identity defect. 
For conformal boundary conditions, this flow pattern has been observed already in \cite[Eqn.\,(9.4)]{Dorey:2005ak} by analysing boundary $g$-functions via elastic scattering (fusing a defect flow to a conformal boundary results in a corresponding boundary flow \cite{Bachas:2004sy,Runkel:2007wd}).
\end{remark}

Let us now turn to the proof of Proposition~\ref{prop:one-solution}. In the rest of this subsection, we will assume that the conditions~\eqref{eq:M2q-assumptions-for-uniqueness} hold. 
We will start by showing that \eqref{eq:one-solution} indeed solves \eqref{eq:Mpq-asymp-funrel_Hirota} and \eqref{eq:Mpq-asymp-funrel_bndcond}. In fact, we will show in Lemma~\ref{lem:one-Tsystem} that \eqref{eq:one-solution} solves the T-system functional equation \eqref{eq:Mpq-asymp-funrel_Tsys}, as well as  the Hirota equation~\eqref{eq:Mpq-asymp-funrel_Hirota}.
Afterwards, we show uniqueness.

\medskip

The inequality $q>2p>0$ gives the following inequalities for the constants in  \eqref{eq:Mpq-alpha} and \eqref{eq:mpq-gamma-beta-delta},
\begin{equation}\label{eq:abc-bounds}
    \tfrac12<\alpha<1
    ~~,\quad
    0<\beta<\tfrac{\pi}2
    ~~,\quad
    0<\gamma<\tfrac{\pi}2
    \ .
\end{equation}
Note also that $\alpha(\pi-\gamma) = \frac{\pi}{2}$, or equivalently $2\pi\alpha = \pi + 2\beta$.
We say the angle $\theta \in \mathbb{R}$ is
\begin{itemize}
    \item in the \textit{bulk wedge} if $|\theta|<\pi-\gamma$, and
    \item in a \textit{boundary wedge} if $\pi-\gamma<|\theta|<\pi$.
\end{itemize}
In analysing the functional equations, we frequently have to distinguish these two cases. We need the following auxiliary lemma.

\begin{lemma}[Shift identity]
\label{lem:one-shift}
For $c(\theta)=\cos(\alpha\princ{\theta})$ and $\theta\in(-\pi,\pi]$,
\begin{equation}\label{eq:one-boundary-cases}
c(\theta-\gamma)+c(\theta+\gamma) = 
\begin{cases}
2\cos(\beta)\,c(\theta) &;~  
\theta~\text{in the bulk wedge} \ ,
\\ 
0 &;~  
\theta~\text{in a boundary wedge} \ .
\end{cases}
\end{equation}
Moreover $c(\pm(\pi-\gamma))=0$, $c(\theta)>0$ in the bulk wedge, and $c(\theta)<0$ in the two boundary wedges.
\end{lemma}

\begin{proof}
Inside the bulk wedge both shifted arguments stay in $\theta\in(-\pi,\pi)$ and the cosine addition formula yields
\begin{equation}
\label{eq:one-shift-proof-bulk}
\cos(\alpha(\theta-\gamma))+\cos(\alpha(\theta+\gamma))
=
2\cos(\alpha\gamma)\cos(\alpha\theta)
=
2\cos(\beta)\,c(\theta) \ .
\end{equation}
On the boundary wedge $\pi-\gamma<\theta<\pi$ the shifted argument $\theta+\gamma$ lies outside $\theta\in(-\pi,\pi]$, and $\princ{\theta+\gamma}=\theta+\gamma-2\pi$. Using $2\pi\alpha=\pi+2\alpha\gamma$,
\begin{equation}
\label{eq:one-shift-proof-boundary}
\cos\bigl(\alpha(\theta+\gamma-2\pi)\bigr)
=
\cos\bigl(\alpha(\theta+\gamma)-\pi-2\alpha\gamma\bigr)
=
-\cos\bigl(\alpha(\theta-\gamma)\bigr),
\end{equation}
and the two terms cancel. The argument for the wedge $-\pi<\theta<-\pi+\gamma$ is analogous. The sign statements follow from $c(\pm(\pi-\gamma))=\cos(\pi/2)=0$ and $c(0)=1$.
\end{proof}

Note that the only place where $\gamma < \frac{\pi}2$ is needed in the above lemma is to ensure that $c(\theta)<0$ on the two boundary wedges.

\begin{lemma}[T-system and Hirota]
\label{lem:one-Tsystem}
For $p=2$, the solution \eqref{eq:one-solution} satisfies the boundary condition \eqref{eq:Mpq-asymp-funrel_bndcond}, the T-system \eqref{eq:Mpq-asymp-funrel_Tsys},  the reflection relation \eqref{eq:Mpq-asymp-funrel_reflect}, and the Hirota system \eqref{eq:Mpq-asymp-funrel_Hirota}.
\end{lemma}

\begin{proof}
For $p=2$ we have $\alpha=q/(2(q-2))$, $\beta = \pi /(q-2)$, $\gamma=2\pi/q$,
the shift $\mathsf{S}_{\pi p} = \mathsf{S}_{2\pi}$ is the identity, and $\cX_{q-1} = \Id$. Hence \eqref{eq:Mpq-asymp-funrel_bndcond} is immediate as $a_1=0=a_{q-1}$, and \eqref{eq:Mpq-asymp-funrel_reflect} reduces to $a_{q-s} = a_s$, which holds because $(q-s-1)\beta = \pi - (s-1)\beta$.

\medskip
\noindent
\textit{T-system \eqref{eq:Mpq-asymp-funrel_Tsys}:} Applying $\mathsf{S}_{-s\gamma}$ to \eqref{eq:Mpq-asymp-funrel_Tsys} gives
\begin{equation}\label{eq:shifted-T-sys}
    (\mathsf{S}_{-s\gamma}\cX_2)\,\cX_s = \mathsf{S}_{\gamma}\cX_{s-1} + \mathsf{S}_{-\gamma}\cX_{s+1} \quad
    \text{for}
    ~~ 2 \le s \le q-2 \ ,
\end{equation}
in which only one shift depends on $s$. 
As the asymptotic defect (recall Definition \ref{def:asymptotic-semiring}) in each $\cX_s$ is $\Id$, the above relations hold if and only if for all but finitely many $\theta \in (-\pi,\pi)$,
\begin{subequations}
\label{eq:one-Tsys-maxplus}
\begin{align}
    &a_2\,c(\theta-s\gamma) + a_s\,c(\theta)
    \,=\,
    \max\big\{ a_{s-1}\,c(\theta+\gamma) \,,\, a_{s+1}\,c(\theta-\gamma) \big\} \ ,
    \label{eq:one-Tsys-maxplus-eqn}
    \\
    &a_{s-1}\,c(\theta+\gamma) \neq a_{s+1}\,c(\theta-\gamma) \ .
    \label{eq:one-Tsys-maxplus-neq}
\end{align}
\end{subequations}
The equation in the first line amounts to the asymptotic exponent in \eqref{eq:shifted-T-sys}. The inequality is needed since, in the case where these two terms are equal, the asymptotic defect on the right hand side of \eqref{eq:shifted-T-sys} becomes $\Id + \Id$, which does not match the $\Id$ on the left hand side.

That \eqref{eq:one-Tsys-maxplus-neq} is true for all but finitely many $\theta$ is clear. It remains to check \eqref{eq:one-Tsys-maxplus-eqn}. To start with, one verifies the following trigonometric identities for all $\theta \in \mathbb{R}$ and $s$:
\begin{subequations}
\label{eq:one-Tsys-identities}
\begin{align}
    a_2\cos(\alpha\theta-s\beta) + a_s\cos(\alpha\theta)
    &\,=\, a_{s+1}\cos(\alpha\theta-\beta) \ ,
\label{eq:one-Tsys-id-A}
\\
    -\,a_2\cos\big(\alpha\theta-(s-2)\beta\big) + a_s\cos(\alpha\theta)
    &\,=\, a_{s-1}\cos(\alpha\theta+\beta) \ ,
\label{eq:one-Tsys-id-B}
\\
    a_{s+1}\cos(\alpha\theta-\beta) - a_{s-1}\cos(\alpha\theta+\beta)
    &\,=\, 2a_2 \,\cos\beta \, \cos\big(\alpha\theta-(s-1)\beta\big) \ .
\label{eq:one-Tsys-id-C}
\end{align}
\end{subequations}
The last identity is the difference of the first two. 
We can now verify \eqref{eq:one-Tsys-maxplus-eqn}, depending on whether $\theta$ is in the bulk wedge or in the boundary wedges.

\smallskip

\noindent
\textit{Bulk wedge.} Here $-\pi+\gamma <\theta < \pi-\gamma$, and so $c(\theta\pm\gamma) = \cos(\alpha\theta \pm \beta)$.
We need to check for which of \eqref{eq:one-Tsys-id-A} and \eqref{eq:one-Tsys-id-B} the right hand side is larger. This is determined by \eqref{eq:one-Tsys-id-C}. Indeed, using $\alpha\gamma=\beta$, by our constraint on $\theta$ we have $-\alpha\pi-(s-2)\beta<\alpha\theta-(s-1)\beta<\alpha\pi-s\beta$. As $2 \le s \le q-2$ and $q \ge 5$, we have $-\frac{3\pi}2<-\alpha\pi-(s-2)\beta$ and $\alpha\pi-s\beta<\frac{\pi}2$. In this range, the cosine changes sign precisely once, at $\alpha\theta-(s-1)\beta=-\frac{\pi}2$, or, equivalently, at $\theta - s \gamma = - \pi$.

For $\theta - s \gamma > - \pi$ we have $c(\theta-s\gamma) = \cos(\alpha\theta - s \beta)$ and the maximum on the right hand side of \eqref{eq:one-Tsys-maxplus-eqn} is given by \eqref{eq:one-Tsys-id-A}, showing that \eqref{eq:one-Tsys-maxplus-eqn} holds. If $\theta - s \gamma < - \pi$, we have, using $2 \pi \alpha = \pi + 2 \beta$ as in \eqref{eq:one-shift-proof-boundary},
\begin{equation}\label{eq:Tsys-lemma-aux1}
c(\theta-s\gamma) = \cos(\alpha(\theta+2\pi) - s \beta) = -\cos(\alpha\theta - (s-2) \beta) \ .
\end{equation}
So in this case, \eqref{eq:one-Tsys-maxplus-eqn} reduces precisely to \eqref{eq:one-Tsys-id-B}.

\smallskip

\noindent
\textit{Boundary wedges.} Suppose $\pi-\gamma<\theta<\pi$. 
Then $\theta-\gamma$ is in the bulk wedge and by Lemma~\ref{lem:one-shift} we have $\cos(\alpha\theta-\beta) = c(\theta-\gamma)>0$. 
Again by Lemma~\ref{lem:one-shift}, $c(\theta-\gamma) + c(\theta+\gamma) = 0$, and so $c(\theta+\gamma)<0$. The maximum on the right hand side of \eqref{eq:one-Tsys-maxplus-eqn} is hence given by $a_{s+1}\,c(\theta-\gamma)$.
Finally, as also $\pi-(s+1)\gamma<\theta-s\gamma\le\pi-s\gamma$ and $(s+1)\gamma<2\pi$, we get 
$c(\theta-s\gamma) = \cos(\alpha\theta-s\beta)$. Hence \eqref{eq:one-Tsys-maxplus-eqn} reduces to \eqref{eq:one-Tsys-id-A}.

Suppose next that $-\pi<\theta<-\pi+\gamma$. The argument is analogous, except that now $\theta+\gamma$ is in the bulk wedge, so that $\cos(\alpha\theta+\beta)=c(\theta+\gamma)>0$, and $c(\theta-s\gamma) = -\cos(\alpha\theta - (s-2) \beta)$ as in \eqref{eq:Tsys-lemma-aux1}.
Hence \eqref{eq:one-Tsys-maxplus-eqn} reduces to \eqref{eq:one-Tsys-id-B}.

\medskip
\noindent
\textit{Hirota system \eqref{eq:Mpq-asymp-funrel_Hirota}:}
From $a_s = a_2 \sin((s-1)\beta)/\sin\beta$ we get $a_{s-1}+a_{s+1} = 2\cos(\beta)\,a_s$ for $2 \le s \le q-2$ (note that $a_1=0=a_{q-1}$).
Hence, using Lemma~\ref{lem:one-shift},
\begin{align}
(\mathsf{S}_\gamma \cX_s)(\mathsf{S}_{-\gamma} \cX_s)
  &= \big[\Id,\, a_s\big(c(\theta-\gamma)+c(\theta+\gamma)\big)\big] \ ,
  \nonumber
  \\
  1 + \cX_{s-1}\cX_{s+1}
  &= [\Id,0] + \big[\Id,\, 2\cos(\beta)\, a_s\, c(\theta)\big] \ .
\end{align}
Recall that $a_s>0$ for $2\le s\le q-2$, and $\cos\beta>0$ as $\beta=\pi/(q-2)<\pi/2$.
In the bulk wedge, the left hand side is $[\Id,\,2\cos(\beta)\,a_s\,c(\theta)]$ and, as $c(\theta)>0$, the second summand on the right hand side is the dominant one.
In a boundary wedge, the left hand side is $[\Id,0]$ and, as $c(\theta)<0$, the summand $[\Id,0]$ is dominant on the right hand side.
Hence in both cases the two sides agree.
\end{proof}

Next we gather the ingredients we need to show that the solution in Proposition~\ref{prop:one-solution} is  unique under the assumptions stated.

\begin{lemma}\label{lem:positivity}
The $a_s$ for $s=2,\cdots, q-2$ are real and positive, and satisfy the recursion
\begin{equation}\label{eq:as-recursion}
    a_{s+1} + a_{s-1} = 2 \cos(\beta)\,a_s   
\ ,
\quad a_1=0=a_{q-1} \ .
\end{equation}
\end{lemma}

\begin{proof}
The Hirota equation \eqref{eq:Mpq-asymp-funrel_Hirota} implies the following identity for the asymptotic exponent, which has to hold for all but finitely many $\theta$ and $s=2,\dots,q-2$,
\begin{equation}\label{eq:hirota-2nd-entry}
    \mathrm{Re}\big(a_s \big(e^{i \alpha [\theta+\gamma]}+e^{i \alpha [\theta-\gamma]}\big)\big)
    =
    \max\!\Big( 0 \,,\, \mathrm{Re}\big(a_{s-1} e^{i \alpha [\theta]}\big) +
    \mathrm{Re}\big(a_{s+1} e^{i \alpha [\theta]}\big)
    \Big) \ .
\end{equation}
Take $\theta$ in the bulk wedge.
Writing $a_s = x+iy$ and $s(\varphi) = \sin(\alpha[\varphi])$, the left hand side becomes
\be
\begin{split}
    \text{(lhs)} &= 
    x \big(c(\theta+\gamma)+c(\theta-\gamma)\big) - y \big(s(\theta+\gamma)+s(\theta-\gamma)\big) \ ,
 \\
    &= 2 \cos(\beta) \big( x c(\theta) - y s(\theta) \big) \ .
\end{split}\ee
In the second step we used Lemma~\ref{lem:one-shift} and the corresponding trigonometric identity for $s(\varphi)$. Suppose now that $x<0$. Then for $\theta$ small enough, the left hand side of \eqref{eq:hirota-2nd-entry} is negative, while the right hand side is always $\ge 0$. Hence we must have $x \ge 0$. Note that for $x=0$ we must also have $y=0$ or else the left hand side can again be negative for small $\theta$. 
Altogether, so far we know $\mathrm{Re}(a_s) \ge 0$ and $\mathrm{Re}(a_s)= 0 \Rightarrow a_s=0$.

Suppose $a_s = 0$ for some $s=3,\dots,q-2$ (recall that $a_2>0$ by assumption). Let $s$ be minimal with this property. Then the left hand side of \eqref{eq:hirota-2nd-entry} is zero. For the right hand side, note that $\mathrm{Re}(a_{s-1})>0$ by assumption, and $\mathrm{Re}(a_{s+1}) \ge 0$ by the previous argument. Hence for small enough $\theta$, the right hand side is positive, which is a contradiction. Hence $\mathrm{Re}(a_s) > 0$ for all $s=2,\dots,q-2$.

It follows that
for small enough $\theta$, the right hand side of \eqref{eq:hirota-2nd-entry} is always positive, and we need to solve, for $s=2,\dots,q-2$,
\be\begin{split}
& 2 \cos(\beta) \big( \mathrm{Re}(a_s) c(\theta) -\mathrm{Im}(a_s) s(\theta) \big)
 \\
&=
\mathrm{Re}(a_{s-1}) c(\theta) -\mathrm{Im}(a_{s-1}) s(\theta)
+
\mathrm{Re}(a_{s+1}) c(\theta) -\mathrm{Im}(a_{s+1}) s(\theta)
\ .
\end{split}
\ee
Comparing coefficients of $c(\theta)$ and of $s(\theta)$, we see that $\mathrm{Re}(a_s)$ and $\mathrm{Im}(a_s)$ satisfy the same recursion relation, for $s=2,\dots,q-2$,
\begin{equation}
    z_{s+1} = 2 \cos(\beta) z_s - z_{s-1}  
\ ,
\quad z_1=0=z_{q-1} \ .
\end{equation}
But for $z_s = \mathrm{Im}(a_s)$, by assumption \eqref{eq:M2q-assumptions-for-uniqueness} we have $z_2 = \mathrm{Im}(a_2)=0$, and the recursion then implies $z_s=0$ for all $s$. Hence all $a_s$ are real and positive, and satisfy the recursion \eqref{eq:as-recursion}.
\end{proof}  

\begin{lemma}
\label{lem:recursionclosed}    
The recursion relation 
$a_{s+1} + a_{s-1} = 2 \cos(\beta)\,a_s$ has a solution with $a_1=0=a_{q-1}$ and $a_s>0$ for $s=2,\dots,q-2$ if and only if $p=2$. In that case, the solutions are parametrised by $a_2>0$ and are given by
    \begin{equation}\label{eq:PF-eigenvector}
a_{s}=a_2\frac{\sin((s-1)\beta)}{\sin\beta} \ . 
    \end{equation}
\end{lemma}
\begin{proof}
The recursion relation can be rewritten as the eigenvalue equation for the adjacency matrix of the $A_{q-3}$ Dynkin diagram: $A_{q-3}\,v = 2\cos\beta\,v$ with $v=(a_2,\dots,a_{q-2})$. As all $a_s>0$, $v$ must be the Frobenius-Perron eigenvector of $A_{q-3}$. This eigenvector is unique up to a positive scalar and is given by \eqref{eq:PF-eigenvector}. Correspondingly, $2 \cos(\beta)$ must be equal to the Frobenius-Perron eigenvalue of $A_{q-3}$:
\begin{equation}
\label{eq:one-pf}
2\cos\beta = 2\cos\left(\frac{\pi}{q-2}\right)\ .
\end{equation}
Both arguments of the cosine lie in $[0,\frac{\pi}2]$, and cosine is injective in this range. Hence $\beta = \frac{\pi}{q-2}$, or equivalently $p=2$, cf.\ \eqref{eq:mpq-gamma-beta-delta}.
\end{proof}

\begin{lemma}
\label{lem:asymp-def-M2p} We have $\cX_s=\bigl[D_s^\infty,\,a_s\,c(\theta)\bigr]$, with $a_s$ as in \eqref{eq:PF-eigenvector}. The asymptotic defects satisfy $D_{s+1}^\infty D_{s-1}^\infty=\Id$ for $s=2,\dots,q-2$.
\end{lemma}

\begin{proof}
The form for $\cX_s$ follows from the initial ansatz in Proposition~\ref{prop:one-solution}, together with Lemmas~\ref{lem:positivity} and~\ref{lem:recursionclosed}.
The equations for the asymptotic defects arise from evaluating \eqref{eq:Mpq-asymp-funrel_Hirota} in the bulk and boundary wedges, respectively.

\smallskip

\noindent
\textit{Bulk wedge.} The $\cX_{s-1} \cX_{s+1}$ term dominates on the right hand side of \eqref{eq:Mpq-asymp-funrel_Hirota}, and so $(D_s^\infty)^2=D_{s+1}^\infty D_{s-1}^\infty$ for $s=2,\dots,q-2$.

\smallskip

\noindent
\textit{Boundary wedges.} The $[\Id,0]$ term dominates on the right hand side of \eqref{eq:Mpq-asymp-funrel_Hirota}, and so $(D_s^\infty)^2=\Id$ for $s=2,\dots,q-2$.
\end{proof}

\begin{proof}[Proof of Proposition~\ref{prop:one-solution}]
Lemma~\ref{lem:one-Tsystem} implies that the Hirota system \eqref{eq:Mpq-asymp-funrel_Hirota} with boundary conditions \eqref{eq:Mpq-asymp-funrel_bndcond} is solved by \eqref{eq:one-solution} (for $p=2$ and $q\ge 5$ odd).

We now show that the condition on $p,q$ is forced and the solution \eqref{eq:one-solution} is unique.
The condition $p=2$ was derived in Lemma~\ref{lem:recursionclosed}. Since $p,q$ are coprime and $q>2p$ by assumption (stated as $\gamma<\pi/2$), we have $q\geq5$ odd. That the asymptotic exponent is of the form stated in \eqref{eq:one-solution} was shown in Lemma~\ref{lem:asymp-def-M2p}. 
It remains to see that the asymptotic defects satisfy $D^\infty_s=\Id$. 
As $D^\infty_1 = \Id$ by \eqref{eq:Mpq-asymp-funrel_bndcond} and $D_{s+1}^\infty D_{s-1}^\infty=\Id$ by Lemma~\ref{lem:asymp-def-M2p}, we conclude that $D^\infty_s=\Id$ for $s$ odd. Again by \eqref{eq:Mpq-asymp-funrel_bndcond}, we have $D^\infty_{q-1} = D_{(1,q-1)} = \Id$ as $p=2$. As $q$ is odd, $q-1$ is even, and $D_{s+1}^\infty D_{s-1}^\infty=\Id$ implies that $D^\infty_s=\Id$ also for $s$ even.
\end{proof}

\subsection{Solution in asymptotic semiring of \texorpdfstring{$M(p,q)$}{M(p,q)}}
\label{sec:pq-solution}

In Proposition~\ref{prop:one-solution} we saw that the condition that each perturbed defect only has a single anti-Stokes ray (along the negative real axis) forces $p=2$. 
For $p>2$ at least one perturbed defect must have more than one anti-Stokes ray, and this opens the possibility that the flow endpoint depends on the angle along which the RG flow approaches infinity.
In this subsection, we give -- for every coprime $p$, $q$ with $q>2p \ge 4$ -- a solution of the functional relations \eqref{eq:Mpq-asymp-funrel} which exhibits this behaviour (Proposition~\ref{prop:pq-solution} and Remark~\ref{rem:pq-Tsys}). 
We will not address uniqueness of this solution, but spot-checks by TCSA support that this solution is indeed realised, see \cite[Sec.\,6]{phys-paper}.
The solution is also defined for $p=2$, where it reduces to \eqref{eq:one-solution}. 

The leading defects will be the entries $D_{(r,1)} \in \mathrm{Gr}(\mathcal{C})_{\ge 0}$, $1 \le r \le p-1$, of the first column of the Kac table. It will be convenient to set
\begin{equation}\label{eq:pq-truncation}
    D_{(r,1)} = 0 \quad \text{for} ~~ r \le 0 ~~\text{and}~~ r \ge p \ .
\end{equation}
These labels are closed under fusion. We have $D_{(1,1)} = \Id$, and $D_{(p-1,1)}$ is tensor-invertible with $D_{(p-1,1)} D_{(r,1)} = D_{(p-r,1)}$. The only fusion rule we will need is
\begin{equation}\label{eq:pq-qdet}
    D_{(r,1)}\,D_{(r,1)} \,=\, \Id + D_{(r-1,1)}\,D_{(r+1,1)}
    \qquad \text{for} ~~ 1 \le r \le p-1 \ ,
\end{equation}
which by \eqref{eq:pq-truncation} reads $D_{(r,1)}^2 = \Id$ at the two ends $r=1$ and $r=p-1$. Note also that $(1,q-1) \sim (p-1,1)$, so that the boundary condition \eqref{eq:Mpq-asymp-funrel_bndcond} can equivalently be stated as $\cX_{q-1} = [D_{(p-1,1)},0]$.

The asymptotic exponents will be built from the shifted copies
\begin{equation}\label{eq:pq-channels}
    c_r(\theta) \,:=\, c\big(\theta-(r-1)\pi\big)
    \qquad \text{where} ~~ c(\theta) = \cos\big(\alpha\princ{\theta}\big) 
    \ .
\end{equation}
We have $c_r(\theta) = c(\theta)$ for $r$ odd and $c_r(\theta) = c(\theta-\pi)$ for $r$ even.
Finally, for $1 \le s \le q-1$ let $n_s$ be the unique integer with
\begin{equation}\label{eq:pq-ns}
    (n_s-1)\,\pi \,<\, s\gamma \,<\, n_s\,\pi \ ,
\end{equation}
which exists because $s\gamma/\pi = ps/q$ is not an integer for $0<s<q$. As $s$ increases by one, $s\gamma$ increases by $\gamma<\pi$. Hence, $n_s$ is monotonically increasing from $n_1 = 1$ to $n_{q-1}=p$ with $n_{s+1} \in \{n_s,n_s+1\}$.

\begin{proposition}
\label{prop:pq-solution}
Let $p,q$ be coprime with $q>2p \ge 4$, and let $a_2>0$. For $1 \le s \le q-1$ set
\begin{equation}
\label{eq:pq-solution}
    \cX_s \,=\, 
    \delta_{n_s>1} \,
    \big[\, D_{(n_s-1,1)} \,,\; A_{n_s-1}(s)\, c_{n_s-1}(\theta) \,\big]
    \,+\,
    \delta_{n_s<p} \, 
    \big[\, D_{(n_s,1)} \,,\; A_{n_s}(s)\, c_{n_s}(\theta) \,\big]
    \ ,
\end{equation}
where
\begin{equation}
\label{eq:pq-ampl}
    A_r(s) \,=\, a_2\,\frac{\sin\!\big(\alpha\big(\,(s-1)\gamma-(r-1)\pi\,\big)\big)}{\sin\beta} \ .
\end{equation}
Then $\cX_1 = \Id$ and $\cX_{q-1} = [D_{(p-1,1)},0]$, and the $\cX_s$ solve the Hirota system \eqref{eq:Mpq-asymp-funrel_Hirota} and the reflection relation \eqref{eq:Mpq-asymp-funrel_reflect}.
\end{proposition}

We refer to the two summands in \eqref{eq:pq-solution} as the \textit{channels} $n_s-1$ and $n_s$ of $\cX_s$. 
For $n_s=1$ the $\delta$-function in front of the first summand is zero and this channel is absent. Note that for $n_s=1$, the term $[D_{(0,1)},A_0(s)c_0(\theta)]$ is not actually an element of the asymptotic semiring $\mathcal{A}(\mathrm{Gr}(\mathcal{C})_{\ge 0})$ as $D_{(0,1)}=0$ by \eqref{eq:pq-truncation} but the only element in $\mathcal{A}(\mathrm{Gr}(\mathcal{C})_{\ge 0})$ with zero in the first entry is $[0,-\infty]$, cf.\ Lemma~\ref{lem:Dmax-def}. For $n_s=p$ accordingly the second $\delta$-function gives zero. 
In particular, for $p=2$ only channel $1$ is ever present, $A_1(s) = a_2\sin((s-1)\beta)/\sin\beta$, and \eqref{eq:pq-solution} is the solution \eqref{eq:one-solution}.

Define 
\begin{equation}\label{eq:pq-Stokes}
    \theta^\ast_s \,=\, \big|\princ{s\gamma}\big| \ .
\end{equation}
By Remark~\ref{rem:theta-star-proof} below, the channel $r$ of $\cX_s$ with $r$ odd dominates for $|\theta|<\theta^\ast_s$ and the one with $r$ even dominates for $\pi>|\theta|>\theta^\ast_s$ (provided the corresponding $\delta$-function is nonzero).
If both $\delta$-functions are nonzero,
the dominant summand of $\cX_s$ changes at the two angles $\theta=\pm\theta^\ast_s$, and so these describe the angles of the anti-Stokes rays.

\begin{remark}
\label{rem:pq-Tsys}
\begin{itemize}
    \item[i)] As in Lemma~\ref{lem:one-Tsystem} for $p=2$, the solution \eqref{eq:pq-solution} also satisfies the T-system \eqref{eq:Mpq-asymp-funrel_Tsys}. The proof is an adaptation of the one above and we do not reproduce it here.
    \item[ii)] In Proposition~\ref{prop:pq-solution} we have assumed that $q>2p$, and in this solution the perturbation of $D_{(1,2)}$ has a single anti-Stokes ray. We can also find a solution in the regime $\frac{4p}{3} < q <2p$, corresponding to the range $0 < h_{(1,3)}< 1/2$ (cf.\ point 1 below \eqref{eq:semiring-hom-mpq}), but in this case the perturbation of $D_{(1,2)}$ develops two anti-Stokes rays, rather than just one. 
    While we do not discuss the details in this paper, we present a solution of the functional relations in this regime in Remark~\ref{rem:solutionless2p} below. 
\end{itemize}

\end{remark}

\begin{figure}[tbp]
    \centering
    \includegraphics[width=0.92\linewidth]{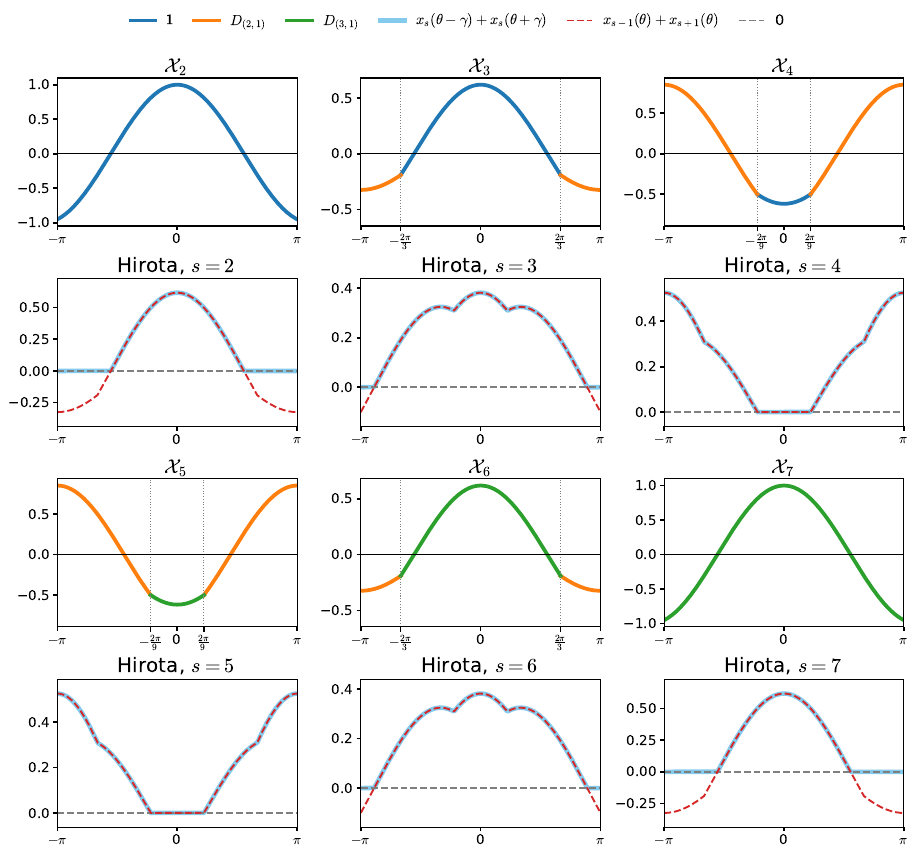}
    \caption{Asymptotic defects/exponents and Hirota equation for $s=2,\dots,7$ in $M(4,9)$. 
In the Hirota plots we give the asymptotic exponentials of the left hand side (solid light blue)  and right hand side (maximum of the dashed red line and the dashed gray zero) of the Hirota equation.
 Write $x_s(\theta)$ for the asymptotic exponent of $\cX_s$. For $s=4,5$ and $|\theta|<\frac{2\pi}9$ the dashed curve $x_{s-1}+x_{s+1}$ lies exactly on $0$: there the summand $\Id$ in \eqref{eq:Mpq-asymp-funrel_Hirota} is neither dominant nor subdominant, and the relation holds by the fusion rule $D_{(2,1)}^2 = \Id+D_{(3,1)}$ (the case ``Two images in $I$'' in the proof of Proposition~\ref{prop:pq-solution}).} 
    \label{fig:pq-M49}
\end{figure}

\begin{figure}[tb]
    \centering
\includegraphics[width=0.92\linewidth]{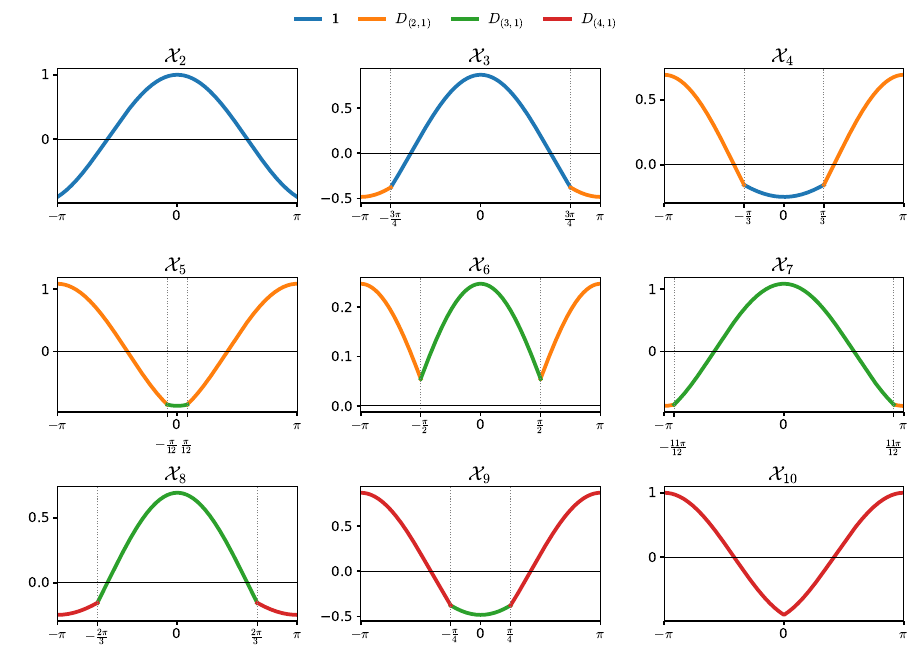}
    \caption{Asymptotic defects/exponents in $M(5,12)$.}    \label{fig:M512}
\end{figure}

\begin{example}
\label{ex:pq-examples}
For $r=1,2,3,4$ the amplitudes \eqref{eq:pq-ampl} are, using $\alpha\gamma=\beta$ and $\alpha\pi = \frac\pi2+\beta$,
\begin{align}
\label{eq:pq-ampl-explicit}
    A_1(s) &= a_2\frac{\sin((s-1)\beta)}{\sin\beta}
    ~,&
    A_2(s) &= -a_2\frac{\cos((s-2)\beta)}{\sin\beta}
    ~,
    \nonumber\\
    A_3(s) &= -a_2\frac{\sin((s-3)\beta)}{\sin\beta}
    ~,&
    A_4(s) &= a_2\frac{\cos((s-4)\beta)}{\sin\beta} \ .
\end{align}
Let us consider as an example the minimal model $M(4,9)$. Here $\alpha=\frac{9}{10}$, $\beta = \frac{2\pi}5$, $\gamma=\frac{4\pi}9$, and
$n_s = 1,1,2,2,3,3,4,4$ for $s=1,\dots,8$. Since $c$ has period $2\pi$ we have
$c_3(\theta) = c_1(\theta) := c(\theta)$ and $c_2(\theta) = c(\theta-\pi)$, and since $5\beta = 2\pi$ the
amplitudes \eqref{eq:pq-ampl-explicit} take only the values listed below:
{\allowdisplaybreaks
\begin{align}
    \cX_1 &= [\,\Id\,,\,0\,] \ ,
    \nonumber\\
    \cX_2 &= [\,\Id\,,\, a_2\,c(\theta)\,] \ ,
    \nonumber\\
    \cX_3 &= [\,\Id\,,\, 2\cos\beta\; a_2\,c(\theta)\,]
          + [\,D_{(2,1)}\,,\, -\tfrac{\cos\beta}{\sin\beta}\; a_2\, c(\theta-\pi)\,] \ ,
    &
    \theta^\ast_3 &= \tfrac{6\pi}{9} \ ,
    \nonumber\\
    \cX_4 &= [\,\Id\,,\, -2\cos\beta\; a_2\,c(\theta)\,]
          + [\,D_{(2,1)}\,,\, -\tfrac{\cos 2\beta}{\sin\beta}\; a_2\, c(\theta-\pi)\,] \ ,
    &
    \theta^\ast_4 &= \tfrac{2\pi}{9} \ ,
    \nonumber\\
    \cX_5 &= 
    [\,D_{(3,1)}\,,\, -2\cos\beta\; a_2\,c(\theta)\,]      + 
    [\,D_{(2,1)}\,,\, -\tfrac{\cos 2\beta}{\sin\beta}\; a_2\, c(\theta-\pi)\,]
  \ ,
    &
    \theta^\ast_5 &= \tfrac{2\pi}{9} \ ,
    \nonumber\\
    \cX_6 &= 
    [\,D_{(3,1)}\,,\, 2\cos\beta\; a_2\,c(\theta)\,] 
    + 
    [\,D_{(2,1)}\,,\, -\tfrac{\cos\beta}{\sin\beta}\; a_2\, c(\theta-\pi)\,]
    \ ,
    &
    \theta^\ast_6 &= \tfrac{6\pi}{9} \ ,
    \nonumber\\
    \cX_7 &= [\,D_{(3,1)}\,,\, a_2\,c(\theta)\,] \ ,
    \nonumber\\
    \cX_8 &= [\,D_{(3,1)}\,,\,0\,] \ .
\end{align}}
For $\cX_{2}$ and $\cX_{7}$ there is a single anti-Stokes ray on the negative real axis. For $\cX_{3,4,5,6}$ there are two anti-Stokes rays at $\theta = \pm \theta_s^*$: For $|\theta|<\theta_s^*$, the first summand is dominant, and for $\pi>|\theta|>\theta_s^*$ the second summand is dominant. 
The reflection relation $\cX_8\cX_s = \mathsf{S}_{4\pi}\cX_{9-s}$ in \eqref{eq:Mpq-asymp-funrel_reflect}
is manifest: the shift acts trivially since $p$ is even and $c$ has period $2\pi$; multiplication with $\cX_8 = [D_{(3,1)},0]$
exchanges $\Id \leftrightarrow D_{(3,1)}$ and fixes $D_{(2,1)}$.
The $\theta$-dependence of the asymptotic defect and exponent of each $\cX_s$ is shown in Figure~\ref{fig:pq-M49}. That figure also shows the asymptotic exponent for each of the Hirota  equations to illustrate that the agreement is quite non-trivial. $M(4,9)$ also exhibits a feature which is absent for $p\le3$, namely that non-invertible
asymptotic defects occur, see Figure~\ref{fig:pq-M49} and Table \ref{tab:49}. As a further example we report the corresponding results for $M(5,12)$ in Table~\ref{tab:49} and Figure~\ref{fig:M512}. 
\end{example}

\begin{table}[bt]
\centering
\qquad
\begin{tabular}{|c|cc|c|}
\hline
$M(4,9)$ & $|\theta|<\theta^\ast_s$ & $|\theta|>\theta^\ast_s$ & $\theta^\ast_s$ \\ \hline
$s=1,2$ & $\Id$ & $\Id$ & -- \\
$s=3$ & $\Id$ & $D_{(2,1)}$ & $6\pi/9$ \\
$s=4$ & $\Id$ & $D_{(2,1)}$ & $2\pi/9$ \\
$s=5$ & $D_{(3,1)}$ & $D_{(2,1)}$ & $2\pi/9$ \\
$s=6$ & $D_{(3,1)}$ & $D_{(2,1)}$ & $6\pi/9$ \\
$s=7,8$ & $D_{(3,1)}$ & $D_{(3,1)}$ & --\\
\hline
\end{tabular}\hfill
\begin{tabular}{|c|cc|c|}
\hline
$M(5,12)$ & $|\theta|<\theta^\ast_s$ & $|\theta|>\theta^\ast_s$ & $\theta^\ast_s$ \\ \hline
$s=1,2$   & $\mathbf{1}$ & $\mathbf{1}$ & $-$ \\
$s=3$     & $\mathbf{1}$ & $D_{(2,1)}$  & $9\pi/12$ \\
$s=4$     & $\mathbf{1}$ & $D_{(2,1)}$  & $4\pi/12$ \\
$s=5$     & $D_{(3,1)}$  & $D_{(2,1)}$  & $\pi/12$ \\
$s=6$     & $D_{(3,1)}$  & $D_{(2,1)}$  & $6\pi/12$ \\
$s=7$     & $D_{(3,1)}$  & $D_{(2,1)}$  & $11\pi/12$ \\
$s=8$     & $D_{(3,1)}$  & $D_{(4,1)}$  & $8\pi/12$ \\
$s=9$     & $D_{(3,1)}$  & $D_{(4,1)}$  & $3\pi/12$ \\
$s=10,11$ & $D_{(4,1)}$  & $D_{(4,1)}$  & $-$\\
\hline
\end{tabular}
\caption{Asymptotic defects in $\cX_s$ depending on the angle $\theta$, in $M(4,9)$ and $M(5,12)$.}
\label{tab:49}
\end{table}

\begin{remark}\label{rem:no-semiring-hom}
The results in Proposition~\ref{prop:pq-solution} and Remark~\ref{rem:pq-Tsys}\,(ii), as well as numerical studies, support the expectation that the topological defects with Kac-labels $(1,s)$ flow to topological defects $(r,1)$. As noted in the introduction, this flow pattern \textit{cannot} be described by a semiring homomorphism from the semiring spanned by the representations $(1,s)$ to the semiring spanned by the representation $(r,1)$. Indeed, there is no such semiring homomorphism in the first place. 
\\
To see this, we first show a corresponding result for $sl(2)$-fusion semirings (see \cite[Sec.\,2.2]{Ambrosino:2026kfu} for a similar argument). Write $F(k)_{\ge0}$ for the semiring (see Example~\ref{ex:semirings}) of truncated $sl(2)$-fusion rules with simple objects $[0],[1],\dots,[k]$.  
Let $\chi_k : F(k)_{\ge0} \to \mathbb{R}_{>0}$ be the one-dimensional representation given by the Frobenius-Perron dimensions, $\chi_k([x]) = \sin(\pi (x+1)/(k+2))/\sin(\pi/(k+2))$,
which in this case are the quantum dimensions. These satisfy $\chi_k([x]) = \chi_k([k-x])$, as well as the following estimates:
\begin{subequations}\label{eq:su2-fusion-estimates}
\begin{align}
&\chi_k([x]) \ge 1 
~~\text{and}~~\chi_k([1])<2
&&\text{for} ~~  k \ge 1 ~~\text{and}~~ x=0,\dots,k \ ,
\label{eq:su2-fusion-estimates-1}
\\
&2 \le \chi_k([2]) \le \chi_k([x])
&&\text{for} ~~  k \ge 4~~\text{and}~~ x=2,\dots,k-2 \ ,
\label{eq:su2-fusion-estimates-2}
\\
&\chi_\ell([1]) < \chi_k([1])
&&\text{for} ~~  1 \le \ell < k \ .
\label{eq:su2-fusion-estimates-3}
\end{align}
\end{subequations}
Suppose there exists a semiring homomorphism $\varphi : F(k)_{\ge0} \to F(\ell)_{\ge0}$ for $1 \le \ell<k$.
Then $\varphi([1]) = \sum_{x=0}^\ell n_x [x]$ for some $n_x \in \mathbb{Z}_{\ge 0}$.
The one-dimensional representation $\chi_\ell \circ \varphi : F(k)_{\ge0} \to \mathbb{R}_{>0}$ is everywhere positive by construction and hence must be equal to the Frobenius-Perron dimensions for $F(k)_{\ge 0}$. Thus
\begin{equation}\label{eq:su2-fusion-PF-condition}
    \chi_k([1]) = \chi_\ell(\varphi([1])) = \sum_{x=0}^\ell n_x \chi_\ell([x])\ .
\end{equation}
As by \eqref{eq:su2-fusion-estimates-1}, $\chi_\ell([x]) \ge 1$ for all $x$ and $\chi_k([1])<2$, at most one $n_x$ can be non-zero and that $n_x$ must be equal to $1$. Hence $\varphi([1])=[x]$ for some $x=0,\dots,\ell$.
If $\ell \ge 4$, by \eqref{eq:su2-fusion-estimates-2} we have $\chi_\ell([x]) \ge 2$ for $x=2,\dots,\ell-2$, and so $x \in \{0,1,\ell-1,\ell\}$. For these $x$, $\chi_\ell([x])$ takes the values $1$ and $\chi_\ell([1])$. By \eqref{eq:su2-fusion-estimates-3} we have $\chi_\ell([x])<\chi_k([1])$, and so \eqref{eq:su2-fusion-PF-condition} has no solution. The remaining three cases also cannot happen: for $\ell=1,2,3$, $\chi_\ell([1])$ is the largest value taken by $\chi_\ell([x])$, and we have $\chi_\ell([1])<\chi_k([1])$.
\\
We now apply the above non-existence result to minimal models.
Let $S_\mathrm{UV}$ be the semiring spanned (over $\mathbb{Z}_{\ge 0}$) by the $D_{(1,s)}$ and $S_\mathrm{IR}$ that spanned by the $D_{(r,1)}$. If $3 \le p < q$, there is a semiring isomorphism $F(q-2)_{\ge 0} \xrightarrow{\sim} S_\mathrm{UV}$, $[x] \mapsto D_{(1,x+1)}$, and analogously $F(p-2)_{\ge 0} \cong S_\mathrm{IR}$. The above result implies there is no semiring homomorphism $S_\mathrm{UV} \to S_\mathrm{IR}$.
This is still true for $2=p<q$. In this case, the above map is still a surjection $F(q-2)_{\ge 0}\to S_\mathrm{UV}$, and $S_\mathrm{IR} = \mathbb{Z}_{\ge 0}$ (generated by $D_{(1,1)}=\mathrm{Id})$). A homomorphism $S_\mathrm{UV} \to S_\mathrm{IR}$ hence induces a homomorphism 
$F(k)_{\ge 0}\to \mathbb{Z}_{\ge 0}$. This would require $\chi_k([1])=1$, which has no solution for $k \ge 3$ (corresponding to $q \ge 5$).
\end{remark}

To prepare the proof of Proposition \ref{prop:pq-solution}, we first establish a series of lemmas. 
Besides the inequalities \eqref{eq:abc-bounds} we shall repeatedly use 
\begin{equation}\label{eq:pq-alpha-ids}
    \alpha\,(2\pi-2\gamma) \,=\, \pi \ ,
    \qquad
    q\gamma \,=\, p\pi \ .
\end{equation}
We extend $A_r(s)$, $c_r$ to all $r,s \in \mathbb{Z}$ by the same formulas, and we set
\begin{equation}\label{eq:pq-Y}
    Y_r(s,\theta) \,:=\, A_r(s)\, c_r(\theta) \ .
\end{equation}
We write $x_s(\theta)$ for the asymptotic exponent of $\cX_s$, i.e.\ for the larger of the $Y_r(s,\theta)$ over the channels $r$ present in $\cX_s$.

We say that $\theta$ is in the \emph{bulk wedge of $c_r$} if $|\princ{\theta-(r-1)\pi}|<\pi-\gamma$, and in the \emph{boundary wedge of $c_r$}
if $|\princ{\theta-(r-1)\pi}|>\pi-\gamma$. Note that for $r$ odd, this agrees with the definition of bulk and boundary wedge below \eqref{eq:abc-bounds}.

\begin{lemma}[Channel identities]
\label{lem:pq-channel}
For all $r,s \in \mathbb{Z}$ and all $\theta$:
\begin{enumerate}
    \item[{\rm(i)}] $A_r(s-1) + A_r(s+1) = 2\cos\beta \, A_r(s)$;
    \item[{\rm(ii)}] 
    \be
c_r(\theta-\gamma) + c_r(\theta+\gamma) = \begin{cases}
    2\cos(\beta)\, c_r(\theta)\,,   & {\rm if}\; |[\theta-(r-1)\pi]| < \pi-\gamma\\
    0\,,  & {\rm if}\; |[\theta-(r-1)\pi]| > \pi-\gamma
\end{cases}\; . 
    \ee
    Moreover $c_r(\theta)>0$ in the first case and $c_r(\theta)<0$ in the second. 

    \item[{\rm(iii)}] $A_r(s-1) = -A_{r+2}(s+1)$, and hence $Y_r(s-1,\theta)+Y_{r+2}(s+1,\theta)=0$;
    \item[{\rm(iv)}] $A_{p-r}(q-s) = A_r(s)$ and $n_{q-s} = p+1-n_s$.
\end{enumerate}
\end{lemma}

\begin{proof}
Write $\varphi_r(s) = \alpha((s-1)\gamma-(r-1)\pi)$ for the argument of the sine in \eqref{eq:pq-ampl}. Then $\varphi_r(s\pm1) = \varphi_r(s)\pm\beta$ because $\alpha\gamma=\beta$, 
and (i) is the addition theorem for the sine. Part (ii) is Lemma~\ref{lem:one-shift} applied to the angle $\theta-(r-1)\pi$. For (iii), $\varphi_{r+2}(s+1)-\varphi_r(s-1) = \alpha(2\gamma-2\pi)= -\pi$ by \eqref{eq:pq-alpha-ids}, so the two sines are opposite, while $c_{r+2}=c_r$. For (iv), $\varphi_{p-r}(q-s)+\varphi_r(s) = \alpha((q-2)\gamma-(p-2)\pi) = \alpha(2\pi-2\gamma) = \pi$ using $q\gamma=p\pi$, and so 
    $\sin \varphi_{p-r}(q-s) = \sin \varphi_r(s)$.
Finally, $(q-s)\gamma = p\pi-s\gamma$ lies in $((p-n_s)\pi,(p-n_s+1)\pi)$ by \eqref{eq:pq-ns}.
\end{proof}

The channel identities are most easily handled with the following device. Fix $\theta \in (0,\pi)$ and call the points $\pm\theta+2m\pi$, $m \in \mathbb{Z}$, the \textit{images} of $\theta$. Each interval $(j\pi,(j+1)\pi)$ contains exactly one image, which we denote by
\begin{equation}\label{eq:pq-images}
    \theta_j \,=\,
    \begin{cases}
        j\pi+\theta &; ~ j ~\text{even} \ ,
        \\
        (j+1)\pi-\theta &; ~ j ~\text{odd} \ ,
    \end{cases}
    \qquad\text{so that}\quad \theta_{j+2} = \theta_j+2\pi \ .
\end{equation}
Two consecutive images $\theta_j,\theta_{j+1}$ are symmetric about $(j+1)\pi$, at
distance $\pi-\theta$ from it if $j$ is even and $\theta$ if $j$ is odd. 
Furthermore, $\theta_j,\theta_{j+1}$ are less than $2\gamma$ apart
precisely if $\theta$ lies in the boundary wedge of $c_{j+1}$, i.e.\ within
$\gamma$ of $(j+1)\pi$ modulo $2\pi$, and more than $2\gamma$ apart precisely if
$\theta$ lies in its bulk wedge.

\begin{lemma}[Pair identities]
\label{lem:pq-pair}
For $\theta \in (0,\pi)$ and all $r,s \in \mathbb{Z}$, with $\varepsilon_r=+1$ for $r$ odd and $\varepsilon_r=-1$ for $r$ even,
\begin{align}
    Y_r(s,\theta)-Y_{r+1}(s,\theta)
    &\,=\,
    \frac{a_2\cos\beta}{\sin\beta}\,\sin\big(\alpha(\theta_r-s\gamma)\big) \ ,
\label{eq:pq-crossing}
\\
    Y_r(s,\theta-\varepsilon_r\gamma)+Y_{r+1}(s,\theta+\varepsilon_r\gamma)
    &\,=\,
    Y_r(s-1,\theta)+Y_{r+1}(s+1,\theta)
    \nonumber
\\
    &\,=\,
    \frac{a_2\cos\beta}{\sin\beta}\,\sin\big(\alpha((s-1)\gamma-\theta_{r-1})\big) \ .
\label{eq:pq-mixed}
\end{align}
\end{lemma}

\begin{proof}
For $r$ odd we have $\theta_{r-2} = (r-1)\pi-\theta$ and $\theta_{r-1}=(r-1)\pi+\theta$, and for $r$ even $\theta_{r-2}=(r-2)\pi+\theta$ and $\theta_{r-1}=r\pi-\theta$. In both cases the addition theorem gives, for $\theta \in (0,\pi)$,
\begin{equation}
\label{eq:pq-master}
    Y_r(s,\theta)
    \,=\,
    \frac{a_2}{2\sin\beta}\Big(
    \sin\big(\alpha((s-1)\gamma-\theta_{r-2})\big)
    +
    \sin\big(\alpha((s-1)\gamma-\theta_{r-1})\big)
    \Big) \ ,
\end{equation}
since for $r$ odd the left hand side is $Y_r(s,\theta) = A_r(s)c(\theta) = A_r(s)\cos(\alpha\theta)$, and for $r$ even it is $A_r(s) c(\theta-\pi) = A_r(s)\cos(\alpha(\theta-\pi))$.
In the difference $Y_r(s,\theta)-Y_{r+1}(s,\theta)$ the terms containing $\theta_{r-1}$ cancel. Writing $Z=\alpha((s-1)\gamma-\theta_{r-2})$, and using $\theta_r = \theta_{r-2}+2\pi$ together with $2\pi\alpha = \pi+2\beta$,
\begin{equation}\label{eq:pq-crossing-aux}
    Y_r(s,\theta)-Y_{r+1}(s,\theta)
    = \frac{a_2}{2\sin\beta}\Big(\sin Z - \sin(Z-\pi-2\beta)\Big)
    = \frac{a_2\cos\beta}{\sin\beta}\,\sin(Z-\beta) \ ,
\end{equation}
and $Z-\beta = \alpha((s-2)\gamma-\theta_{r-2})$ adds up with $\alpha(\theta_r-s\gamma)$ to $\alpha(2\pi-2\gamma)=\pi$, so the sine functions on the right hand sides of \eqref{eq:pq-crossing} and \eqref{eq:pq-crossing-aux} agree.

For the second equality in \eqref{eq:pq-mixed} insert \eqref{eq:pq-master} twice: setting $B = \alpha((s-1)\gamma-\theta_{r-1})$, the $\theta_{r-1}$-terms of $Y_r(s-1,\theta)$ and of $Y_{r+1}(s+1,\theta)$ are $\sin(B-\beta)$ and $\sin(B+\beta)$ and add up to $2\cos\beta \sin B$, while the remaining two terms involve $\theta_{r-2}$ and $\theta_r=\theta_{r-2}+2\pi$ and cancel each other. 
For the first equality note that $\varepsilon_r$ is chosen precisely so that the
left hand side after substitution of $Y_r$ from \eqref{eq:pq-Y} and $c_r$ from \eqref{eq:pq-channels}, the function $c$ has the arguments $\theta-\gamma$ and $\theta+\gamma-\pi$,
for either parity of $r$, while on the right hand side it has arguments $\theta$ and
$\theta-\pi$. 
For $\theta \in (0,\pi)$ all four lie in $(-\pi,\pi)$, so
$\princ{\cdot}$ acts as the identity and $c$ may be replaced by
$\cos(\alpha\,\cdot\,)$ throughout; using $A_{r+1}(s) = -a_2\cos(\varphi_r(s)-\beta)/\sin\beta$ (recall the notation at the beginning of the proof of Lemma~\ref{lem:pq-channel}), both sides agree by trigonometric identities.
\end{proof}

\begin{lemma}[Dominant channel]
\label{lem:pq-dominance}
Let $1 \le s \le q-1$ and $\theta \in (0,\pi)$, and suppose that $s\gamma$ is not an image of $\theta$. Then channel $n_s-1$ dominates $\cX_s$ at $\theta$ if $\theta_{n_s-1}>s\gamma$, and channel $n_s$ dominates if $\theta_{n_s-1}<s\gamma$; if one of the two channels is absent, the other one dominates. Equivalently: the dominant channel is the number
\begin{equation}
\label{eq:pq-count}
    \nu_s(\theta) \,:=\, \#\big\{\, \text{images of $\theta$ in } (0,s\gamma) \,\big\}
    ~\in~ \{n_s-1,n_s\}
\end{equation}
whenever $1 \le \nu_s(\theta) \le p-1$, and it is $1$ if $\nu_s(\theta)=0$ and $p-1$ if $\nu_s(\theta)=p$.
\end{lemma}

\begin{proof}
Both $\theta_{n_s-1}$ and $s\gamma$ lie in $((n_s-1)\pi,n_s\pi)$, so $|\theta_{n_s-1}-s\gamma|<\pi<2\pi-2\gamma$ by \eqref{eq:abc-bounds}, and by \eqref{eq:pq-alpha-ids} the sign of $\sin(\alpha(\theta_{n_s-1}-s\gamma))$ is that of $\theta_{n_s-1}-s\gamma$. By \eqref{eq:pq-crossing} with $r=n_s-1$, channel $n_s-1$ dominates channel $n_s$ precisely when $\theta_{n_s-1}>s\gamma$. For the reformulation note that the images in $(0,s\gamma)$ are $\theta_0,\theta_1,\dots$ up to $\theta_{n_s-2}$, together with $\theta_{n_s-1}$ if and only if $\theta_{n_s-1}<s\gamma$; so $\nu_s(\theta) \in \{n_s-1,n_s\}$ is the dominant channel, up to the two channels that may be absent.
\end{proof}

As we just saw in the proof, since the $\theta_j$ are strictly increasing in $j$, for
$\theta\in(0,\pi)$ the images of $\theta$ in $(0,s\gamma)$ are
$\theta_0,\theta_1,\dots,\theta_{\nu_s(\theta)-1}$, and if $s\gamma$ is not an
image of $\theta$,
\begin{equation}\label{eq:nus-vs-sgamma}
    \theta_{\nu_s(\theta)-1} \,<\, s\gamma \,<\, \theta_{\nu_s(\theta)} \ ,
\end{equation}
where $\theta_{-1} = -\theta$ in the case $\nu_s(\theta)=0$.

\begin{remark}\label{rem:theta-star-proof}
The anti-Stokes angle \eqref{eq:pq-Stokes} can now be read off. We just saw in the above proof that the two channels of $\cX_s$ exchange
dominance precisely when $\theta_{n_s-1}=s\gamma$. 
For $n_s-1$ even, this becomes $\theta + (n_s-1)\pi = s \gamma$. As both $\theta_{n_s-1}$ and $s\gamma$ are contained in $((n_s-1)\pi,n_s\pi)$, this is equivalent to $[\theta]=[s\gamma]$. 
The channel $n_s-1$ is dominant for $[\theta]>[s\gamma]$.
For $n_s-1$ odd, we get $-\theta + n_s\pi = s \gamma$, or equivalently $[-\theta]=[s\gamma]$. 
The channel $n_s-1$ is dominant for $[-\theta]>[s\gamma]$. As in this case $-\pi<[-\theta],[s\gamma]<0$, we can equivalently say $|[\theta]|<|[s\gamma]|$.
The absolute value in \eqref{eq:pq-Stokes} covers both cases.
\end{remark}

\medskip
\begin{proof}[Proof of Proposition~\ref{prop:pq-solution}]
\emph{Boundary values \eqref{eq:Mpq-asymp-funrel_bndcond}:} 
Recall the convention $D_{(0,1)}=0=D_{(p,1)}$ from \eqref{eq:pq-truncation}.
We have $n_1=1$ and $A_1(1)=0$, so that $\cX_1 = [\Id,0] = 1$. Similarly $n_{q-1}=p$, and, using \eqref{eq:pq-alpha-ids}, $\alpha((q-2)\gamma-(p-2)\pi) = \alpha(2\pi-2\gamma)=\pi$ gives $A_{p-1}(q-1)=0$, so that $\cX_{q-1} = [D_{(p-1,1)},0]$.

\smallskip

\noindent
\emph{Reflection \eqref{eq:Mpq-asymp-funrel_reflect}:} Multiplying \eqref{eq:pq-solution} by $[D_{(p-1,1)},0]$ replaces the labels $D_{(r,1)}$ by $D_{(p-r,1)}$ and leaves the asymptotic exponents unchanged. By Lemma~\ref{lem:pq-channel}\,(iv) the channels of $\cX_{q-s}$ are $n_{q-s}-1 = p-n_s$ and $n_{q-s} = p-n_s+1$, and channel $p-r$ of $\mathsf{S}_{p\pi}\cX_{q-s}$ has exponent $A_{p-r}(q-s)\,c_{p-r}(\theta+\pi p) = A_{p-r}(q-s)\,c(\theta+p\pi-(p-r-1)\pi) = A_r(s)\,c_r(\theta)$, as $c$ is even and $2\pi$-periodic, and $\alpha((q-s-1)\gamma-(p-r-1)\pi) = -\alpha((s-1)\gamma-(r-1)\pi) + \pi$. This proves \eqref{eq:Mpq-asymp-funrel_reflect}.

\smallskip

\noindent
\emph{Hirota equation \eqref{eq:Mpq-asymp-funrel_Hirota}:} All $c_r$ are even and $2\pi$-periodic, so it suffices to verify \eqref{eq:Mpq-asymp-funrel_Hirota} for $2\le s \le q-2$ and $\theta \in (0,\pi)$. We further assume that none of $(s-1)\gamma$, $s\gamma$, $(s+1)\gamma$ is an image of $\theta$; this excludes finitely many $\theta$. Consider the interval
\begin{equation}
    I \,=\, \big(s\gamma-\gamma,\, s\gamma+\gamma\big) \ .
\end{equation}
It has length $2\gamma<\pi$, so it contains at most two images of $\theta$, and if two then they are consecutive and less than $2\gamma$ apart. We shall use the following three counts, which follow directly from \eqref{eq:pq-count}: 
\begin{enumerate}
    \item[{\rm(a)}] $\nu_{s+1}(\theta)-\nu_{s-1}(\theta) \ge 0$ is the number of images of $\theta$ in $I$;
    \item[{\rm(b)}] $\nu_s(\theta-\gamma)-\nu_{s-1}(\theta) \ge 0$ is the number of images of the form $\theta+2m\pi$ in $I$, and
    \\
    $\nu_s(\theta+\gamma)-\nu_{s-1}(\theta) \ge 0$ is the number of images of the form $-\theta+2m\pi$ in $I$.
\end{enumerate}
For (b) note that the images of $\theta-\gamma$ are the points $\theta+2m\pi-\gamma$ and $-\theta+2m\pi+\gamma$; those in $(0,s\gamma)$ correspond to the points $\theta+2m\pi$ in $(\gamma,(s+1)\gamma)$ and $-\theta+2m\pi$ in $(-\gamma,(s-1)\gamma)$. For $\theta>\gamma$ these are the images of $\theta$ of the respective form in $(0,(s+1)\gamma)$ and in $(0,(s-1)\gamma)$, and for $\theta<\gamma$ the first count misses the image $\theta$ while the second gains the point $-\theta$, which cancels. The count for $\theta+\gamma$ is analogous, without any correction.

Let $r_-$ and $r_+$ be the channels dominating $\cX_s$ at $\theta-\gamma$ and at $\theta+\gamma$, and $r_{s-1}$, $r_{s+1}$ the channels dominating $\cX_{s-1}$ and $\cX_{s+1}$ at $\theta$. By Lemma~\ref{lem:pq-dominance} these are $r_{\mp} = \nu_s(\theta\mp\gamma)$ and $r_{s \mp 1}=\nu_{s\mp1}(\theta)$, except for when the corresponding count $\nu$ is zero, in which case we have to take $r_x=1$, or when the count is $p$, in which case $r_x=p-1$ ($x=\mp,s\mp1$); we say that \textit{truncation occurs} in that case. 
With this notation, \eqref{eq:Mpq-asymp-funrel_Hirota} reads
\begin{align}
&\big[\,D_{(r_-,1)}D_{(r_+,1)},\; Y_{r_-}(s,\theta-\gamma)+Y_{r_+}(s,\theta+\gamma)\,\big]
\nonumber \\
&=~ 
[\Id,0] + \big[\,D_{(r_{s-1},1)}D_{(r_{s+1},1)},\; Y_{r_{s-1}}(s-1,\theta)+Y_{r_{s+1}}(s+1,\theta)\,\big] \ .
\label{eq:pq-two-sides}
\end{align}
In the remainder of the proof we will establish \eqref{eq:pq-two-sides} by case distinction. We first consider the case that truncation occurs, and then the case that truncation does not occur. Each of these cases is again split into various subcases.

\smallskip

\noindent
\textbf{Truncation occurs:}
By (a) and (b), if any count is zero, then $\nu_{s-1}(\theta)=0$.
Thus $0<(s-1)\gamma<\theta<\pi$, so by \eqref{eq:pq-ns} we have $n_{s-1}=1$ and
$\cX_{s-1}(\theta)=[\Id,Y_1(s-1,\theta)]$. 
Recall the inequalities \eqref{eq:abc-bounds}.
We have $A_1(s)>0$, since $0<\alpha(s-1)\gamma<\pi$.
As $(s+1)\gamma<\pi+2\gamma<2\pi$, the only possible images in $I$
are $\theta$ and $2\pi-\theta$. Thus (a) and (b) give
\be
 (r_{s-1},r_-,r_+,r_{s+1})=
 \begin{cases}
 (1,1,1,1),&\text{if $I$ contains at most one image, or if $p=2$},\\
 (1,1,1,2),&\text{if $I$ contains both images and $p\ge3$}.
 \end{cases}
\ee
\textit{Configuration $(1,1,1,1)$:}
Set $R=2\cos\beta\,A_1(s)c_1(\theta)$.
By Lemma~\ref{lem:pq-channel}\,(i), the RHS of \eqref{eq:pq-two-sides} is equal to $[\Id,\max\{0,R\}]$.
For the LHS, note that 
Lemma~\ref{lem:pq-channel}\,(ii) gives that $Y_1(s,\theta-\gamma)+Y_1(s,\theta+\gamma)$ is equal to $R$ for $\theta$ in the bulk wedge and equal to $0$ for $\theta$ in the boundary wedge. As $A_1(s)>0$, we also have that $R>0$ (resp.\ $R<0$) in the bulk (resp.\ boundary) wedge. Altogether, also the LHS equals $[\Id,\max\{0,R\}]$.

\smallskip

\noindent
\textit{Configuration $(1,1,1,2)$:}
As images lie in $I$, we have $\theta>(s-1)\gamma$ and $2\pi - \theta < (s+1)\gamma$. This implies
$2\pi-2\theta<2\gamma$, i.e.\ $\theta$ is in the boundary wedge, and so the left side of \eqref{eq:pq-two-sides} is $[\Id,0]$.
Since $0<\theta-(s-1)\gamma<\pi$, the crossing identity
\eqref{eq:pq-crossing} gives $Y_0(s-1,\theta)-Y_1(s-1,\theta)>0$.
Together with Lemma~\ref{lem:pq-channel}\,(iii), for the asymptotic exponent on the right hand side of \eqref{eq:pq-two-sides} we find
\be
 Y_1(s-1,\theta)+Y_2(s+1,\theta)
 =Y_1(s-1,\theta)-Y_0(s-1,\theta)<0 \ .
\ee
Thus the identity term dominates on the right hand side of \eqref{eq:pq-two-sides}, giving $[\Id,0]$ as well. 

\medskip

Instead, if any count equals $p$, then $\nu_{s+1}(\theta)=p$.
Put $s'=q-s$ and $\phi=\theta$ for even $p$, or $\phi=\pi-\theta$
for odd $p$. Reflecting the $p$ images in $(0,p\pi)$
gives $\nu_{s'-1}(\phi)=p-\nu_{s+1}(\theta)=0$.
The already proven reflection relation identifies Hirota at $(s,\theta)$
with that at $(s',\phi)$: each product acquires two factors
$G=[D_{(p-1,1)},0]$, and $G^2=1$, and so the count-$p$ case follows from the count-$0$ case.

\medskip

\noindent
\textbf{No truncation occurs:} If none of the four numbers $\nu_s(\theta\pm\gamma),\nu_{s\pm1}(\theta)$ is zero or $p$, we have $r_\pm = \nu_s(\theta\pm\gamma)$ and $r_{s\pm 1}=\nu_{s\pm1}(\theta)$. We will use the inequality \eqref{eq:nus-vs-sgamma} in the following form:
\begin{equation}\label{eq:theta-bounds-aux}
    \theta_{r_{s-1}-1}<(s-1)\gamma 
    \quad \text{and}\quad 
    (s+1)\gamma < \theta_{r_{s+1}} \ .
\end{equation}
The proof of the case with no truncation is divided into three subcases, corresponding to the number of images in $I$ being $0$, $1$, or $2$.

\smallskip

\noindent
\emph{No image in $I$:}
By (a), $\nu_{s-1}(\theta)=\nu_{s+1}(\theta)$, and together
with (b) this gives
\be
r_{s-1}=r_-=r_+=r_{s+1}=:r_0.
\ee
Comparing to \eqref{eq:theta-bounds-aux}, we see that $\theta_{r_0-1}<(s-1)\gamma$ and $(s+1)\gamma<\theta_{r_0}$. 
In particular,
$\theta_{r_0-1}$ and $\theta_{r_0}$ are more than $2\gamma$
apart and $\theta$ lies in the bulk wedge of $c_{r_0}$.
Moreover, $(r_0-1)\pi<\theta_{r_0-1}<(s-1)\gamma$ and
$(r_0+1)\pi>\theta_{r_0}>(s+1)\gamma$ give
$(s-1)\gamma-(r_0-1)\pi\in(0,2\pi-2\gamma)$,
i.e.\ $A_{r_0}(s)>0$ by \eqref{eq:pq-alpha-ids}. 
Then, by Lemma~\ref{lem:pq-channel}\,(i) and (ii), the asymptotic
exponents in \eqref{eq:pq-two-sides} are
$2\cos\beta\,Y_{r_0}(s,\theta)$ and
$\max\{0,\,2\cos\beta\,Y_{r_0}(s,\theta)\}$.
Since $A_{r_0}(s)>0$ and $\theta$ lies in the bulk wedge,
$Y_{r_0}(s,\theta)>0$. Thus both exponents agree,
$\Id$ is subdominant, and both asymptotic defects are
$D_{(r_0,1)}^2$.

\smallskip

\noindent
\emph{One image in $I$:} Let $\theta_r$ be this image.
By (a), $r_{s-1}=r$ and $r_{s+1}=r+1$, while by (b)
$\{r_-,r_+\}=\{r,r+1\}$. In particular, $r_-\ne r_+$.  By (b), $r_-=r+1$ if $\theta_r$ is of the form $\theta+2m\pi$, i.e.\ if $r$ is even, and $r_-=r$ if $r$ is odd; in other words channel $r$ dominates at $\theta-\varepsilon_r\gamma$ and channel $r+1$ at $\theta+\varepsilon_r\gamma$. Hence \eqref{eq:pq-mixed} shows that the two exponents in \eqref{eq:pq-two-sides} agree,
\begin{equation*}
    Y_{r_-}(s,\theta-\gamma)+Y_{r_+}(s,\theta+\gamma)
    \,=\, Y_r(s-1,\theta)+Y_{r+1}(s+1,\theta)
    \,=\, \frac{a_2\cos\beta}{\sin\beta}\,\sin\big(\alpha((s-1)\gamma-\theta_{r-1})\big) \ ,
\end{equation*}
and this is strictly positive: From \eqref{eq:theta-bounds-aux} we read off $\theta_{r-1}<(s-1)\gamma$ and $\theta_{r+1}=\theta_{r-1}+2\pi>(s+1)\gamma$, so that $(s-1)\gamma-\theta_{r-1} \in (0,2\pi-2\gamma)$. So $\Id$ is subdominant and both leading defects are $D_{(r,1)}D_{(r+1,1)}$.

\smallskip

\noindent
\emph{Two images in $I$:} Write the two consecutive images
as $\theta_{r_0-1}$ and $\theta_{r_0}$.
By  \eqref{eq:nus-vs-sgamma} we have $(s-1)\gamma<\theta_{r_{s-1}}$ and $\theta_{r_{s+1}-1}<(s+1)\gamma$.
By (a), $r_{s+1}-r_{s-1}=2$, and so we must have $r_{s-1}=r_0-1$ and $r_{s+1}=r_0+1$, while
by (b), $r_-=r_+=r_0$.
Since the two images are less than $2\gamma$ apart,
$\theta$ lies in the boundary wedge of $c_{r_0}$. The left hand side of \eqref{eq:Mpq-asymp-funrel_Hirota} is therefore $[D_{(r_0,1)}^2,0]$ by Lemma~\ref{lem:pq-channel}\,(ii). On the right hand side, $\cX_{s\pm 1}$ is dominated at
$\theta$ by channel $r_{s \pm 1} = r_0\pm 1$. Since no truncation occurs,
both channels are present, i.e.\ $2\le r_0\le p-2$.
Then Lemma~\ref{lem:pq-channel}\,(iii) gives
$Y_{r_0-1}(s-1,\theta) + Y_{r_0+1}(s+1,\theta) = 0$
so the right hand side of \eqref{eq:pq-two-sides}
is $[\Id+D_{(r_0-1,1)}D_{(r_0+1,1)},0]$, which is the
left hand side by the fusion rule \eqref{eq:pq-qdet}.

\smallskip

This completes the proof of \eqref{eq:Mpq-asymp-funrel_Hirota}.
\end{proof}

\begin{remark}\label{rem:solutionless2p} As anticipated in Remark~\ref{rem:pq-Tsys}, we also found a solution of the functional relations in the regime $\frac{4p}{3}<q<2p$, differing from the one in the regime $q>2p$ in having two anti-Stokes rays for all of the perturbed defects $D_{(1,s)}(\eta_s \lambda)$. We retain the definitions of $A_r(s)$, $c_r$ and $\nu_s$ from  \eqref{eq:pq-ampl}, \eqref{eq:pq-channels}, and \eqref{eq:pq-count},  replace $a_2>0$ by $a_2 \cos\beta /\sin\beta >0$, then a solution of all the relations \eqref{eq:Mpq-asymp-funrel} is: 
\be \label{eq:pq-positive-count}
    \cX_s(\theta)
=\big[D_{(\nu_s(\theta),1)},\,
       A_{\nu_s(\theta)}(s)c_{\nu_s(\theta)}(\theta)\big]
    \qquad (2\le s\le q-2) \ .
\ee
This solution exhibits exactly two anti-Stokes rays per defect. 
Let us sketch how the argument above gets modified in this case. We reuse the image-count proof of
Proposition~\ref{prop:pq-solution} with the following changes.
The trigonometric equalities in Lemmas~\ref{lem:pq-channel} and
\ref{lem:pq-pair} remain valid, but the boundary-wedge sign assertion in
Lemma~\ref{lem:pq-channel}\,(ii) is no longer used.
The interval $I=((s-1)\gamma,(s+1)\gamma)$ still contains at most two images,
since $2\gamma<2\pi=\theta_{j+2}-\theta_j$.
In the zero-image case, replace the separate positivity of $A_r(s)$ and
$\cos\beta$ by $\cos\beta\,A_r(s)>0$.
The sine-argument bounds in this case and in the one-image case still give
 arguments in $(0,\pi)$, so $a_2 \cos\beta /\sin\beta >0$ gives the required positivity. {Since $\frac\pi2<\beta<\frac{3\pi}{2}$ for $\frac{4p}{3}<q<2p$, this forces $u_2(0) <0$ when $\frac{3p}{2}<q<2p$, and $u_2(0)>0$ for $\frac{4p}3<q<\frac{3p}2$.}
The two-image fusion check is unchanged.
All internal counts lie in $\{1,\ldots,p-1\}$; a zero count in the Hirota
check can only occur at $s=2$, with $\theta>\gamma$.
There the shifted product is $[\Id,0]$ and 
$x_3(\theta)=-a_2 \cos\beta /\sin\beta\sin(\alpha(\theta-\gamma))<0$, so the identity dominates
the right hand side. The count-$p$ case follows by reflection, whose
verification is unchanged.

Finally, writing $x_s$ for the exponent in \eqref{eq:pq-positive-count},
put $f(u)=a_2c(u+\pi)/(2\sin\beta)$.
Then $x_s(\theta)=f(\theta+s\gamma)+f(\theta-s\gamma)$, and for
$3\le s\le q-3$ the scalar T-system reduces to
\be    f(\theta)+\min\{f(\theta+2\gamma),f(\theta-2\gamma)\}=0 \ .
\ee
This follows from $\alpha(2\pi-2\gamma)=\pi$ and $a_2 \cos\beta /\sin\beta >0$.
The corresponding coefficient checks are multiplication by $\Id$ when
one summand dominates and
$D_{(2,1)}D_{(r,1)}=D_{(r-1,1)}+D_{(r+1,1)}$ when they tie.
The boundary T-equations follow from shifted Hirota and reflection.

Whether $u_2(0)<0$ or $u_2(0)>0$ is related to the sign of the perturbing field. 
Recall that the rotation $\mathsf S_\pi$ by $\pi$ is a semiring automorphism and commutes with all angular shifts appearing in the relations.  When the solution exhibited above has $u_2(0)<0$,  the solution where $\mathcal{X}_2$ is positive along the positive real axis can be obtained  by replacing $\mathcal{X}_s(\theta)\to\mathcal{X}_s(\theta+\pi)$.
\begin{figure}
    \centering
    \includegraphics[width=\linewidth]{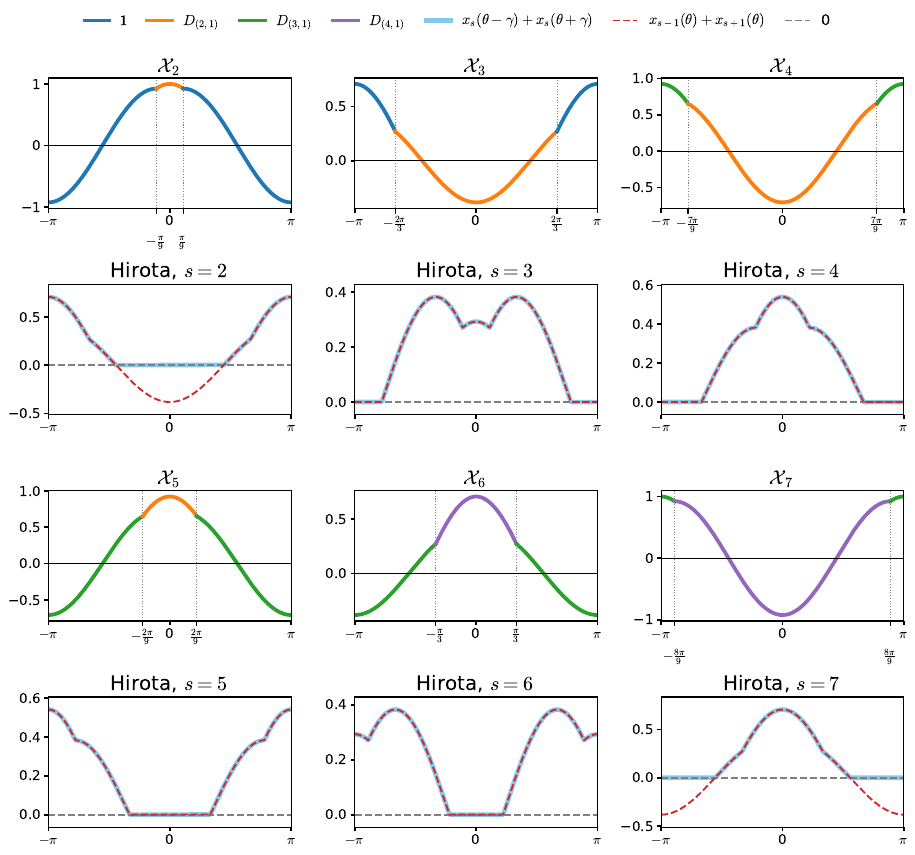}
    \caption{Solution to the functional relations in $M(5,9)$ such that $\mathcal{X}_2$ is positive on the positive real axis. }
\label{fig:M59solandhirota}
\end{figure}
In Figure~\ref{fig:M59solandhirota} we report, as an example, the solution~\eqref{eq:pq-positive-count}   for which  $\mathcal{X}_2$  is positive at $\theta=0$ for the functional relations of  $M(5,9)$. For the actual defect flows in $M(p,q)$ minimal models, the sign of $u_2(0)$ is not arbitrary but fixed by the choice of structure constants as in \cite{phys-paper}.
\end{remark}

\section{Outlook}\label{sec:outlook}

There are many follow-up questions which would be interesting to pursue in the future, and here we list some of them:
\begin{itemize}

\item The main general result of this paper is that under suitable conditions, renormalisation group flows of holomorphically perturbed topological defects are described by a semiring homomorphism between two semirings obtained from Grothendieck rings of certain categories, cf.\ \eqref{eq:classify-hom-classify-rg}. As already discussed in Remark~\ref{rem:categorify}, this strongly hints at a categorified version of this result which states that the above flows are described by a suitable monoidal functor. That would be extremely interesting, as the functor will contain information about compatibility of associators -- i.e.\ it will allow one to compare various anomalies between the UV and IR of the flows. This information is not visible if one just looks at the underlying semiring structure.

\item In the Virasoro minimal model examples in Section~\ref{sec:minmod-13}, we only considered A-series (diagonal) models, and not D- and E-type minimal models, and we only considered perturbations by a holomorphic (1,3)-field. We expect similar constraints from functional relations for the other generically integrable perturbations, namely by (1,2)-, (2,1)-, and (1,5)-fields. For (1,2)-perturbations we report results in the Ising and Tricritical Ising models in \cite{phys-paper}, but a comprehensive study of minimal models is still open.

\item So far our methods rely on convergence of the perturbed defect operators without the need for regularisation and counterterms. For example, in the case of minimal models with (1,3)-perturbations this rules out all unitary models. It is known that counterterms can indeed modify the asymptotic behaviour (see \cite{Gaiotto:2020fdr,Ambrosino:2026kfu} and the end of Section 4 of \cite{Ambrosino:2025pjj}).
It would be interesting to develop a version of the formalism presented here (and in \cite{Ambrosino:2025myh,Ambrosino:2025pjj}), which can treat perturbations which need regularisation.

\item In a rational CFT, which is what we have been considering in this paper, the topological defects that are explicitly accessible are those which preserve the full rational chiral symmetry. Apart from minimal models, it is not known what all topological defects are. Holomorphically perturbed defects flow to topological defects, and we expect that these endpoints will typically no longer preserve the full rational chiral symmetry. The semiring homomorphisms studied here hence offer a window into the fusion rules of such more general topological defects. A particularly interesting case would be that of perturbed topological defects in WZW models. 

\item In the present paper we only consider defect perturbations with unperturbed ambient CFT. As explained in \cite{Buecher:2012ma,Ambrosino:2025myh,Ambrosino:2025pjj}, the algebraic structure describing such perturbations is a special case of the one describing compatible bulk and defect perturbations. It would therefore be very interesting to generalise the algebraic description of defect flows developed in this paper to such combined bulk and boundary flows. This would allow us to address the question asked in Figure 3 of \cite{Ambrosino:2025pjj}: Does including holomorphically perturbed defects make it possible to detect more of the topological symmetry present in the IR fixed point of a bulk flow?
\end{itemize}

\paragraph*{Acknowledgments.}
We thank Michael Borinsky, Kevin Costello, Alexei Davydov, Davide Gaiotto, Nikita Grygoryev, Alastair King, Stefano Negro, Cris Negron, and Gregor Schaumann for interesting discussions. 
AK, IR and GW would like to thank the Isaac Newton Institute for Mathematical Sciences, Cambridge, for support and hospitality during the programme \textit{Quantum field theory with boundaries, impurities, and defects}, where work on this paper was undertaken
and supported in part by EPSRC grant EP/Z000580/1.
FA, AK and IR would like to thank GGI and the organisers of the programme \textit{Defects and Extended Excitations in Quantum Field Theory, Quantum Matter and Statistical Models} for the hospitality where part of this work was conducted.\\
I.R.\ acknowledges support by the Deutsche Forschungsgemeinschaft (DFG, German Research Foundation) under Germany's Excellence Strategy - EXC 2121 ``Quantum Universe'' - 390833306 and the Collaborative Research Centre - SFB 1624 ``Higher structures, moduli spaces and integrability'' - 506632645. Research at Perimeter Institute is supported in part by the Government of Canada through the Department of Innovation, Science and Economic Development Canada and by the Province of Ontario through the Ministry of Colleges and Universities.

\paragraph*{Statement of AI use.}
At various stages in the preparation of this manuscript the authors have employed generative AI (gAI) tools (Claude Opus 5, Fable 5.1; GPT 5.6 Sol, 6 Astra) to aid the generation of code aimed to perform various tests on some specific examples,  as well as to proof-read the paper, and to improve  some figures. All the changes and corrections suggested by these gAI tools have been verified and implemented by the authors, and we take full responsibility for the content of this paper.

\bibliographystyle{utphys}
\begin{spacing}{0.7}
\begingroup
\hypersetup{urlcolor=blue}
\bibliography{refsmath}
\endgroup\end{spacing}

\end{document}